\documentclass[a4paper, 10pt, conference]{ieeeconf}      
\usepackage{FG2021}

\FGfinalcopy 

\IEEEoverridecommandlockouts                              
\def\FGPaperID{****} 

\title{\LARGE \bf
Generating Master Faces for Dictionary Attacks with\\ a Network-Assisted Latent Space Evolution}

\usepackage{amsmath,graphbox}
\usepackage{algorithm}[noend]
\usepackage{algpseudocode}
\usepackage{tikz}
\usepackage{multirow}
\usepackage{bbm}
\DeclareMathOperator*{\argmax}{arg\,max}
\DeclareMathOperator*{\argmin}{arg\,min}

\usetikzlibrary{shapes.geometric, arrows}

\usepackage{subfigure}
\usepackage{booktabs} 
\usepackage{thm-restate}

\usepackage{xcolor}
\usepackage[pagebackref=false,breaklinks=true,colorlinks,bookmarks=true]{hyperref}

\usepackage{amsthm}
\usepackage{dsfont}
\usepackage{amssymb}
\usepackage{makecell}

\def\G{{\mathcal{G}}}
\def\M{{\mathcal{M}}}

\author{\parbox{16cm}{\centering
    {\large Ron Shmelkin$^{1*}$, Tomer Friedlander$^{2*}$ and Lior Wolf$^{1}$}\\
    {\normalsize
    $^1$The Blavatnik School of Computer Science, Tel Aviv University\\
    $^2$The School of Electrical Engineering, Tel Aviv University\\}}
    \thanks{*Equal contribution}
}

\begin{document}

\ifFGfinal
\thispagestyle{empty}
\pagestyle{empty}
\else
\author{Anonymous FG2021 submission\\ Paper ID 115\\}
\pagestyle{plain}
\fi
\maketitle

\begin{abstract}

A master face is a face image that passes face-based identity-authentication for a large portion of the population. These faces can be used to impersonate, with a high probability of success, any user, without having access to any user-information. We optimize these faces, by using an evolutionary algorithm in the latent embedding space of the StyleGAN face generator. Multiple evolutionary strategies are compared, and we propose a novel approach that employs a neural network in order to direct the search in the direction of promising samples, without adding fitness evaluations. The results we present demonstrate that it is possible to obtain a high coverage of the LFW identities (over 40\%) with less than 10 master faces, for three leading deep face recognition systems.
\end{abstract}

\section{INTRODUCTION}

In dictionary attacks, one attempts to pass an authentication system by sequentially trying multiple inputs. In real-world biometric systems, one can typically attempt only a handful of inputs before being blocked. However, the matching in biometrics is not exact, and the space of biometric data is not uniformly distributed. This may suggest that with a handful of samples, one can cover a larger portion of the population.

Indeed, as we show for the case of face recognition, there are face images that would be authenticated successfully, using state of the art face recognition systems and acceptable match thresholds, of a large portion of users in a given dataset. In some cases, a single face can cover more than 20\% of the identities in LFW. Following previous work on fingerprints~\cite{roy2017masterprint}, we term such faces ``master faces'', due to the analogy to master keys.

The process of master face generation employs a realistic face generator network, in our case StyleGAN~\cite{styleGan}. Since the objective function, i.e., the number of identities that are similar enough to the face image we optimize, is non-differentiable, black box optimization methods are utilized. Unsurprisingly, we find that optimization methods that are tailored for high dimensional data outperform other methods. We then propose a novel method, in which a trained neural network is used to predict the competitiveness of a given sample.

When attempting to cover an even larger number of faces, we advocate for a greedy method in which one repeats the master face generation process sequentially, each time to cover the identities that were not covered by the previously generated faces. Using such a greedy process, we obtain coverage of 40\%-60\% with nine images (the focus on nine images arises from a different experiment, in which the samples in the latent space are clustered).

The experiments are conducted using three different deep face recognition systems, each with its own processing pipeline, training dataset, objective, and similarity measure. The similarity threshold used is a conservative one that is based on obtaining a standard FAR value of 0.001, or, when available, is the one prescribed by the method itself. 

Overall, our results indicate that performing a dictionary attack of face authentication systems is feasible at high success rates. This is demonstrated for multiple face representations and explored with multiple, state-of-the-art optimization methods.

\section{Related work}
\label{sec:preliminaries}
\paragraph{Face Verification}
\label{subsec:face_verification}
Face Verification is the task of comparing two face thumbnail images and determining whether or not they belong to the same subject. In the past few years, deep neural network (DNN) approaches have dominated the field, e.g.~\cite{fd:dlib09,fd:Facenet,fd:sphereface,fd:deepface}. These DNNs learn embedding face representations, such that the distance under a certain similarity metric (e.g. cosine distance, euclidean distance etc.), corresponds to the face similarity.
Different loss functions have been presented for the purpose of achieving such a behavior, e.g., the triplet loss~\cite{fd:Facenet}, angular loss \cite{ang_loss} and contrastive loss~\cite{contr_loss}. 

Face verification is often assessed by considering the False Acceptance Rate (FAR) and the False Rejection Rate (FRR). FAR is the percentage of embedding representations in which different subjects were incorrectly authorized as the same person, FRR is the percentage of embedding representations in which embedding representations that belong to the same subject were incorrectly rejected. There is a trade-off between these two metrics, which is balanced by a matching distance threshold $\theta$.

\paragraph{Face Image Generation using GAN}
The Generative Adversarial Network (GAN)~\cite{GAN} is a machine learning framework in which two neural networks, the generator $G$ and the discriminator $D$, are trained in the form of min-max game. Given the dataset distribution $X$ and $z\sim Z$ latent input vector, the goal of $G$ is to generate samples $G(z)\sim X$. Given a data sample $x$, $D$ is trained to determine whether $x\sim X$ or $x\sim G(z)$, where $z\sim Z$. On the other hand, $G$ is trained to create samples for which $D$ fails to discriminate between the two. The quality and resolution of data samples generated using GANs have been constantly improving. One of the most notable lines of work is that of the StyleGAN Face Image Generation (FG)~\cite{styleGan,Karras2019stylegan2}.

StyleGan~\cite{styleGan} applies a mapping neural network $f: Z\rightarrow W$ to convert the latent vector $z$ into a more disentangled latent representation $w$ that separates content and style. It is then fed to each convolutional layer of generator through the adaptive instance normalization (AdaIN)~\cite{AdaIn}. This allows better control of the image synthesis process and results in the generation of quality and more detailed images.

\paragraph{Master sample attacks}
\label{sec:previous_work}

A parallel work that we became aware of post-publication~\cite{nguyen2020generating} attempts to generates faces that are similar to a large portion of the faces in a given dataset using StyleGAN but with a different evolutionary strategy, which is one of our baselines (CMA-ES). In the field of finger print verification, Roy et al.~\cite{roy2017masterprint} have suggested to exploiting the lower quality of partial fingerprint systems to generate fingerprint templates that can be matched to a large number of users' fingerprints, without any knowledge of the actual user. This generation does not produce an actual image, which is required for the actual match.

To produce such an image, Bontrager et al. \cite{deepmasterprints} used a deep neural network. Our work performs a similar task for faces. In comparison to fingerprints, faces are characterized by a larger latent space, and we develop an optimizer that is better suitable for large dimensions. In a very extensive set of experiments, we explore various state of the art solutions and demonstrate the effectiveness of our method.

\paragraph{Evolution Strategies}\label{subsec:evolution_strategies}
Covariance Matrix Adaptation Evolution Strategy (CMA-ES)~\cite{hansen2003reducing} is a type of an iterative evolutionary algorithm extensively used for solving non-convex continuous optimization problems with no usage of gradient information. It is considered one of the most powerful stochastic numerical optimizers used for solving difficult black-box problems~\cite{varelas2018comparative}.

At each iteration, CMA-ES generates a population of $\lambda$ candidate solutions, by sampling a multivariate Gaussian distribution, whose mean vector and covariance matrix were estimated in the previous iterations. The $\mu$ fittest candidates in terms of the objective function are selected out of the current population and are used to adapt the last estimate of the model's learnable parameters.

The quadratic time and space-complexity of CMA-ES limit its applicability for high-dimensional problems~\cite{varelas2018comparative} and its performance degrades significantly~\cite{omidvar2010comparative} in larger dimensions. 
Limited-Memory Matrix Adaptation Evolution Strategy (LM-MA-ES)~\cite{loshchilov2017limited} 
reduces the time and storage complexity to $\mathcal{O}(n\log{n})$. This algorithm is reported to perform well on high-dimensional variants of well established benchmark problems. This makes LM-MA-ES an appropriate choice for solving the high-dimensional black-box optimization problem in the current work.

Evolutionary algorithms assisted by an additional machine learning model have been presented in the literature, e.g., surrogate-assisted CMA-ES algorithms \cite{pitra2017overview}. { Some models train a regression model online with the evolutionary algorithm in order to learn the fitness function of the black-box problem \cite{surr_cmaes}. Instead of computing the expensive fitness function itself for all candidates, the fitness scores of some candidates are predicted by this regression model, and as a result, the number of evaluation calls is decreased. Models can be assisted by the surrogate for another purpose of predicting if a given candidate is going to be competitive in terms of its fitness score and evaluate only such candidates. ACM-ES \cite{loshchilov2010comparison} is an example of such a model, which uses a comparison-based surrogate, instead of a regression model.

In our work, we train a neural classifier coupled with the evolutionary algorithm in order to predict a given candidate's probability to be fitter (lower fitness) relative to candidates generated in the few last iterations, without evaluating its fitness score explicitly.
}

\section{METHODOLOGY}
\label{sec:method}
Given a dataset $D=\{x\in \mathbbm{R}^{w \times h\times c}\}$ which contains face images (each of size $w\times h$ and with $c$ channels) with a single image per subject, a deep convolutional face embedding model $\M(x)\in \mathbbm{R}^d$ and a matching threshold $\theta$, we define the \textbf{Master Face $\mathbf{x_{mf}}$} as following:
\begin{equation}\label{eq:x_mf}
\mathbf{x_{mf}} = \argmax_{x_{mf}} \sum_{x \in D}f(\M(x_{mf}),\M(x)),\theta)  
\end{equation}
The binary function $f:\mathbbm{R}^{2d+1} \rightarrow \{0,1\}$ compares two embedding vectors and assigns a value of $1$, if the embeddings are similar by some similarity by the amount determined by the threshold $\theta$. 
We note that our method searches for an optimal face image $x_{mf}$ and for an optimal embedding 
\begin{equation}\label{eq:c_mf}
  c = \argmax_{c} \sum_{x \in D}f(c,\M(x),\theta)  
\end{equation}.

As we show in Sec.~\ref{subsec:cent_inv}, it is not possible to invert $\M$ effectively and obtain 
$\M^{-1}(c)\in \mathbbm{R}^{w \times h\times c}$ which achieves both high visual quality and a high coverage. 

Instead, we suggest to optimize the Face Generator's latent vector $z$ based on its matching score, in order to find a better representation in the image space.
\begin{equation}\label{eq:z_optpre}
  \mathbf{z_{opt}} = \argmax_{z} \sum_{x \in D}f(\M(\G(z)),\M(x)),\theta)
\end{equation}

Since, by convention, evolutionary algorithms minimize the score instead of maximizing it, the above matching score is altered by subtracting it from the total number of face images in the training set $n=|D|$. Moreover, the score is normalized to be in $[0,1]$.

\begin{equation}\label{eq:z_opt}
  \mathbf{z_{opt}} = \argmin_{z} \frac{1}{n} (n-\sum_{x \in D}f(\M(\G(z)),\M(x),\theta))
\end{equation}

\begin{figure}[t]
  \centering
    \includegraphics[scale=0.42]{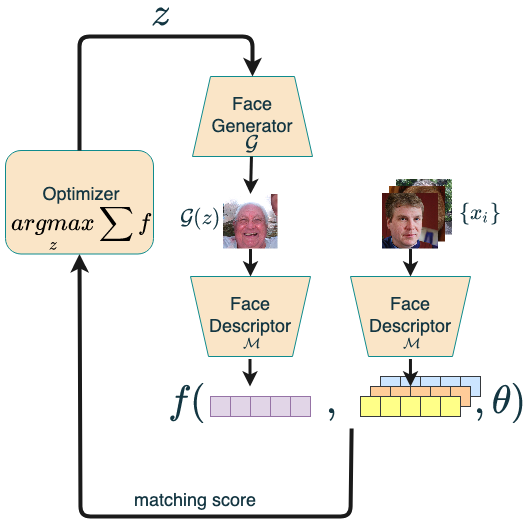}
\caption{An overview of the generation flow for finding a master face.}
\label{fig:alg_flow}
\end{figure}

An overview of the method is given in Fig.~\ref{fig:alg_flow}. Initially, the dataset $D$ is transformed to the embedding space. Denote the embedded dataset by $\widehat{D}=\{\M(x):~\forall x \in D\}$. An evolutionary algorithm is used to find the optimal latent vector $\mathbf{z_{opt}}$, which solves the optimization problem defined in Eq.~\ref{eq:z_opt}. At each iteration of the optimization algorithm, a set of candidate solutions is generated and evaluated by the fitness function. 
Given a candidate latent vector $z$, an image corresponding to $z$ is generated by applying the face generator $\G$ on $z$. The face is then extracted from the generated image $\G(z)$ and is embedded using the face description model $\M$.

\subsection{The evolutionary algorithm}\label{subsec:opt_imp}

\begin{figure*}[]
\centering
\begin{minipage}[b]{0.425\linewidth}
    \centering
    \includegraphics[scale=0.34]{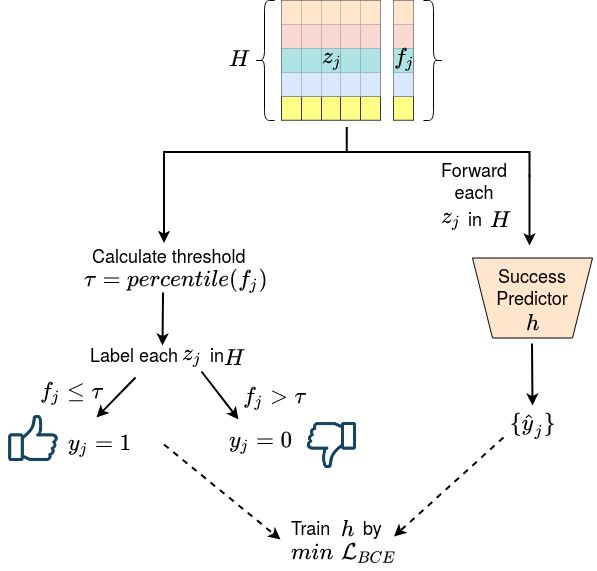}\label{fig:training_predictor}
    \centerline{(a)}
\end{minipage}
\begin{minipage}[b]{0.425\linewidth}
    \centering
    \includegraphics[scale=0.34]{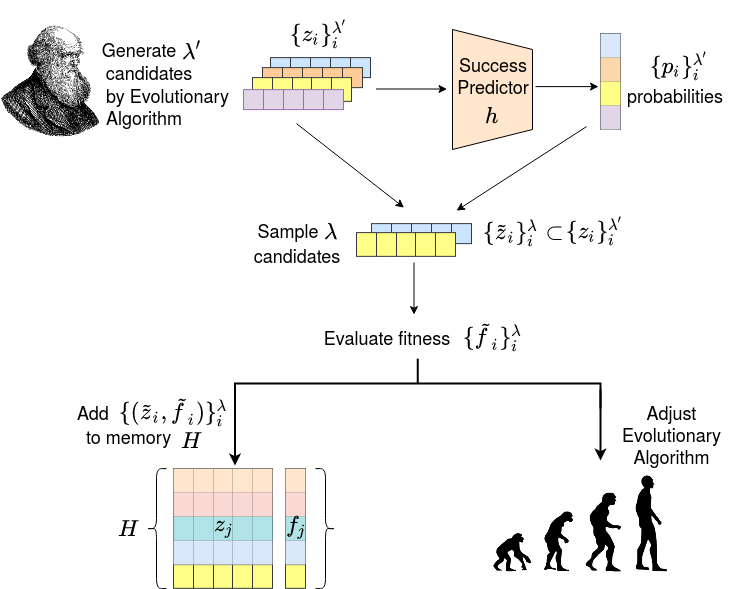}     \label{fig:filtering_by_predictpr}
     \centerline{(b)}
\end{minipage}
\caption{(a) Training the success predictor using samples stored in memory $H$. (b) Filtering generated candidates by the success predictor and updating $H$.}
\label{fig:success_predictor_flow}
\end{figure*}

We present an evolutionary algorithm that is coupled with a neural predictor. The latter estimates the probability of a given candidate to be {\color{black}fitter (lower fitness score)} than the p-percentile of candidates generated during the few last iterations. {These estimated probabilities are used to obtain a set of promising candidates for the purpose of enhancing the latent space optimization.} 

The LM-MA-ES evolutionary algorithm, which is known to perform well on high-dimensional black box optimization problems, is selected as the baseline method for the optimization problem in Eq.~\ref{eq:z_opt}. At each iteration, the original variant of LM-MA-ES generates a population of $\lambda$ candidate solutions and the $\mu$ fittest candidates are selected for adjusting the parameters of the model. Since the candidates are generated randomly by sampling the probability density learned by the LM-MA-ES model, some of the candidates may be unsuccessful in terms of their fitness score. In order to obtain a set of successful candidates in a higher probability, we suggest training a neural classifier for predicting {the probability of a given candidate to be fitter than the p-percentile of stored samples from the few last iterations.} We use this classifier, named the Success Predictor, to filter the more probable candidates to be promising out of a larger generated set, without the need to explicitly evaluate their fitness scores.

Let $h$ be a neural classifier, which gets a vector $z$ as an input. The training process of classifier $h$ is described in Fig \ref{fig:success_predictor_flow}(a) and is performed online in the end of each iteration of the evolutionary algorithm. The training samples are candidates generated during the last few iterations of the evolutionary algorithm, which are stored in a finite memory $H$ with a capacity of $C$ samples. Each new evaluated candidate and its fitness score are added to the memory $H$. {When the memory $H$ becomes over-occupied, excessive samples are removed out of it randomly, without removing the best candidate.} After each update of the memory $H$, each sample $z_i$ in $H$ is defined to be a successful candidate (labeled +1), if its fitness score is better than the p-percentile of the fitness scores of all samples stored in the memory. Otherwise, it is defined as an unsuccessful candidate (labeled 0). Let $(z_i, y_i) \in H$ be the i-th sample in the memory and its corresponding label. The classifier is trained to successfully classify the class of the training samples, by minimizing the binary cross entropy loss between the predicted class and the the true class:

\begin{equation}
\mathcal{L}_{BCE} = -\frac{1}{|H|}\sum_{(z_i,y_i) \in H} [y_i log(h(z_i)) + (1-y_i)log(1-h(z_i))]
\end{equation}

The classifier $h$ is used in the inference mode during the generation step of the evolutionary algorithm, as described in Fig. \ref{fig:success_predictor_flow}(b). Instead of generating only $\lambda$ candidates, $\lambda' > \lambda$ candidates are generated. Prior to evaluating the candidates' fitness values, each candidate is forwarded through the classifier $h$ in order to obtain a {score representing its probability of belonging to the successful class of samples. The scores of all $\lambda'$ candidates are concatenated to a single vector, which is afterwards converted to a probability vector by applying the Softmax operator.} Then, only $\lambda$ candidates are selected to be evaluated, by sampling them out of the total $ \lambda'$ candidates according to the obtained probability scores. The rest of the evolutionary algorithm steps remain unchanged and are performed on these $\lambda$ candidates, in particular, the fittest $\mu$ candidates are chosen out of this already filtered set.

{After evaluating the fitness scores of the new filtered candidates, we can retrospectively determine if these candidates {were correctly classified by the classifier $h$ earlier. In particular, if a candidate which obtained a prediction score higher (lower) than 0.5 by $h$, is indeed (less) fitter than the p-percentile of the samples in the memory $H$.} Correspondingly, the performance of $h$ is tested, by calculating the averaged prediction accuracy on these new candidates. If this averaged accuracy becomes lower than a predefined threshold, $\tau_{acc}$, for a predefined number of $T$ iterations, the learnable parameters of $h$ are re-initialized randomly. Moreover, we use a predefined warm-up period of iterations at the first generations of the evolutionary algorithm for which the classifier is only trained, but is not used for filtering new candidates.}

\subsection{Dataset coverage}\label{subsec:coverage}
Given a set of face images $D$, we strive to find a minimal set of master face images $S=\{x\in \mathbbm{R}^{w\times h\times x}\}$ such that for as many subjects $x\in D$ as possible, there is a least one $x'\in S$ such that $f(\M(x),\M(x'),\theta)$ is one.

A natural choice is to divide the embedding space into clusters, e.g., by using KMeans~\cite{kmeans}, and optimize each member of $S$ to cover a different cluster, e.g., by considering the center of each cluster. However, as shown in~\ref{subsec:cent_inv} it is difficult to invert the cluster centroid point to the image space. Thus, we propose a greedy approach Alg.~\ref{alg:greedy} to find such set of master face images. After each iteration, the dataset $D$ is reduced by already covered face images, namely, images that were already incorrectly authorized {with the generated master face of the current iteration}, are removed from the dataset and the next search iteration is performed on the updated dataset. {\color{black}Therefore, the current generated master face might cover some face images, which were already covered by previously generated master faces, while the opposite direction cannot occur. However, such an intersection does not result in counting covered face images twice in the total coverage percentage calculation, since the current coverage iteration is performed on the reduced dataset.} 

The method uses the function $find\_matched$, which given a face image $x_{mf}^i$ and the dataset $D$ $find\_matched$ returns the set of faces from $D$ that are incorrectly authorized with $x_{mf}^i$. We set the limit of the number of iterations in Alg.~\ref{alg:greedy} to be 
the number of clusters that cover most of the $\widehat{D}$, as presented in Sec.~\ref{subsec:cent_inv}.
\begin{algorithm}[]
  \begin{algorithmic}[1]
    \Function{find coverage}{$\G, D, \M, \theta, max\_iter $}
    \State $center\_imgs \gets []$
    \For {$i = 1..{ max\_iter}$}
        \State $x_{mf}^i \gets\textrm{master face generation}(\G, D, \M, \theta)$ 
        \State $center\_imgs \gets center\_imgs \bigcup x_{mf}^i$ 
        \State $D \gets D \backslash\ find\_matched(D,x_{mf}^i)$
    \EndFor
    \State\Return $center\_imgs$
\EndFunction
    \end{algorithmic}
\caption{$greedy$-Coverage Search}\label{alg:greedy}
\end{algorithm}
\section{Experiments}
\label{sec:experiments}
We evaluated our method with three different CNN based, face descriptors: FaceNet~\cite{fd:Facenet}, SphereFace~\cite{fd:sphereface}, and Dlib~\cite{fd:dlib09}. Each face descriptor is equipped with its own combination of an architecture, a similarity metric and a loss function, thus providing additional validation of our method.

The FaceNet implementation employs an Inception-ResNetV1~\cite{resnetv1}. It obtains a $0.9905$ accuracy on the LFW face verification benchmark. The SphereFace face descriptor is implemented as a deep neural network with 20 convolutional layers. An accuracy of $0.9922$ is measured on the LFW face verification benchmark. In both cases, the face regions are extracted and aligned using MTCNN~\cite{mtcnn}, the embeddings similarity is measured by cosine distance, training takes place on the CasiaWebface~\cite{ds:casia} dataset. The embedding dimension is 512. The thresholds for the FaceNet and SphereFace were chosen to preserve $FAR\sim0.001$.

We also use the Dlib face descriptor implementation~\cite{fd:dlib09}, which employs a ResNet with 29 convolutional layers. This architecture is ResNet-34~\cite{he2015deep} with a reduced number of layers, in which the number of filters per layers is reduced by half. The LFW accuracy reported is $0.9938$. The model trained on the face scrub~\cite{ds:scrub}, VGG~\cite{ds:vgg} and additional face images scraped from the internet. The model embeds the face images in $\mathbbm{R}^{128}$ and employs the euclidean distance with a predefined recommended threshold $\theta=0.6$. The face detection and alignment processes use Dlib's detector~\cite{dlib_fd}.


As a face image generator $\G$ StyleGAN~\cite{styleGan} is used, in which $\G: \mathbbm{R}^{512} \rightarrow \mathbbm{R}^{1024\times1024\times3}$. We use the pre-trained StyleGAN model, trained with FFHQ~\cite{styleGan} dataset.

{
The architecture of Success Predictor $h$ is a feed-forward neural network with three fully connected layers, whose output dimensions are 256, 128 and 1 neurons respectively. The first two hidden layers and the output layer are followed by the ELU \cite{clevert2015fast} activation function and a Sigmoid respectively. The first hidden layer uses the BatchNorm regularization layer \cite{ioffe2015batch}, prior to the activation function. The network is trained using the ADAM \cite{kingma2014adam} optimizer with a learning rate of $0.001$ on mini-batches of size 32. The population size $\lambda$ is set to $22$ according to the calculation in \cite{loshchilov2017limited} and $\lambda'$ is set to 1,000 candidates. The threshold defining the promising class is set to the 5-th percentile. The capacity $C$ of the memory $H$ is set to 5000 samples. The parameters $\tau_{acc}$ and $T$ are set to 0.6 and 20 respectively. The warm-up period of $h$ is set to the first 5\% of the evolutionary algorithm iterations.
}

The method is evaluated on the LFW~\cite{ds:LFWTech} dataset. One image per subject is used. In Sec.~\ref{subsec:gf_eval}, we evaluated our approach on the predefined split $|D_{train}|/|D_{test}| =4038/1711$ while in Sec.~\ref{subsec:dataset_covarage} and Sec.~\ref{subsec:cent_inv} we perform a coverage search on the whole dataset $|D|=5749$.

Note that, since we compare a new generated face to the set of different subjects, we measure the matching score as Mean Set Coverage (MSC). 
$$MSC=\frac{\# \textrm{of incorrect authorization }}{|D|} * 100$$

\subsection{Cluster centroid inversion}\label{subsec:cent_inv}

We first explore the alternative method of clustering the data in the embedding space and then trying to convert the prototypes found to images. This also provides an estimate of the number of target master faces that one can use, in an ideal case in which one can use a dictionary attack with embeddings and not with real-world faces.

In this experiment, we focus on the Dlib face descriptor. An embedding dataset $\widehat{D}$ is created and then clustered using KMeans~\cite{kmeans}. With nine clusters $\sim 91.63\%$ of $\widehat{D}$ was covered, i.e. in the embedding space we were able to find nine center points (cluster centroids) such that over 90\% of the samples in $\widehat{D}$ were in the euclidean distance of less than $\theta=0.6$ away from at least one of these centroids.

Since the actual master-based dictionary attack requires using images, we trained a neural network to generate a face image from the Dlib's embedding representation. Specifically, we trained a neural network face generator $G:\mathbbm{R}^{128}\rightarrow\mathbbm{R}^{64x64x3}$ with five layers. The layers consist of one linear layer and four de-convolutional, similar to the generator's architecture presented in the DC-GAN~\cite{dcgan}. The generator $G$ was trained on the Dlib's embedding representation of the FFHQ\cite{styleGan} dataset. It is trained to minimize MSE loss between the generated image and the original one.
\begin{equation}
\mathcal{L}_{MSE} = \frac{1}{|D_{FFHQ}|}\sum_{x\in D_{FFHQ}}(x - G(\M(x)))^2 
\end{equation}

\begin{figure}[t]
\centering
\begin{tabular}{@{~}c@{~}c@{~}c@{~}c@{~}c@{~}}
    \includegraphics[width=1.3cm]{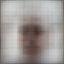} & \includegraphics[width=1.3cm]{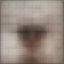} & \includegraphics[width=1.3cm]{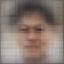} &      \includegraphics[width=1.3cm]{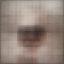} &   
    \includegraphics[width=1.3cm]{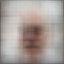} \\ \includegraphics[width=1.3cm]{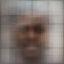} & \includegraphics[width=1.3cm]{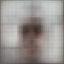} &      \includegraphics[width=1.3cm]{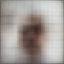} &   
    \includegraphics[width=1.3cm]{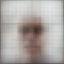} \\
\end{tabular}
\caption{Clusters center in latent space, converted to image space}\label{fig:cent_images}
\end{figure}

While $G$ performs well on embeddings of the real image, the visual quality in the case of cluster centroids is unsatisfactory, see Fig.~\ref{fig:cent_images}. Moreover, only two out of the nine generated faces are detected as faces by the Dlib detector. These generated faces are also ineffective as master faces and the highest MSC score of any of the nine is only $\sim2\%$.

\subsection{Experiments for one Master Face image}\label{subsec:gf_eval}

The generic scheme we present for optimizing a single image, as depicted in Fig.~\ref{fig:alg_flow}, is employed for recovering a single Master Face image. For each face descriptor, each black box optimization method was trained on $D_{train}$ for five runs. Each run differs in its initial random seed. Out of the face images obtained from all five runs, the one which achieved the highest MSC score on the training dataset, was chosen to be reported as the master face obtained for this face descriptor on the training dataset. This way, we eliminate some of the sensitivity of such optimization methods to the random seed, without using any test data.

For the sake of making a fair comparison, all algorithms were trained for an equal number of fitness function calls (26400) with the same set of five seeds. The number of fitness function calls was chosen based on the observation that longer training processes resulted in a negligible improvement. After this selection process, each master face was then evaluated on the test dataset, $D_{test}$. 

We compare the performance of our method, which is denoted by LM-MA-ES (LME) + Success Predictor to the following baselines: (i) A random Search algorithm was used to set the baseline results. We used the version implemented in the Nevergrad package \cite{nevergrad}. (ii) LM-MA-ES~\cite{loshchilov2017limited} is the high-dimensional variant of CMA-ES~\cite{hansen2003reducing}, which our method is based on. (iii) For completeness, the original CMA-ES algorithm is compared, although despite not being suitable for high dimensions. We use the implementation from the pycma package~\cite{hansen2019pycma}.

Differential Evolution (DE) is another very successful family of evolutionary algorithms. In addition to  original (iv) DE \cite{storn1997differential}, we compare to the newer variants (v) LSHADE-RSP~\cite{stanovov2018lshade} and (vi) IMODE~\cite{sallam2020improved}, which achieved the second place in the IEEE CEC'2018 competition and the first place in the IEEE CEC'2020 competition respectively. 

(vii) NGOpt~\cite{liu2020versatile} is an algorithm which automatically selects the right evolutionary algorithm to be trained out of a set of several algorithms, according to the properties of the optimization problem. NGOpt is implemented in the Nevergrad package.
{ (viii) ACM-ES \cite{loshchilov2010comparison} is a surrogate-assisted CMA-ES variant with a comparison-based surrogate (ranking SVM). Similar to our approach, it selects to evaluate only a subset of candidates predicted to be promising according to the surrogate model. The initial larger population and the warm-up period is set like the model assisted by the Success Predictor. We use the implementation from the BOLeRo package \cite{bolero}. }
(ix) lq-CMA-ES \cite{surr_cmaes} is a surrogate-assisted CMA-ES variant, which is assisted by a linear-quadratic regression model (LQ) { It uses the surrogate for decreasing the number of evaluations and not for predicting promising candidates.} We use the implementation from pycma \cite{hansen2019pycma}. (x) For the purpose of comparing our model more directly to another surrogate-assisted model, we add this LQ regression model to the LM-MA-ES algorithm and train it online as done in lq-CMA-ES. This variant is named LME + LQ-Filter. Similar to our method, a larger set of $\lambda'>\lambda$ candidates is generated by the LM-MA-ES. Their fitness score is then predicted by the LQ model and their success probability is estimated by applying the Softmax function on the negated predicted fitness scores. The estimated probabilities are used for sampling a smaller set of $\lambda$ candidates. For the purpose of adjusting the LM-MA-ES's parameters fairly, the actual fitness values of all selected candidates are evaluated, instead of using the predicted values by the LQ model.

Table~\ref{table:ea_comp} presents a comparison between the different optimization algorithms for the master face generation task, in terms of the MSC score. It can be observed that our LM-MA-ES+Success Predictor achieved the highest result among all compared algorithms on the train set for two out of three face descriptors, when considering either the train set or the test set. When not leading, our LM-MA-ES assisted by the Success Predictor achieved the second best result. In comparison to the the original LM-MA-ES, LM-MA-ES assisted by the Success Predictor achieved better results on all three training sets and on two out of three test sets. Moreover, the Success Predictor seems to better improve the LM-MA-ES baseline algorithm than the LQ regression model in all experiments, except for the training set of SphereFace. { In addition, LM-MA-ES assisted by the Success Predictor outperforms the comparison-based surrogate-assisted ACM-ES in all experiments.} An additional observation is that the high MSC results on the training set, are often preserved on the test set.

\begin{figure}
\centering
\begin{tabular}{@{~}l@{~}l@{~}l@{~}l@{~}}
    &\hfil Dlib \hfil & \hfil FaceNet & \hfil SphereFace \\
    (a)&
     \includegraphics[align=c,align=c,width=1.9cm]{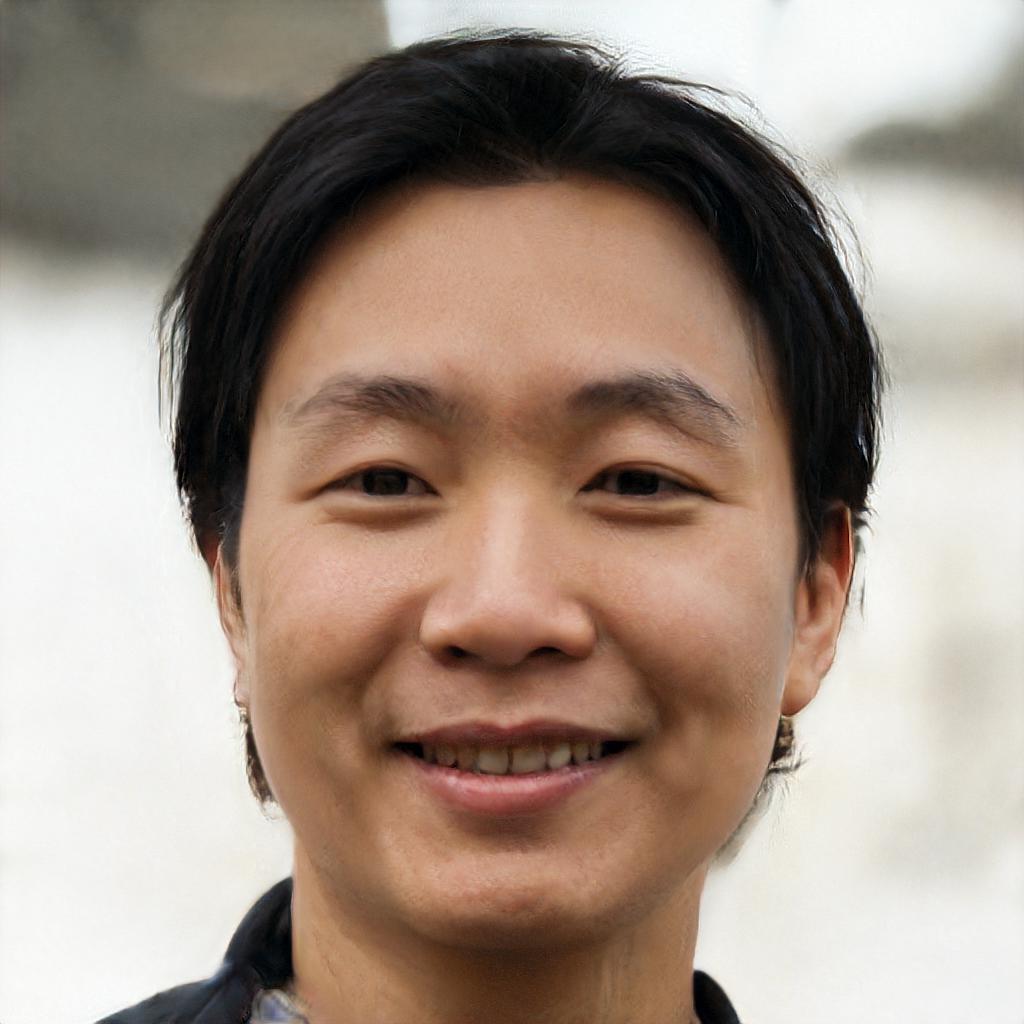} & \includegraphics[align=c,align=c,width=1.9cm]{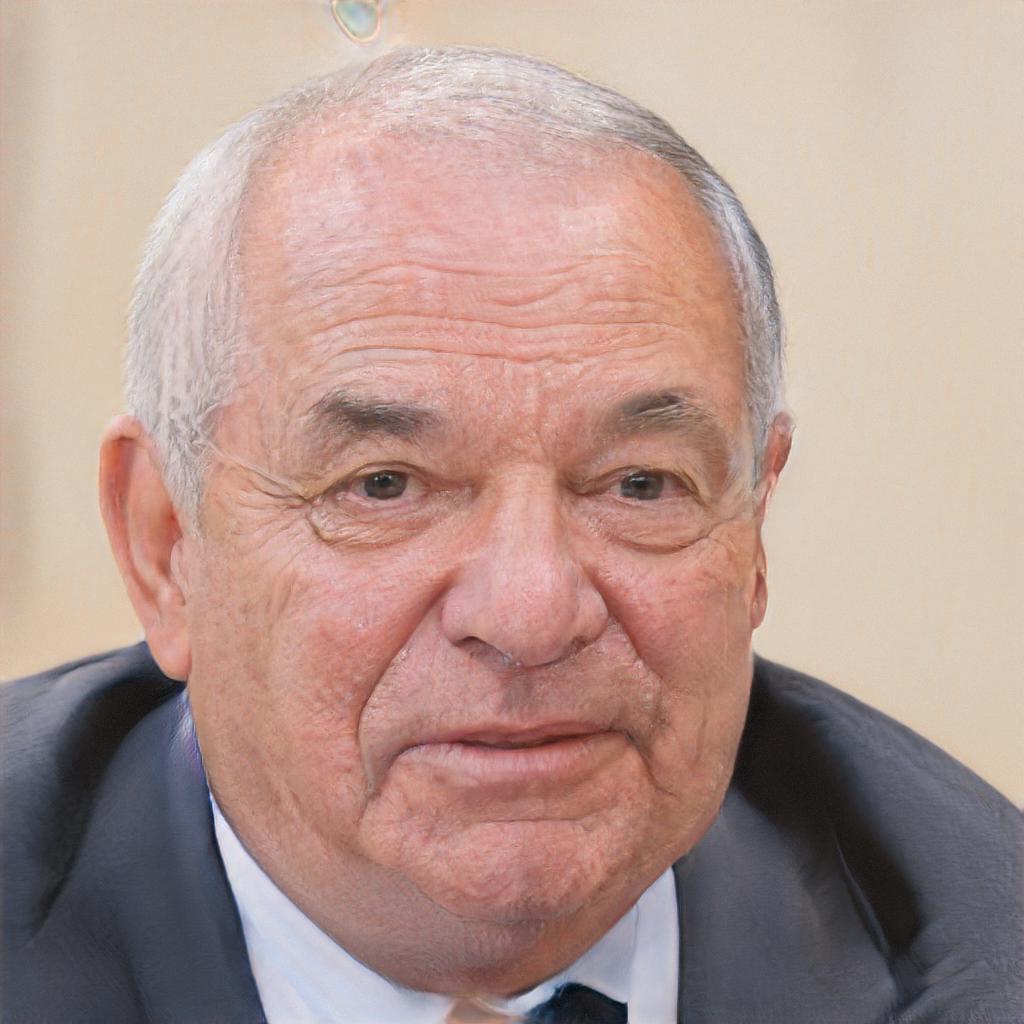} &  
    \includegraphics[align=c,align=c,width=1.9cm]{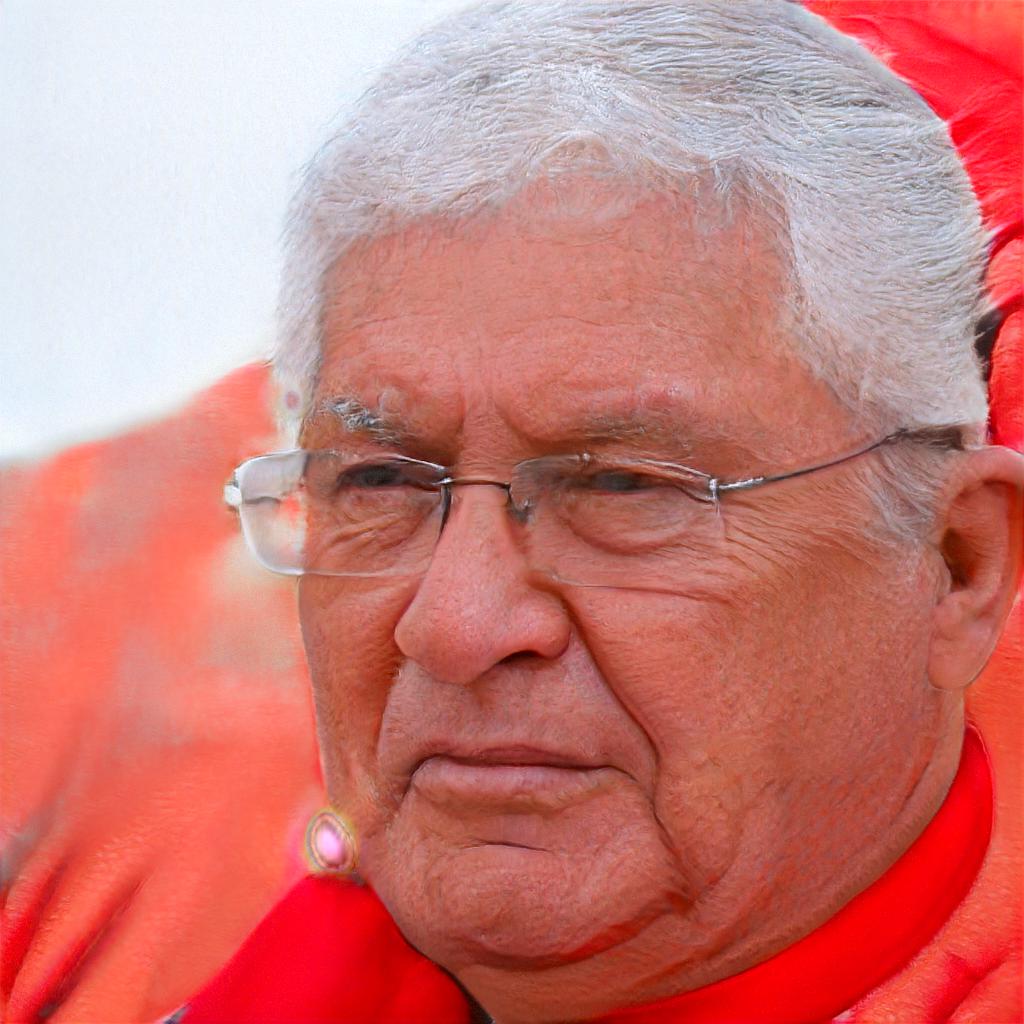} \\
     (b)&
     \includegraphics[align=c,width=1.9cm]{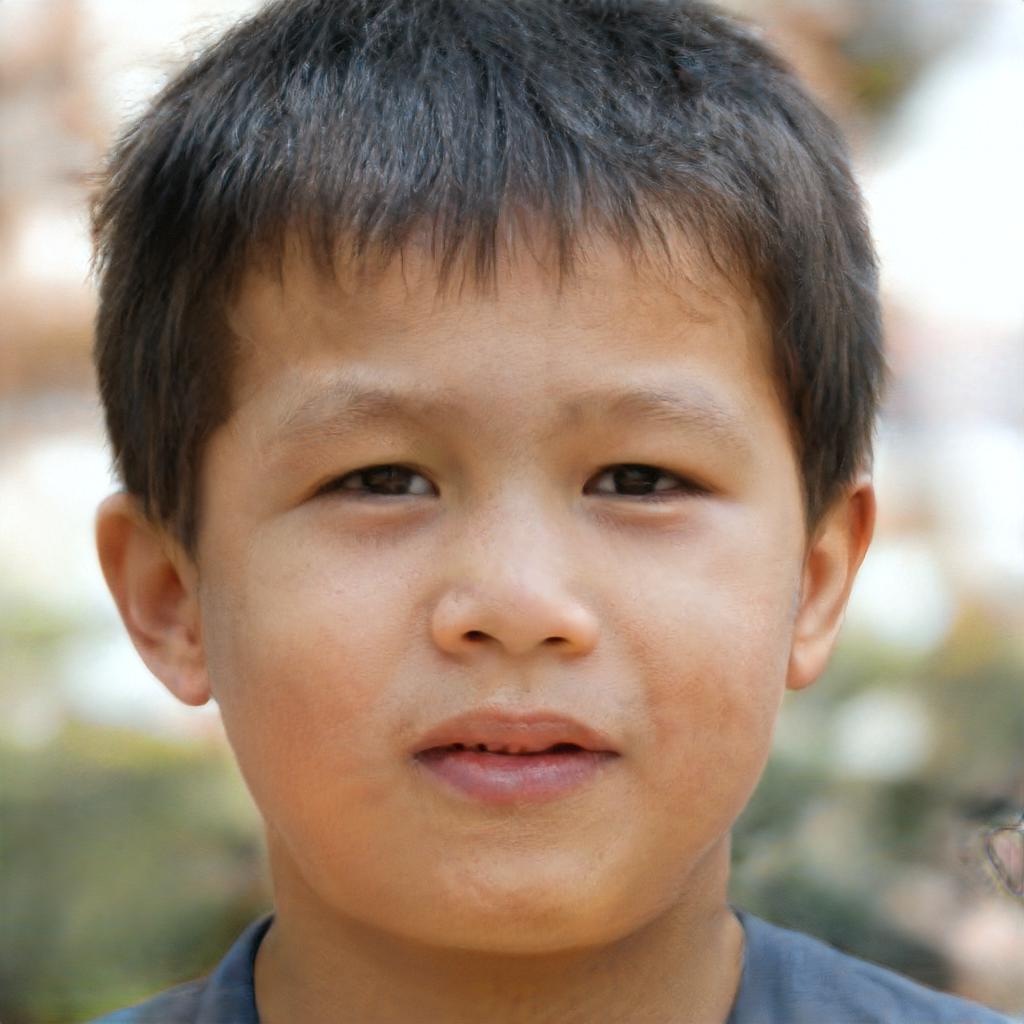} & \includegraphics[align=c,width=1.9cm]{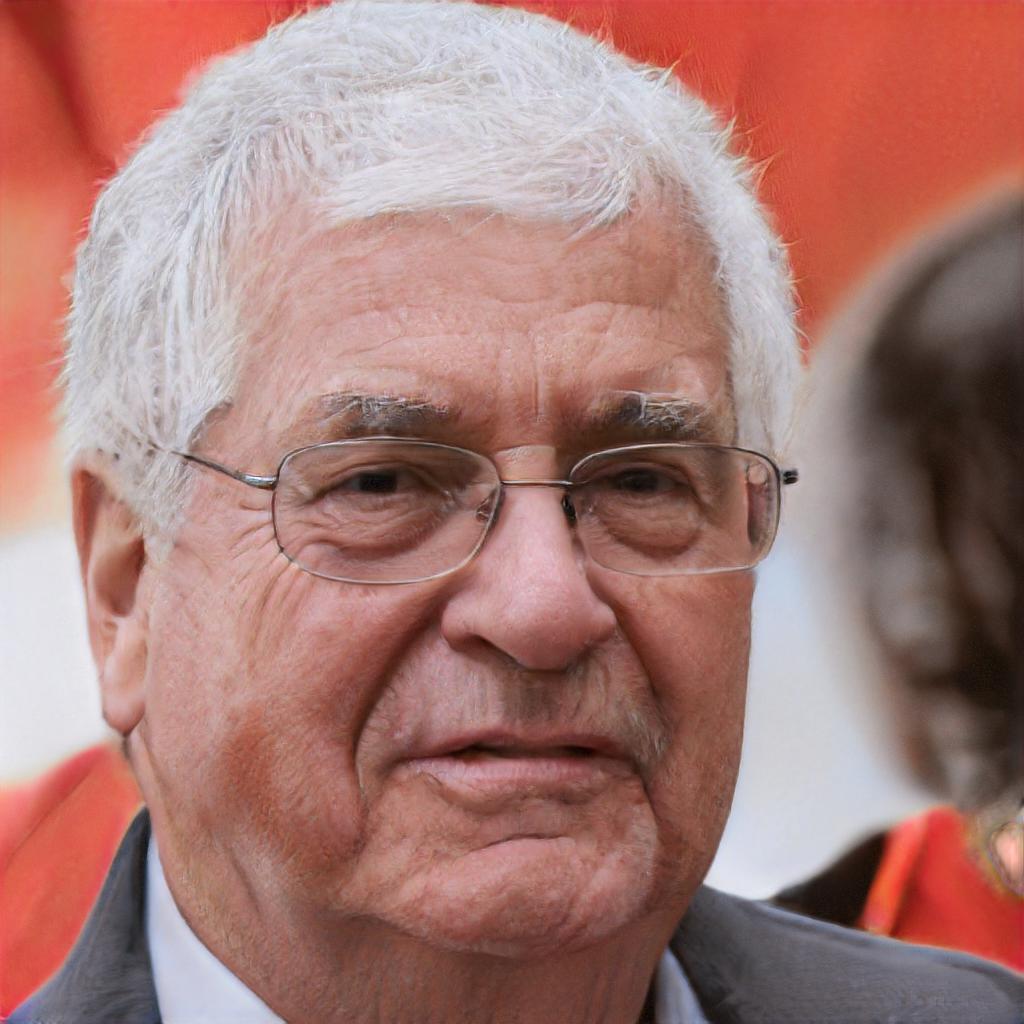} &  
    \includegraphics[align=c,width=1.9cm]{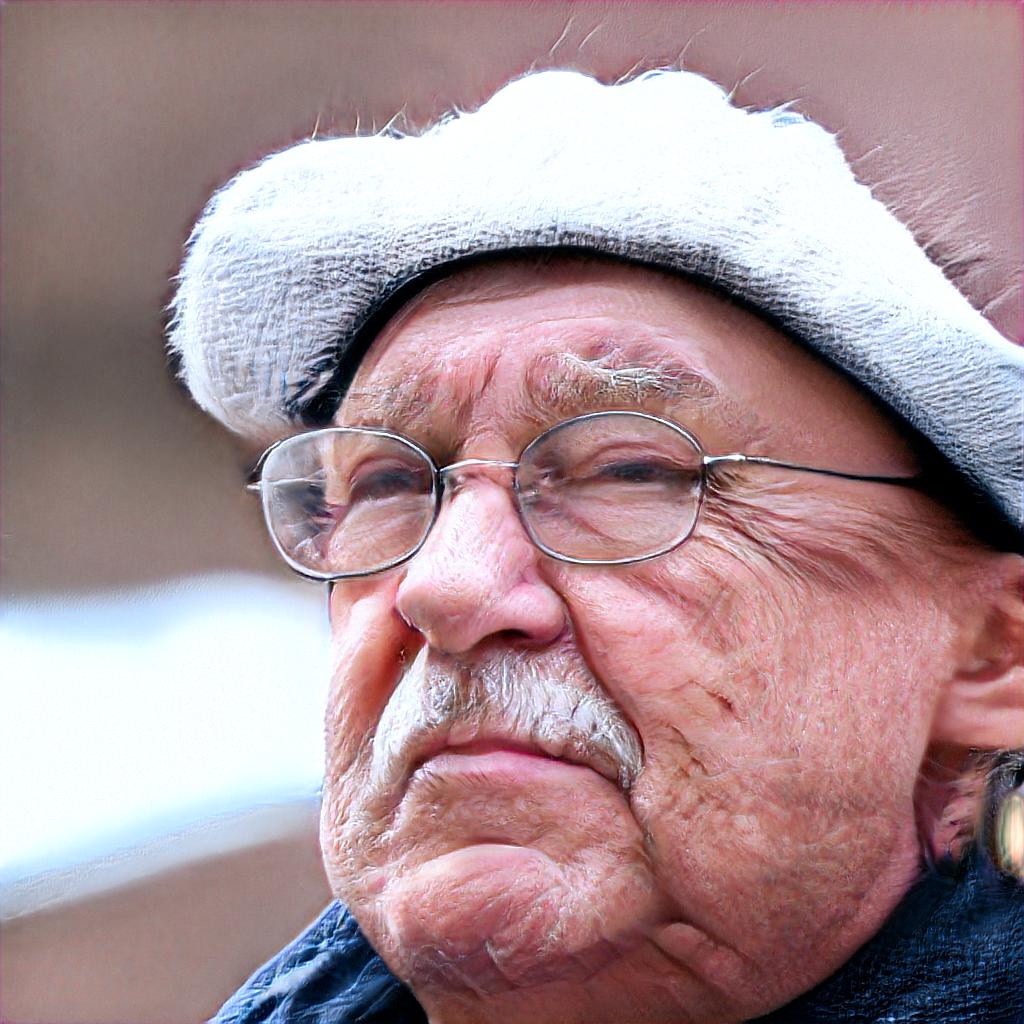} \\
     (c)&
     \includegraphics[align=c,width=1.9cm]{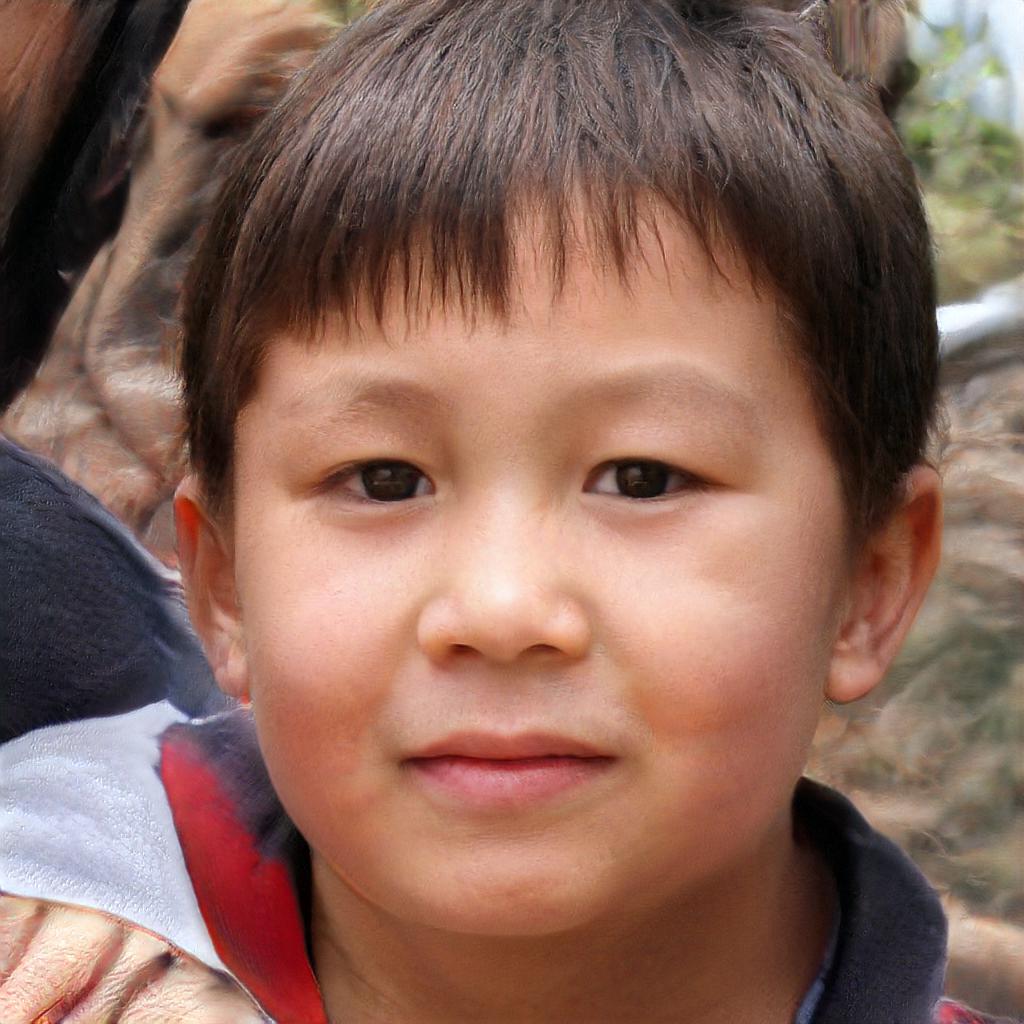} & \includegraphics[align=c,width=1.9cm]{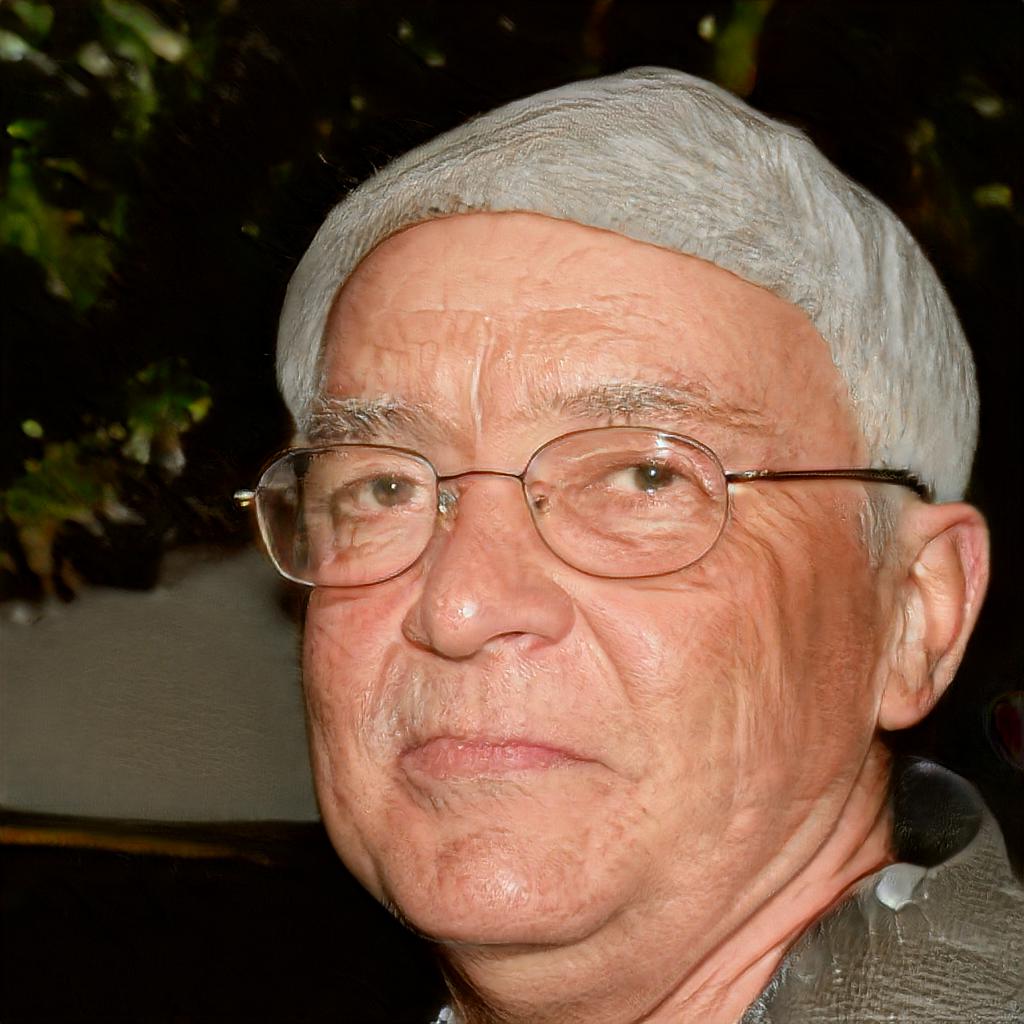} &
    \includegraphics[align=c,width=1.9cm]{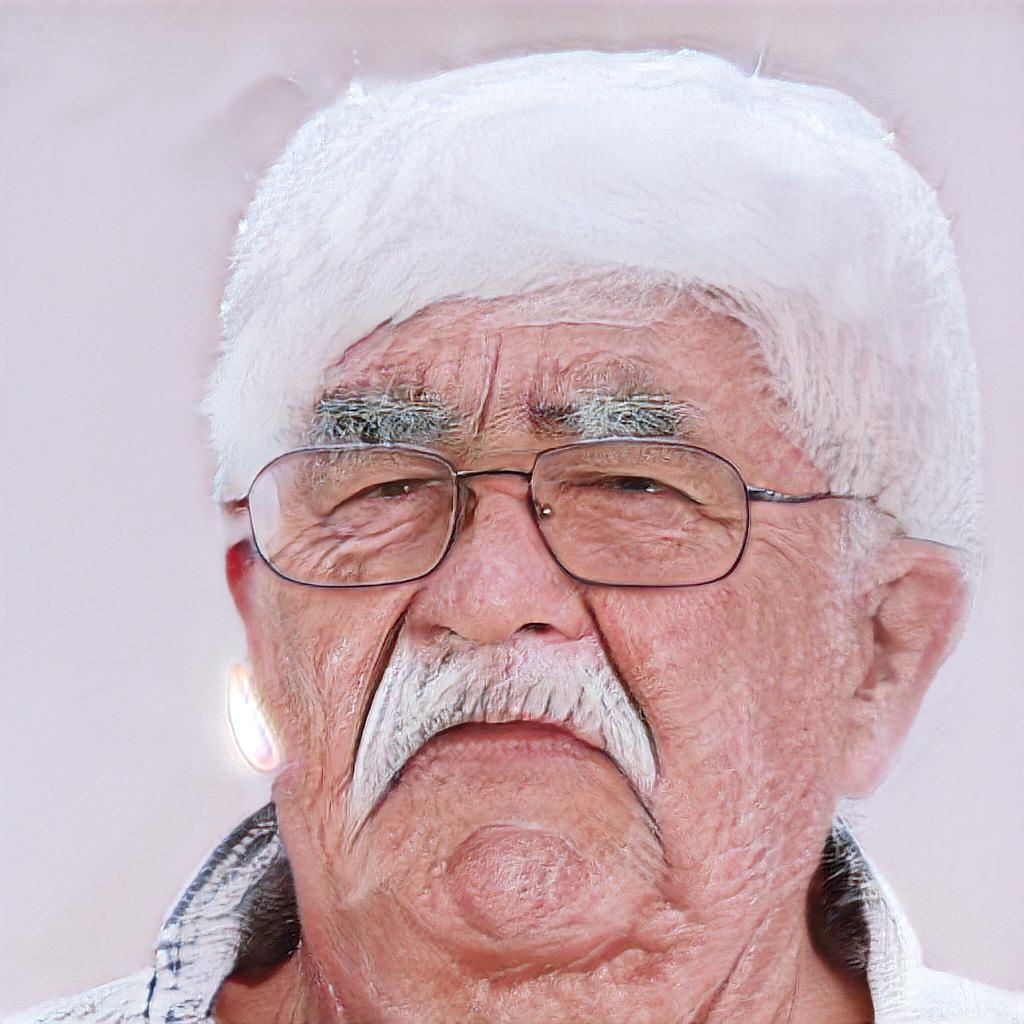} \\
     (d)&
     \includegraphics[align=c,width=1.9cm]{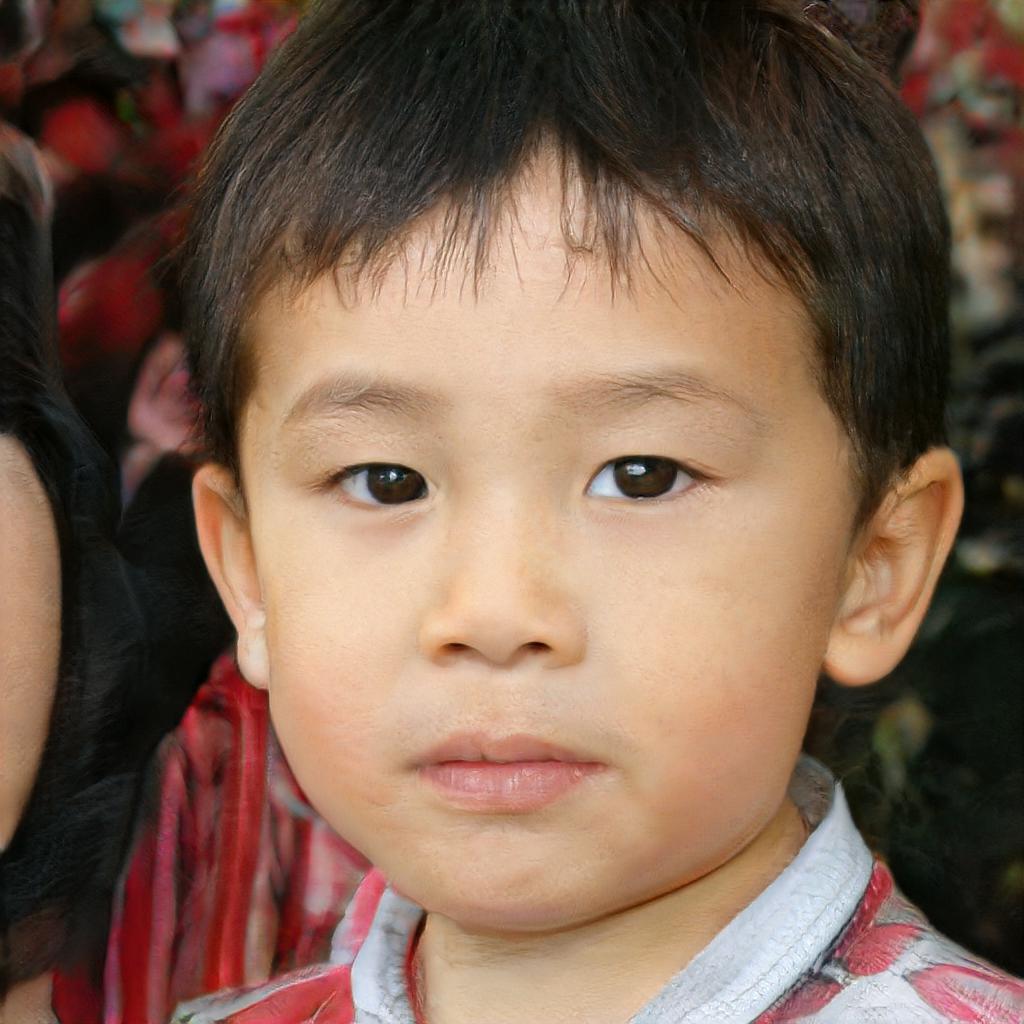} & \includegraphics[align=c,width=1.9cm]{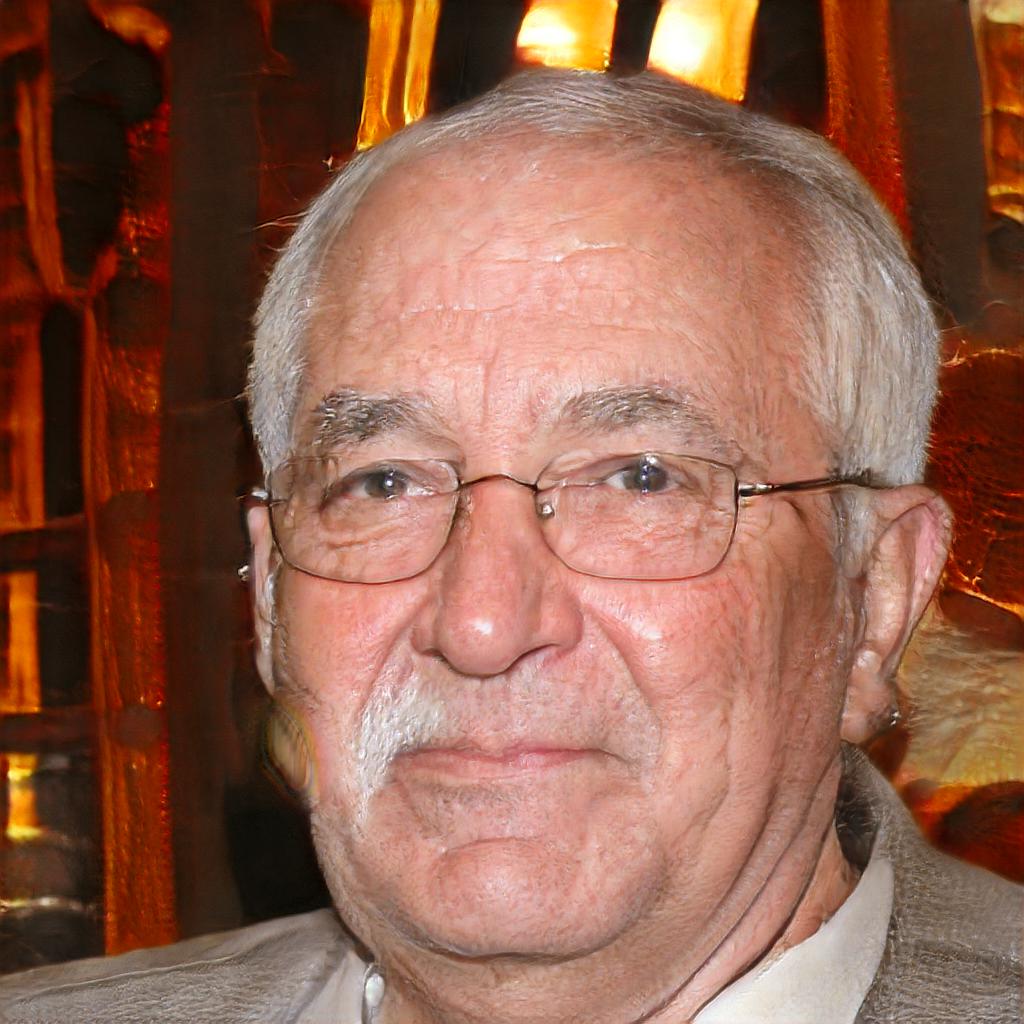} &
    \includegraphics[align=c,width=1.9cm]{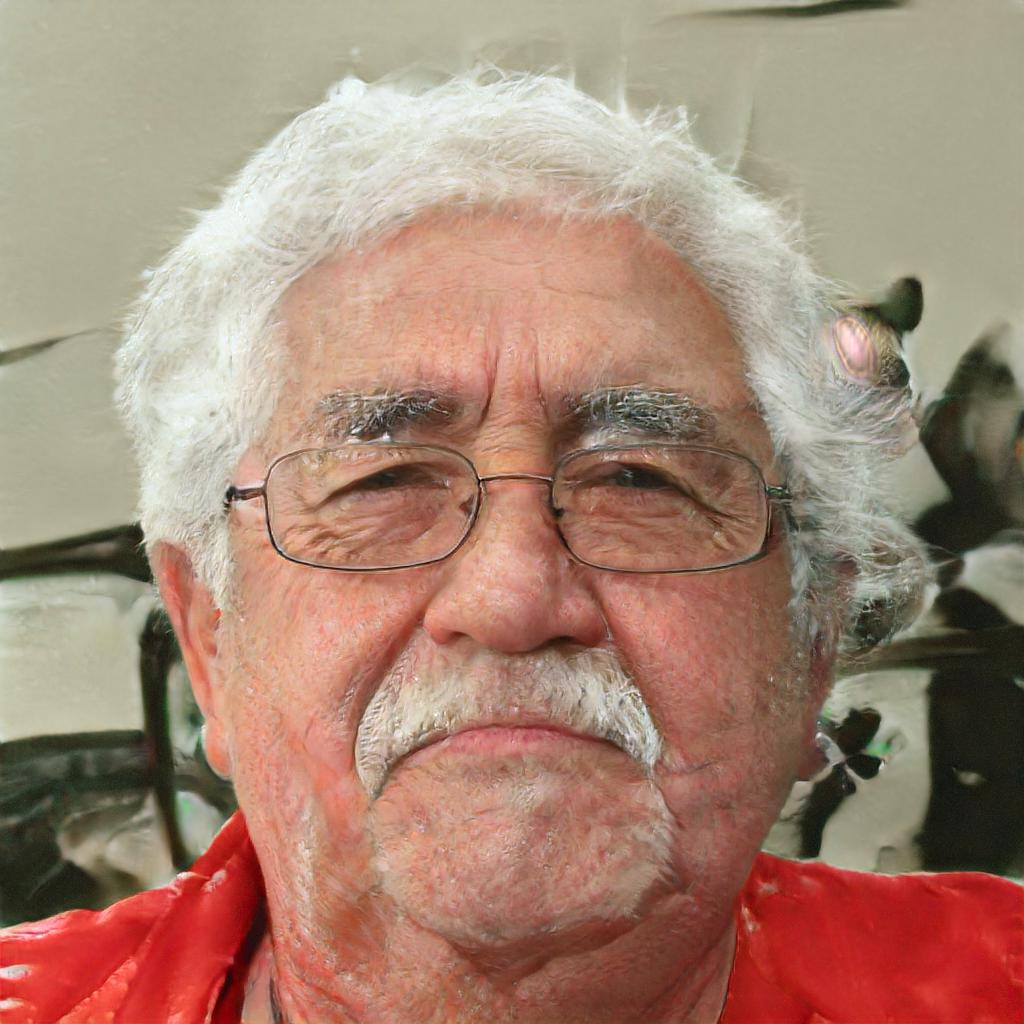} \\
     (e)&
     \includegraphics[align=c,width=1.9cm]{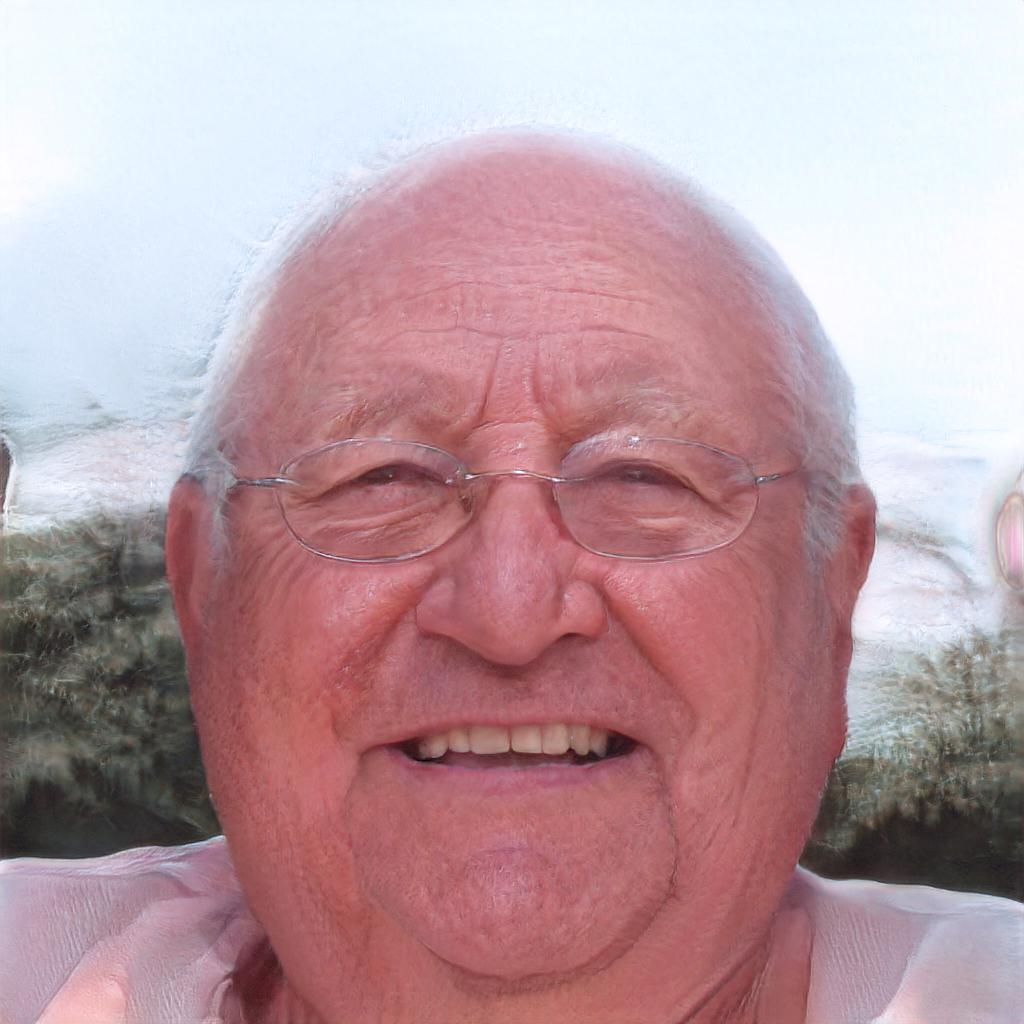} & \includegraphics[align=c,width=1.9cm]{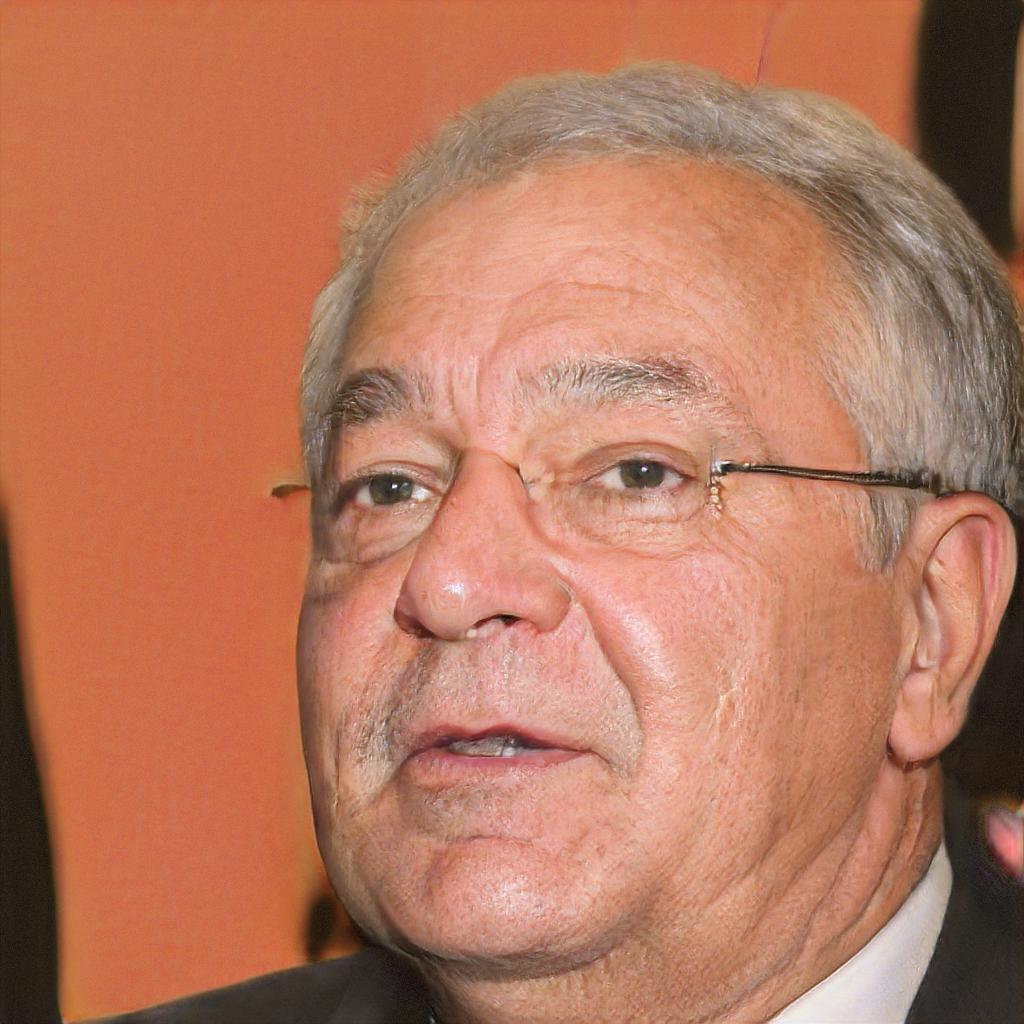} &
    \includegraphics[align=c,width=1.9cm]{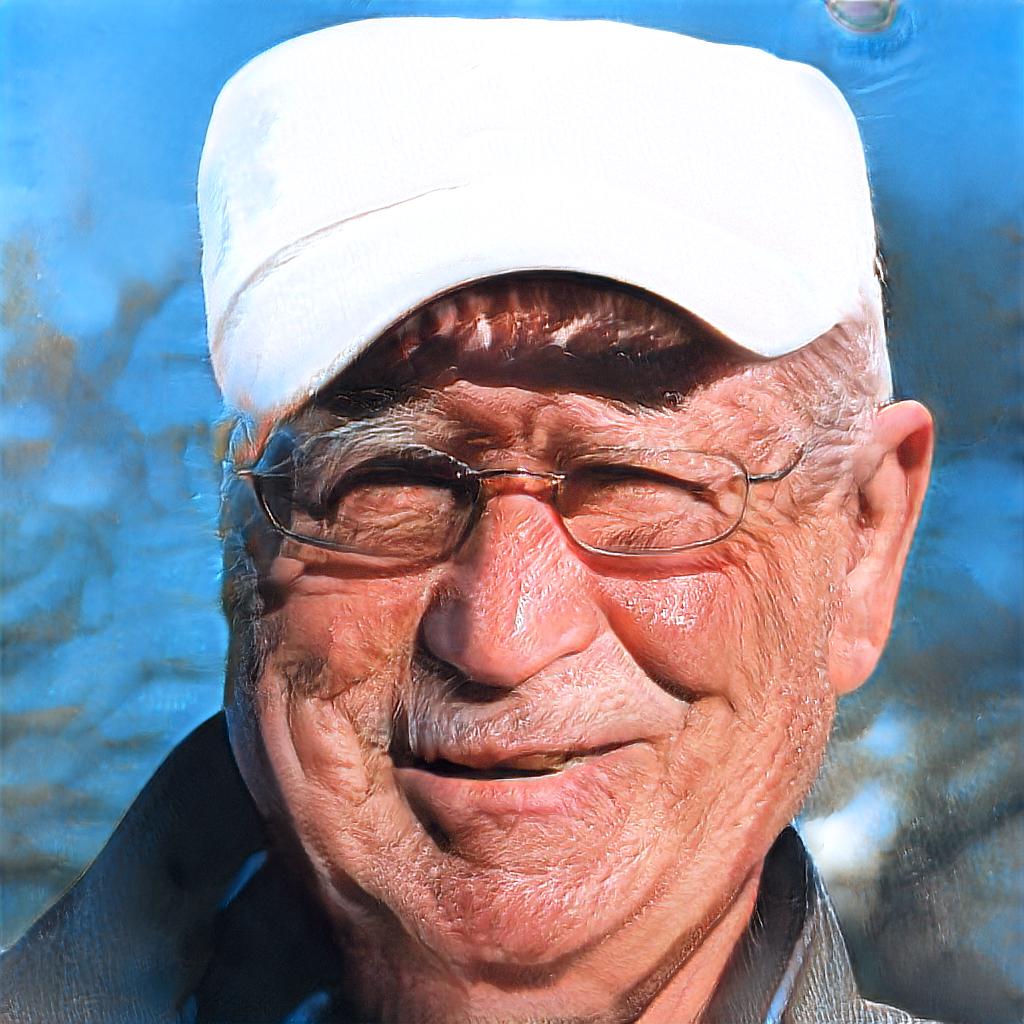} \\
     (f)&
     \includegraphics[align=c,width=1.9cm]{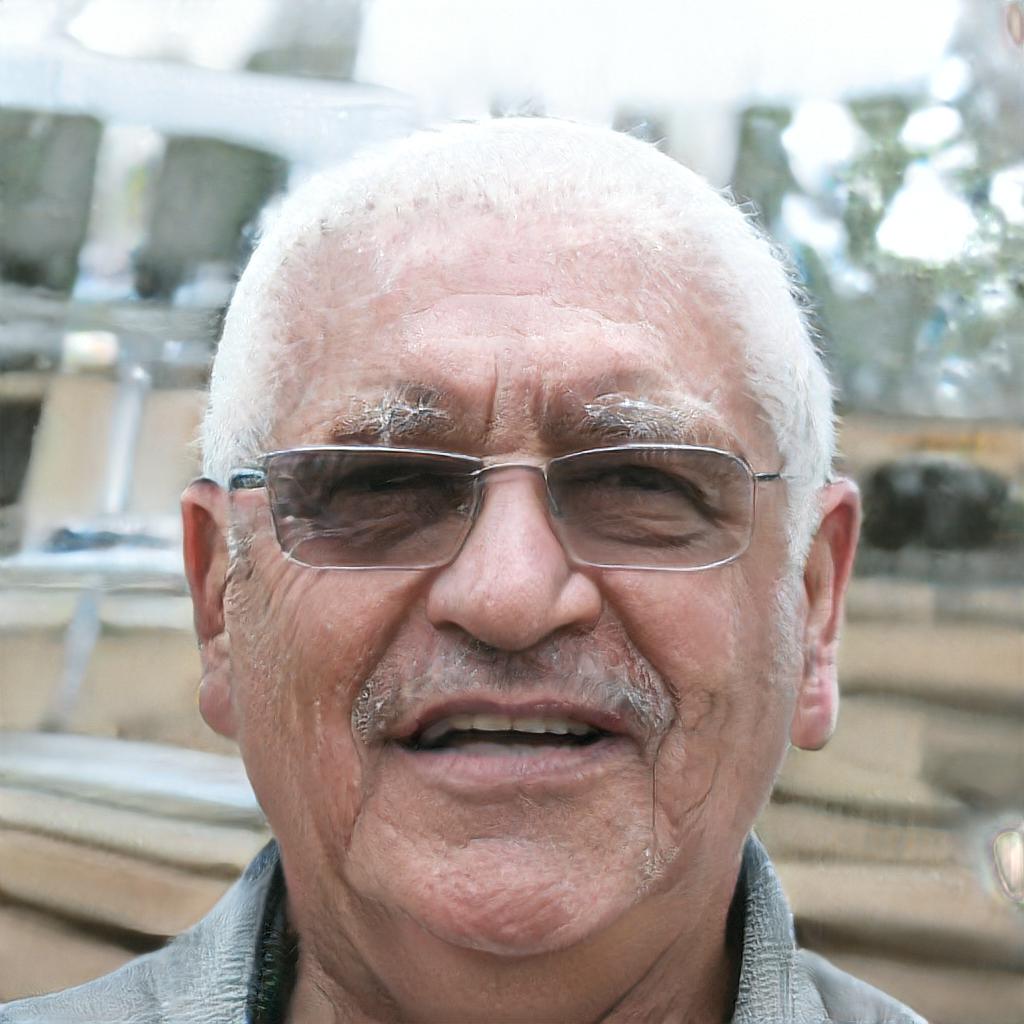} & \includegraphics[align=c,width=1.9cm]{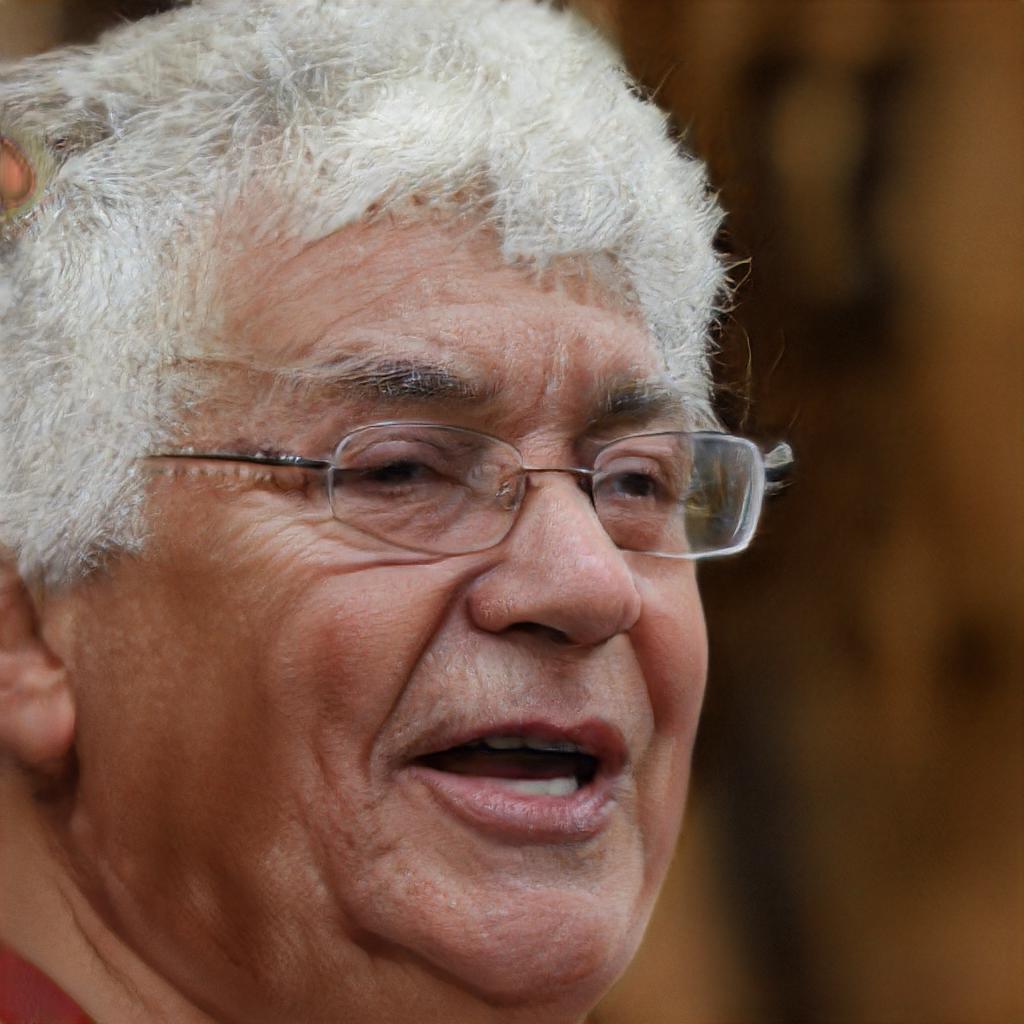} &
    \includegraphics[align=c,width=1.9cm]{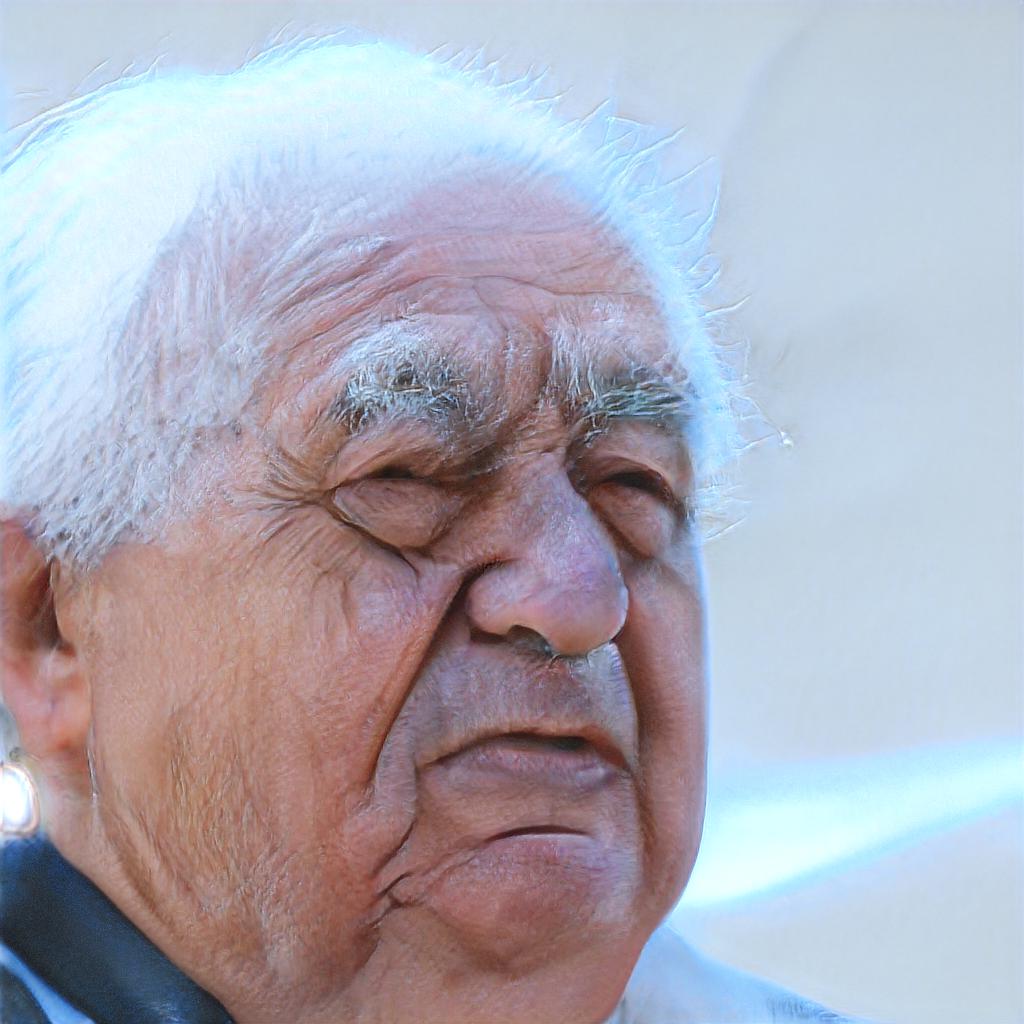} \\
    (g)&
     \includegraphics[align=c,width=1.9cm]{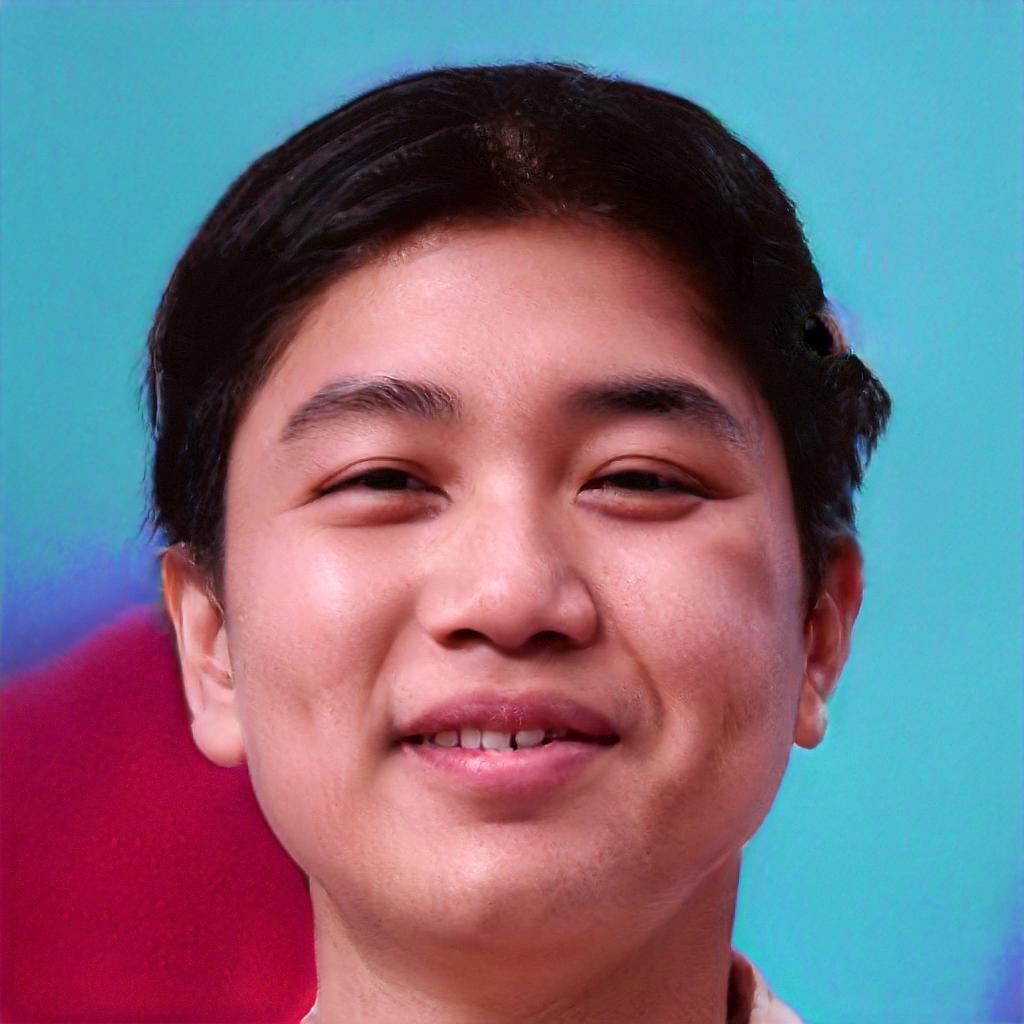} & \includegraphics[align=c,width=1.9cm]{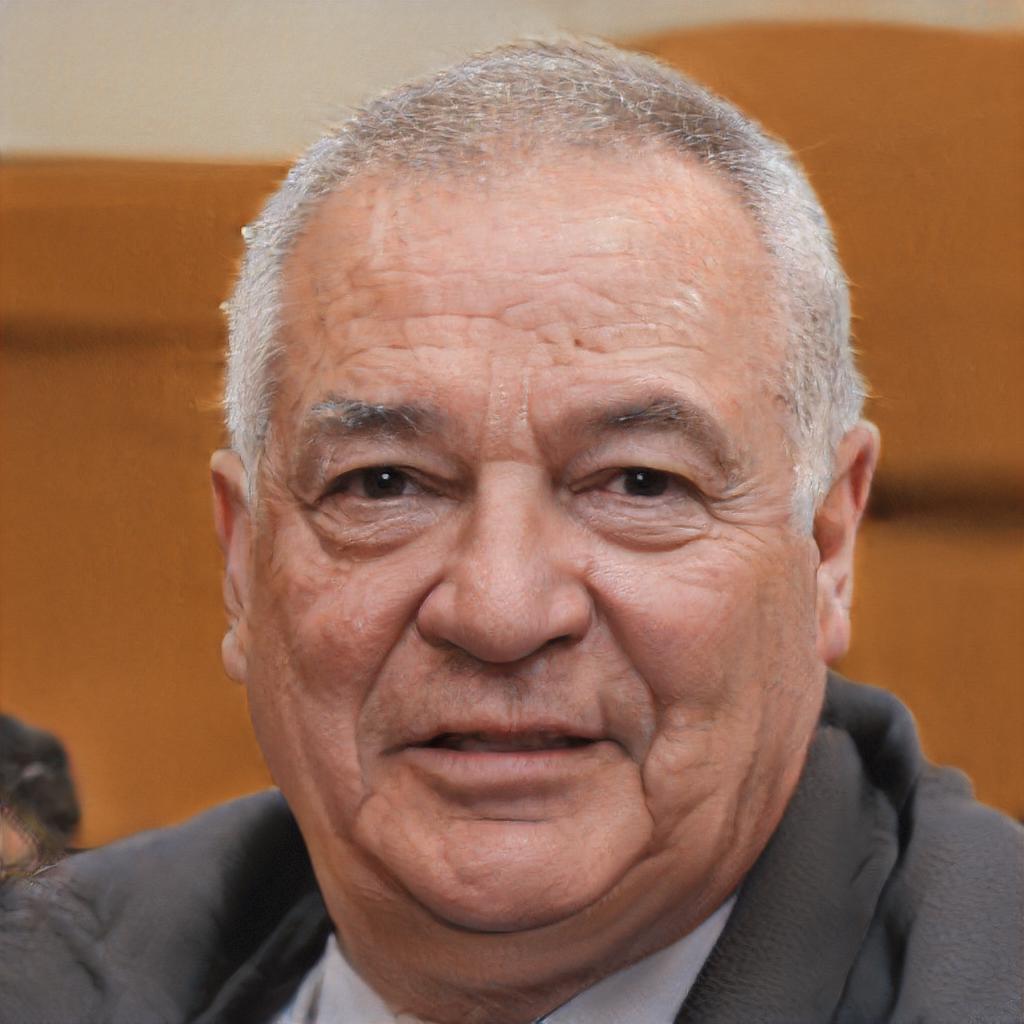}&
    \includegraphics[align=c,width=1.9cm]{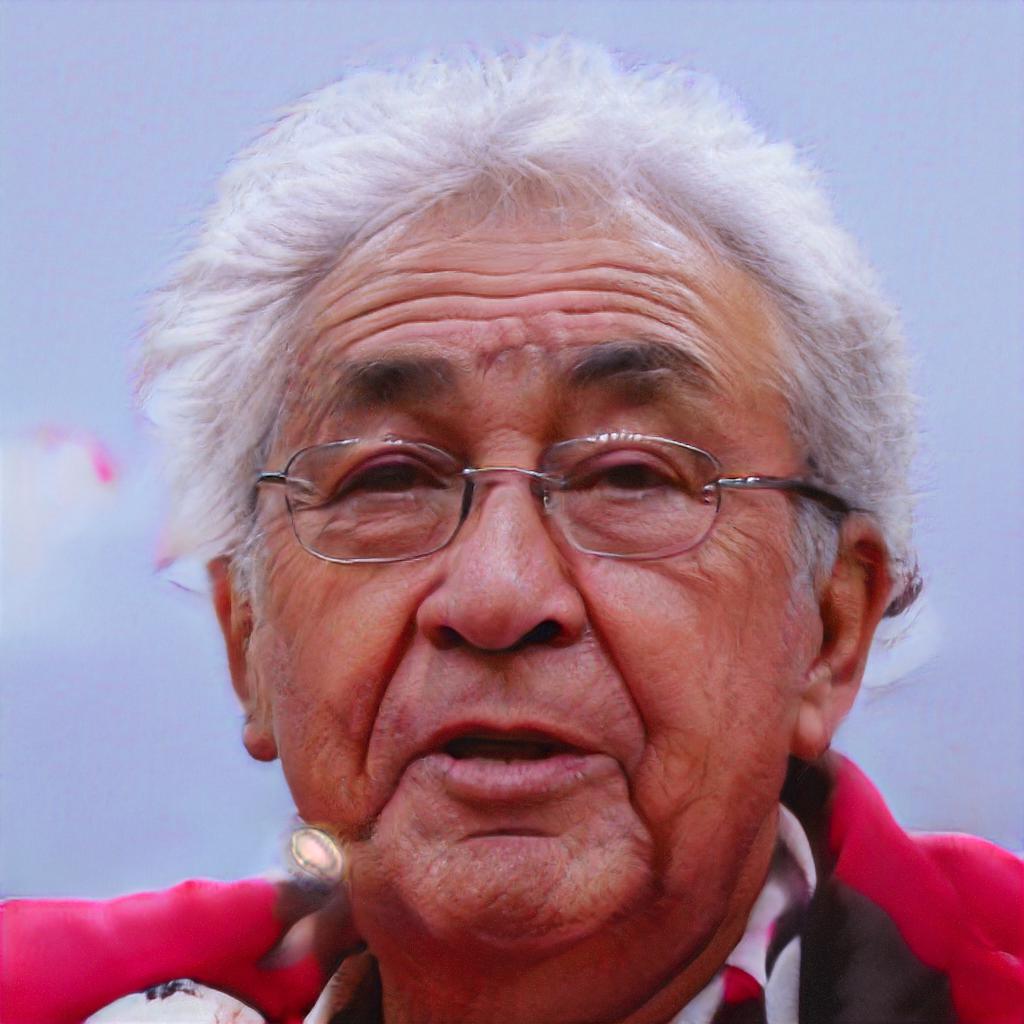} \\
     (h)&
     \includegraphics[align=c,width=1.9cm]{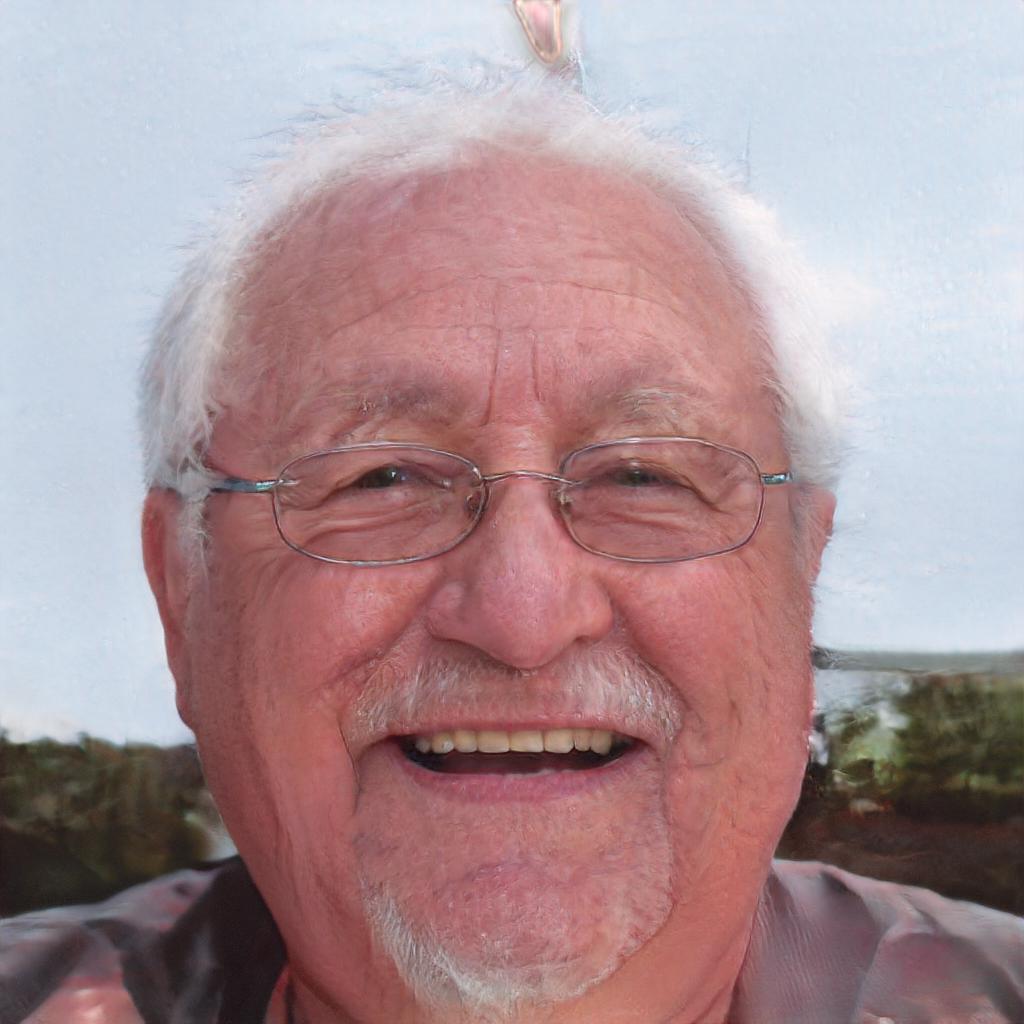} & \includegraphics[align=c,width=1.9cm]{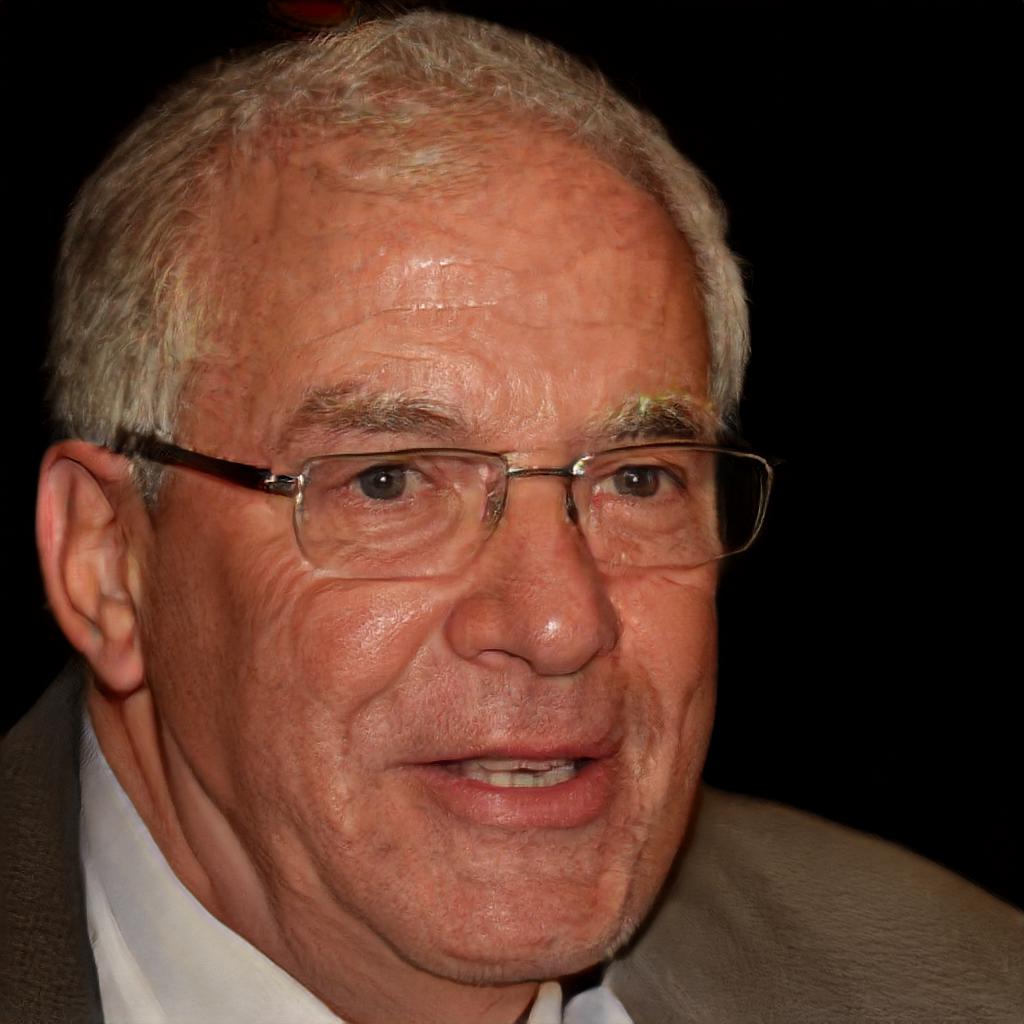} &
    \includegraphics[align=c,width=1.9cm]{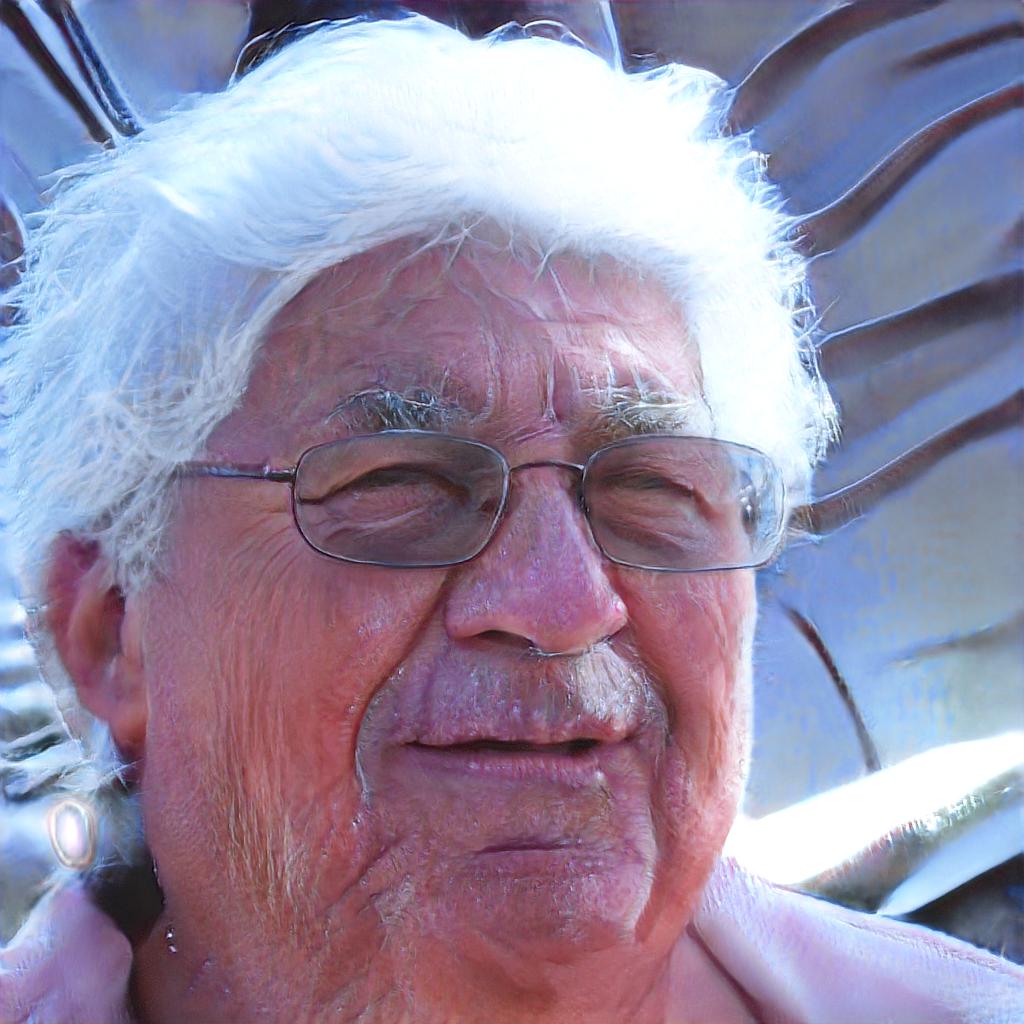} \\
     (i)&
     \includegraphics[align=c,width=1.9cm]{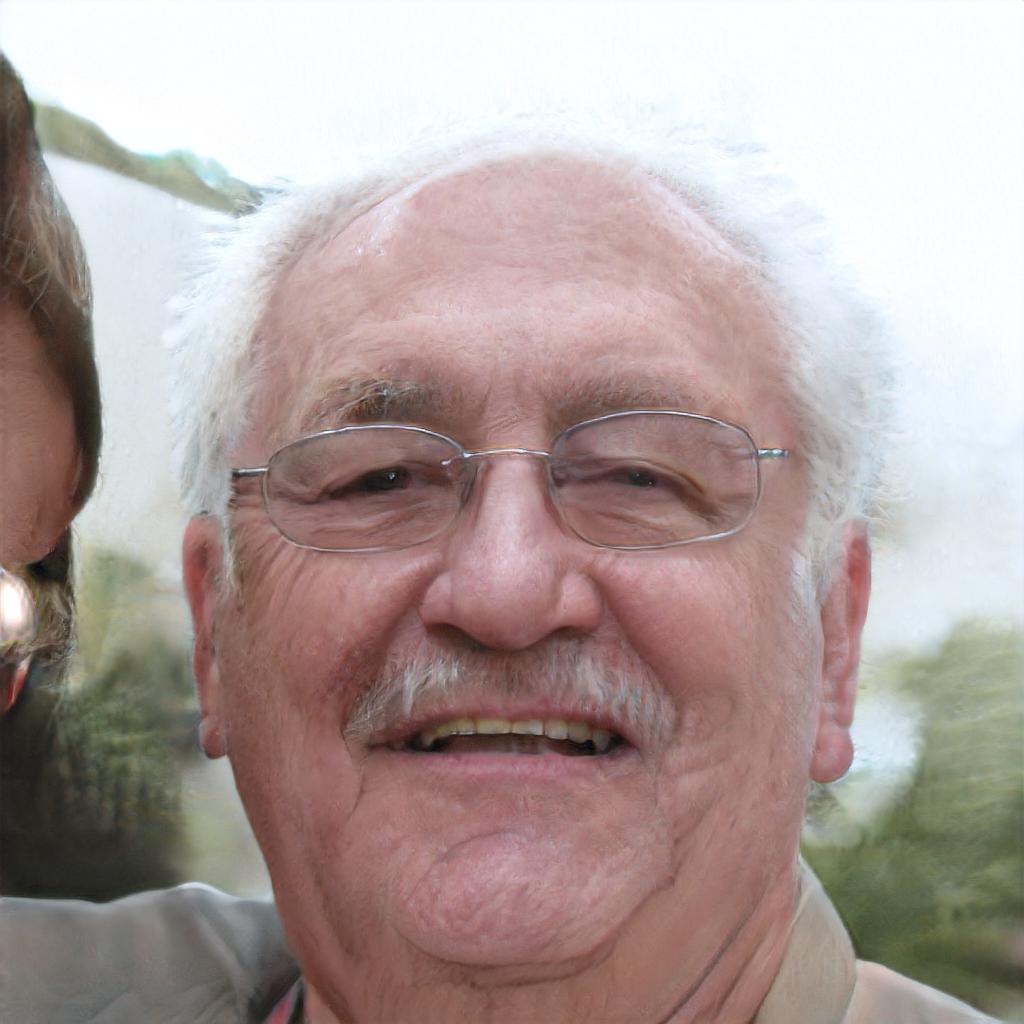} & \includegraphics[align=c,width=1.9cm]{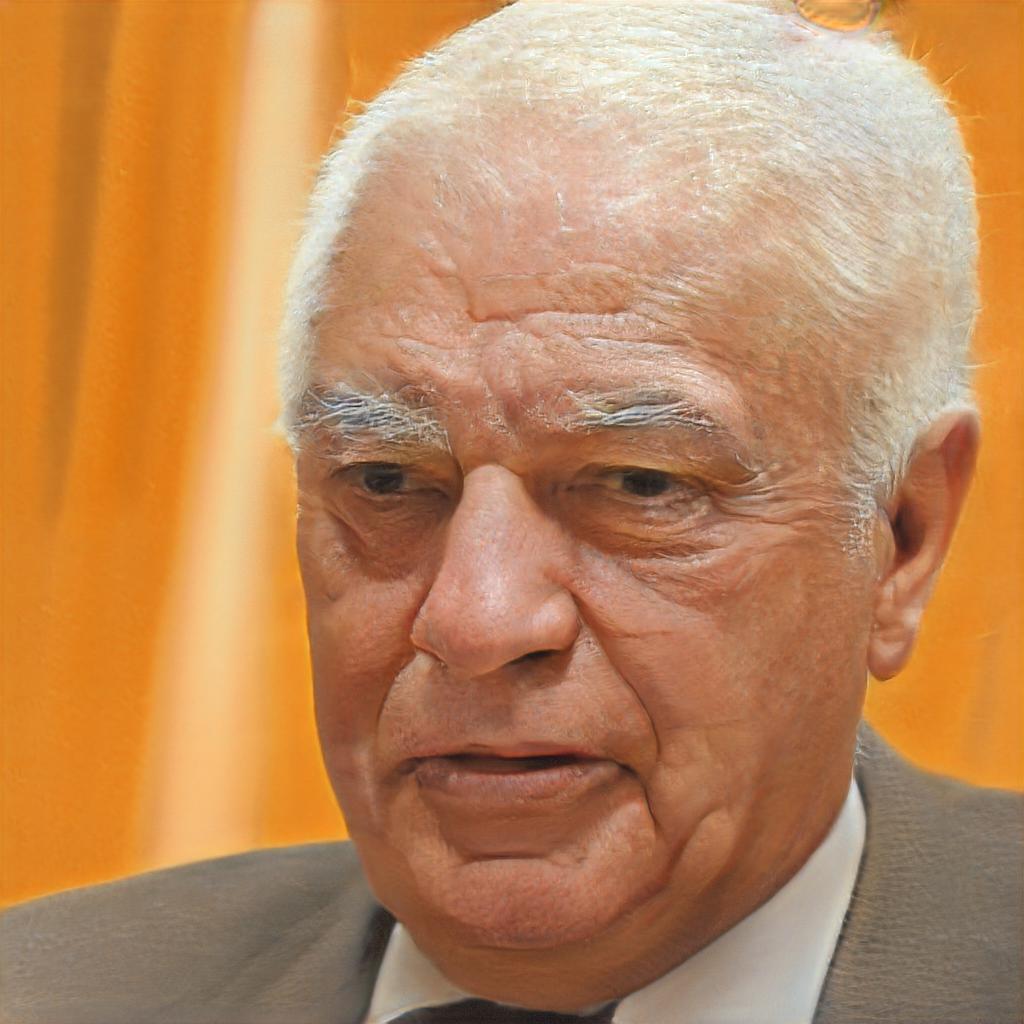} &
    \includegraphics[align=c,width=1.9cm]{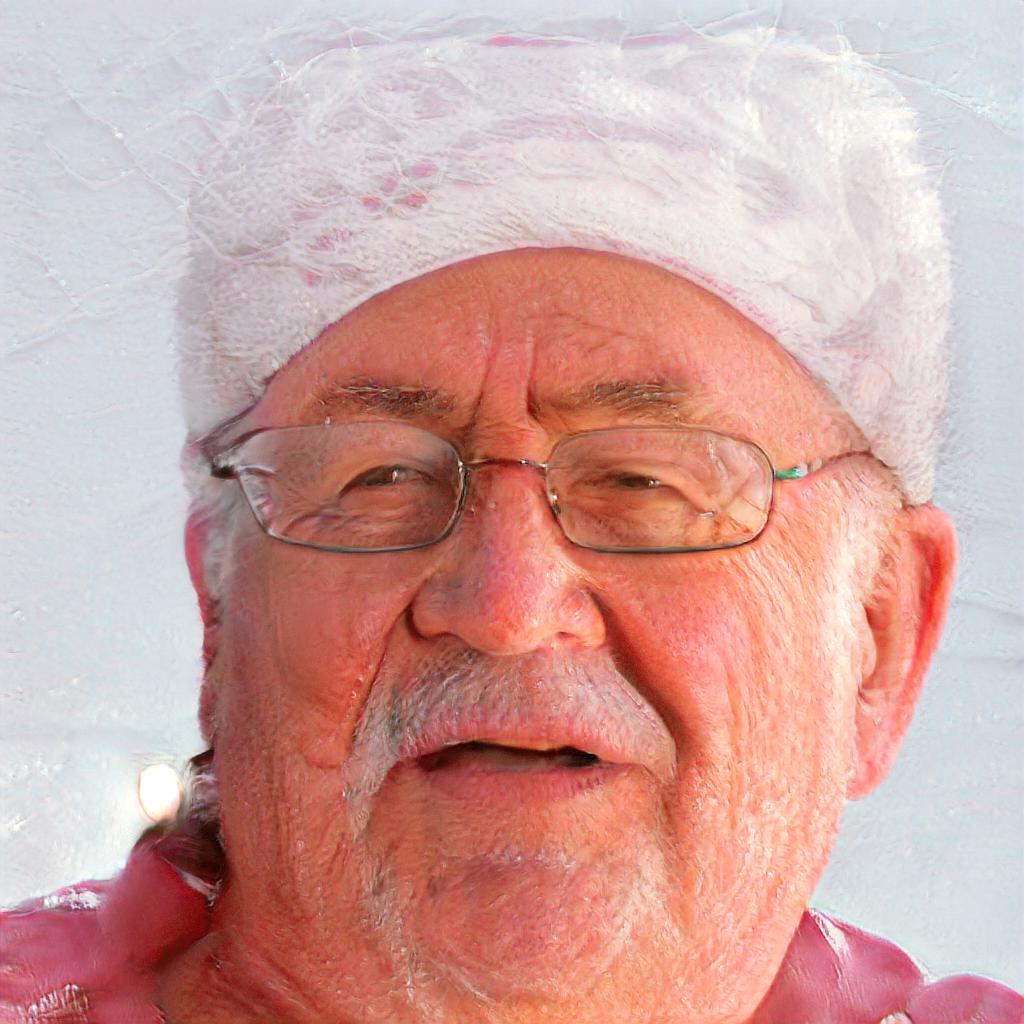} \\
    (j)&
     \includegraphics[align=c,width=1.9cm]{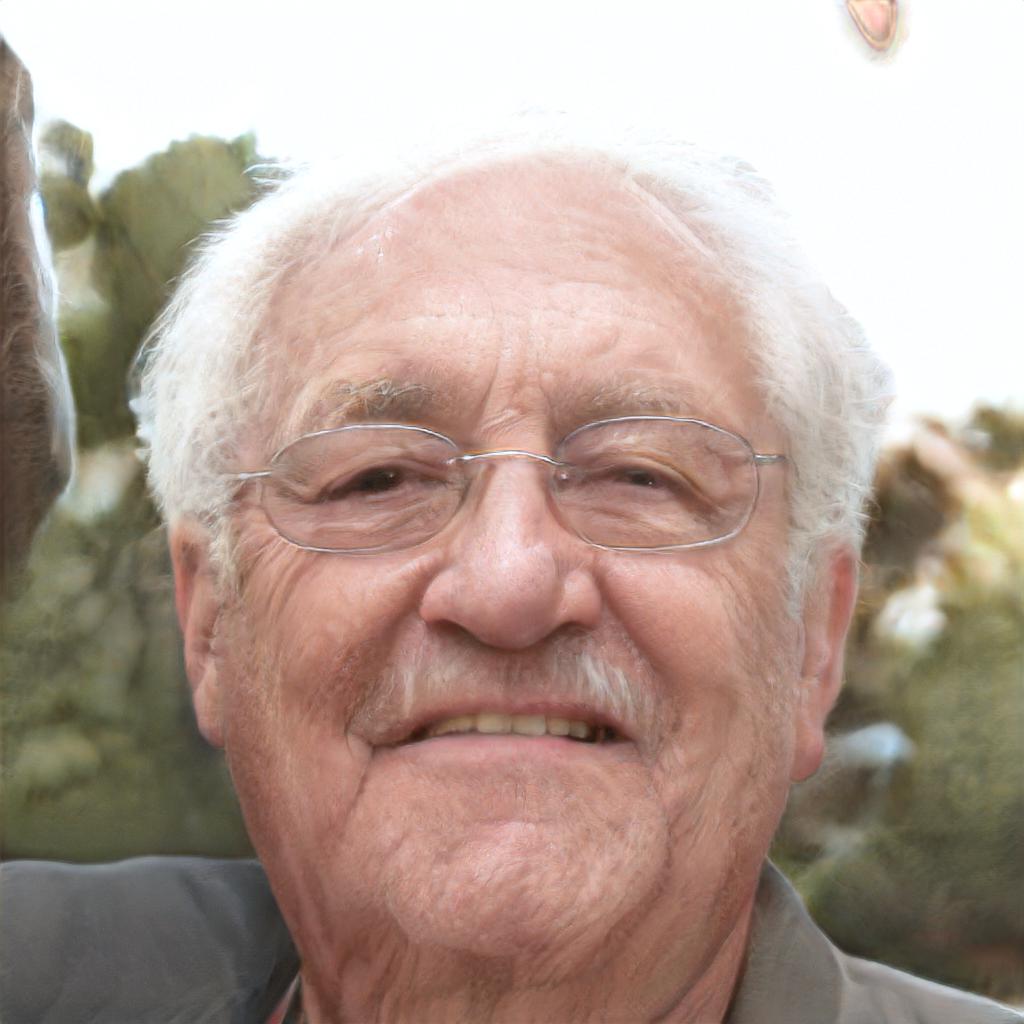} & \includegraphics[align=c,width=1.9cm]{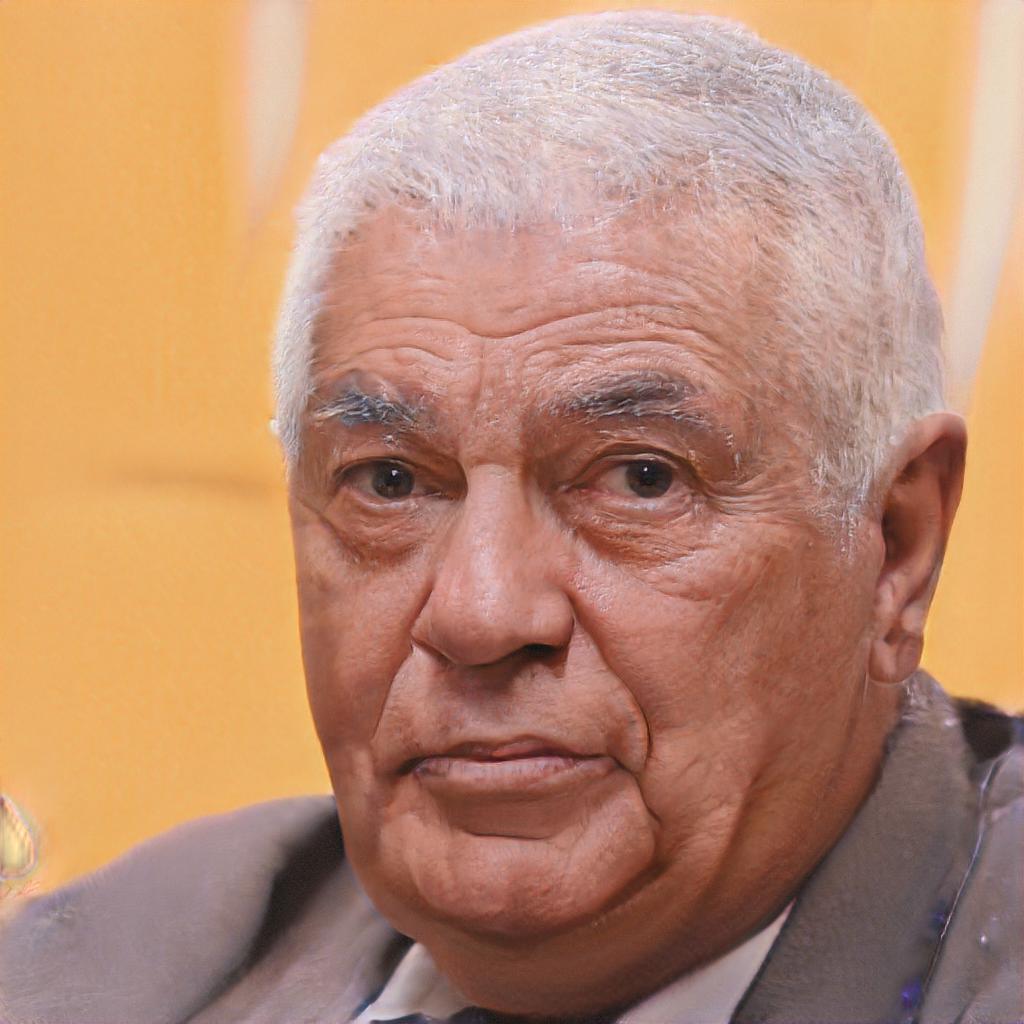} &
    \includegraphics[align=c,width=1.9cm]{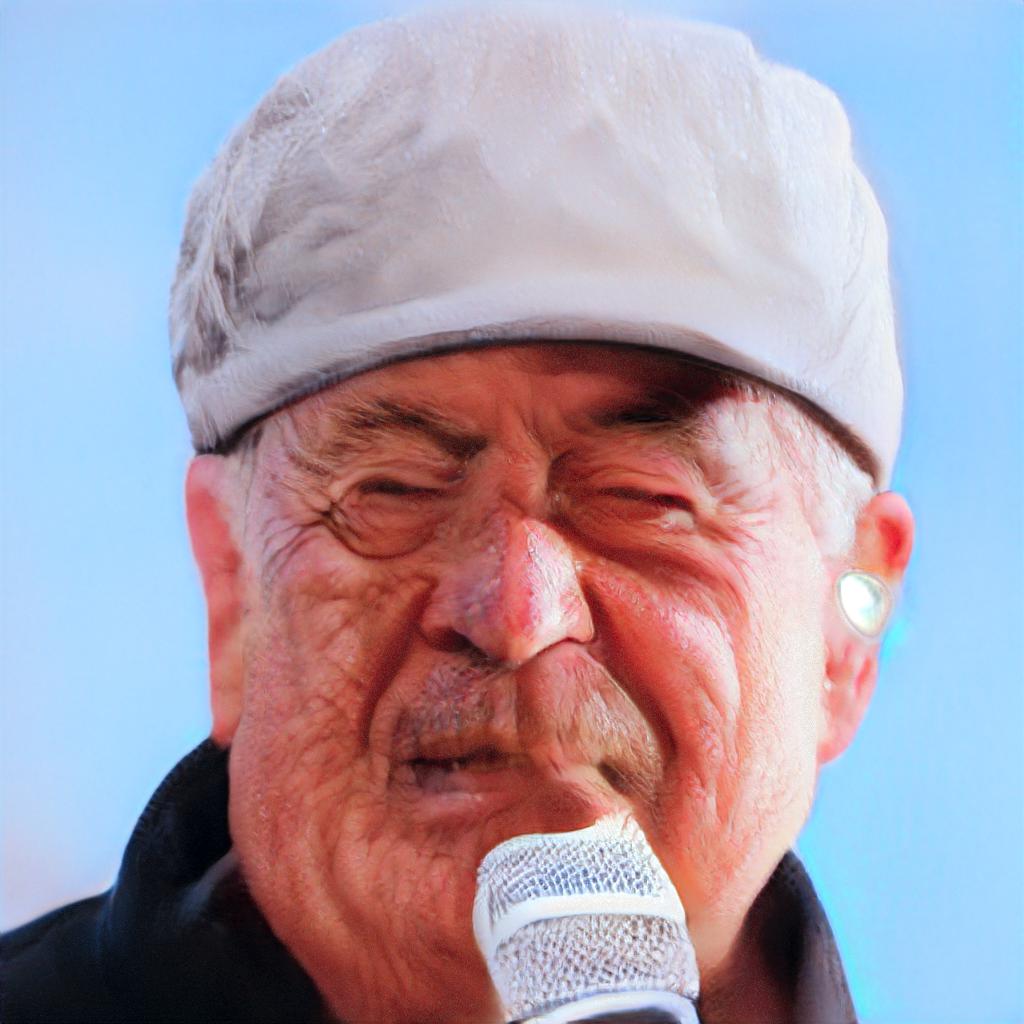} \\
    (k)&
     \includegraphics[align=c,width=1.9cm]{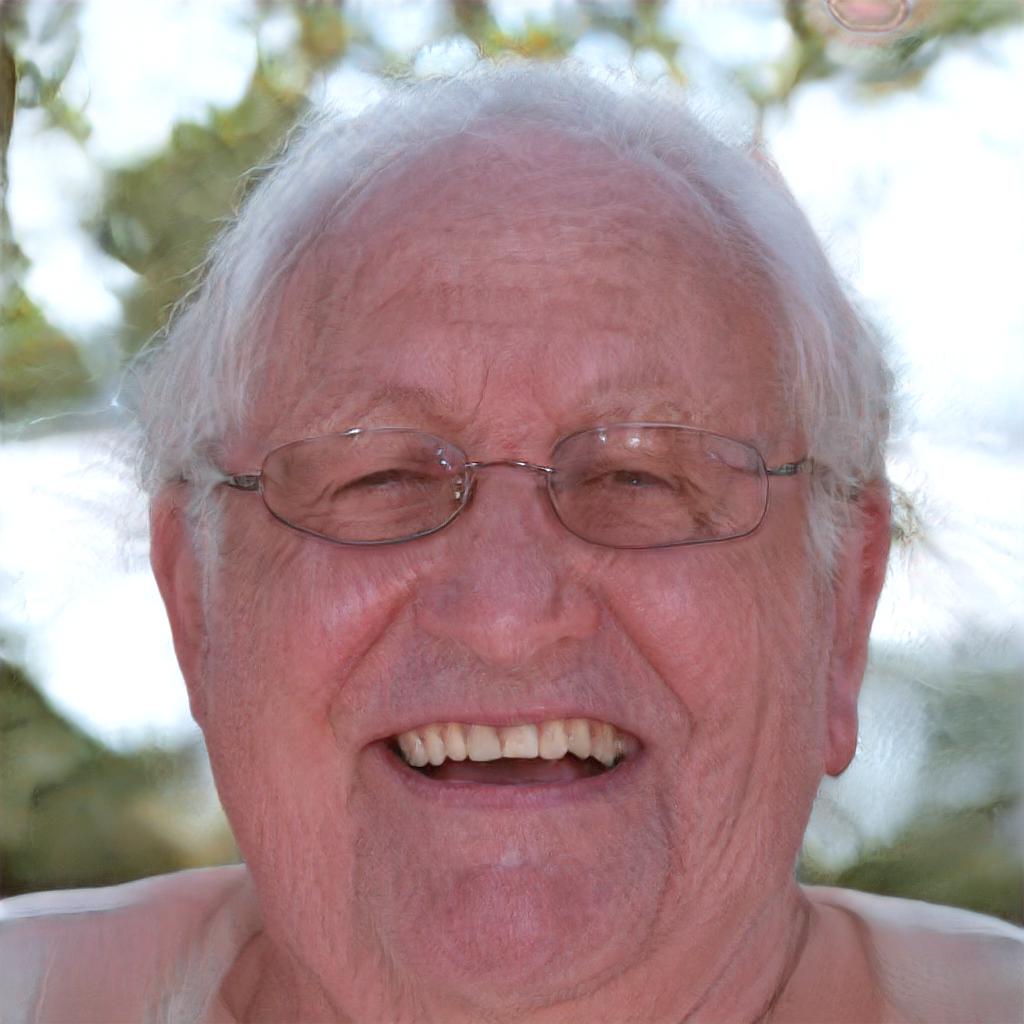} & \includegraphics[align=c,width=1.9cm]{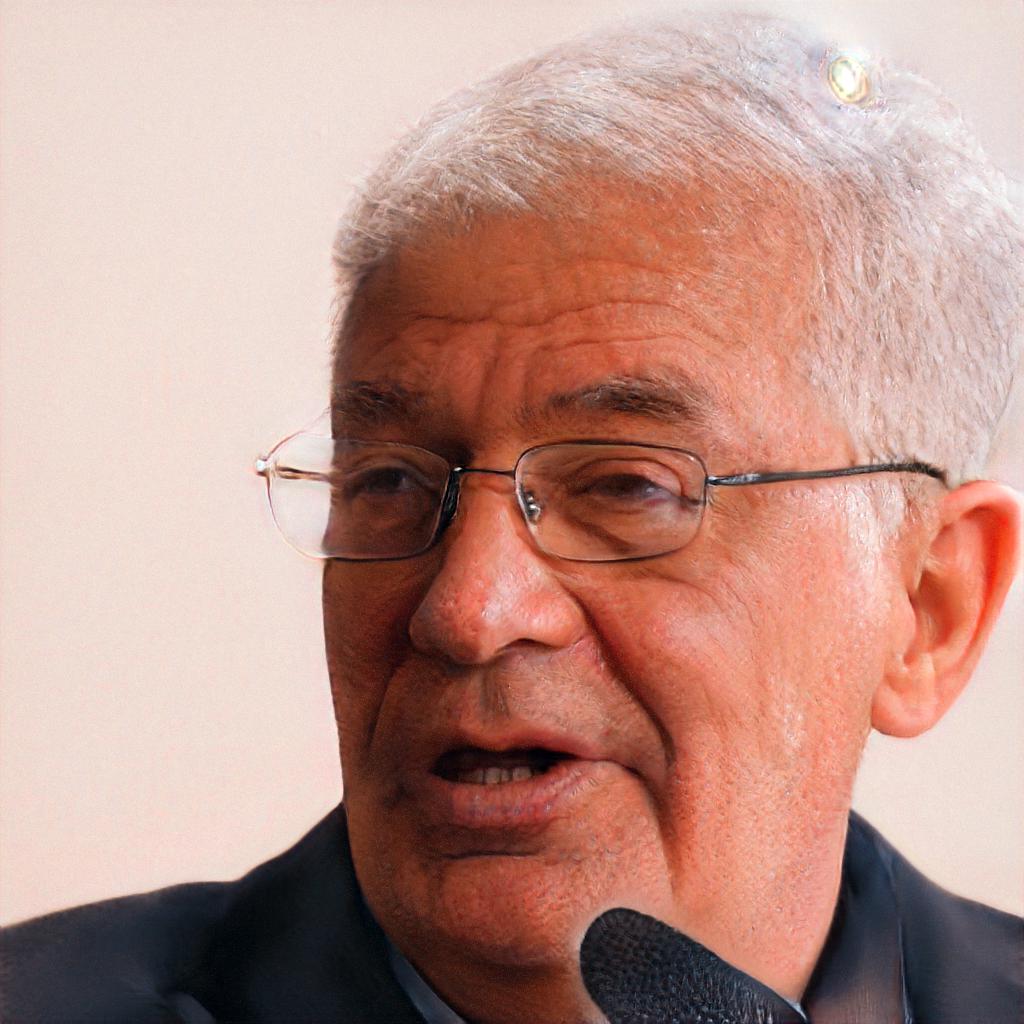} &
    \includegraphics[align=c,width=1.9cm]{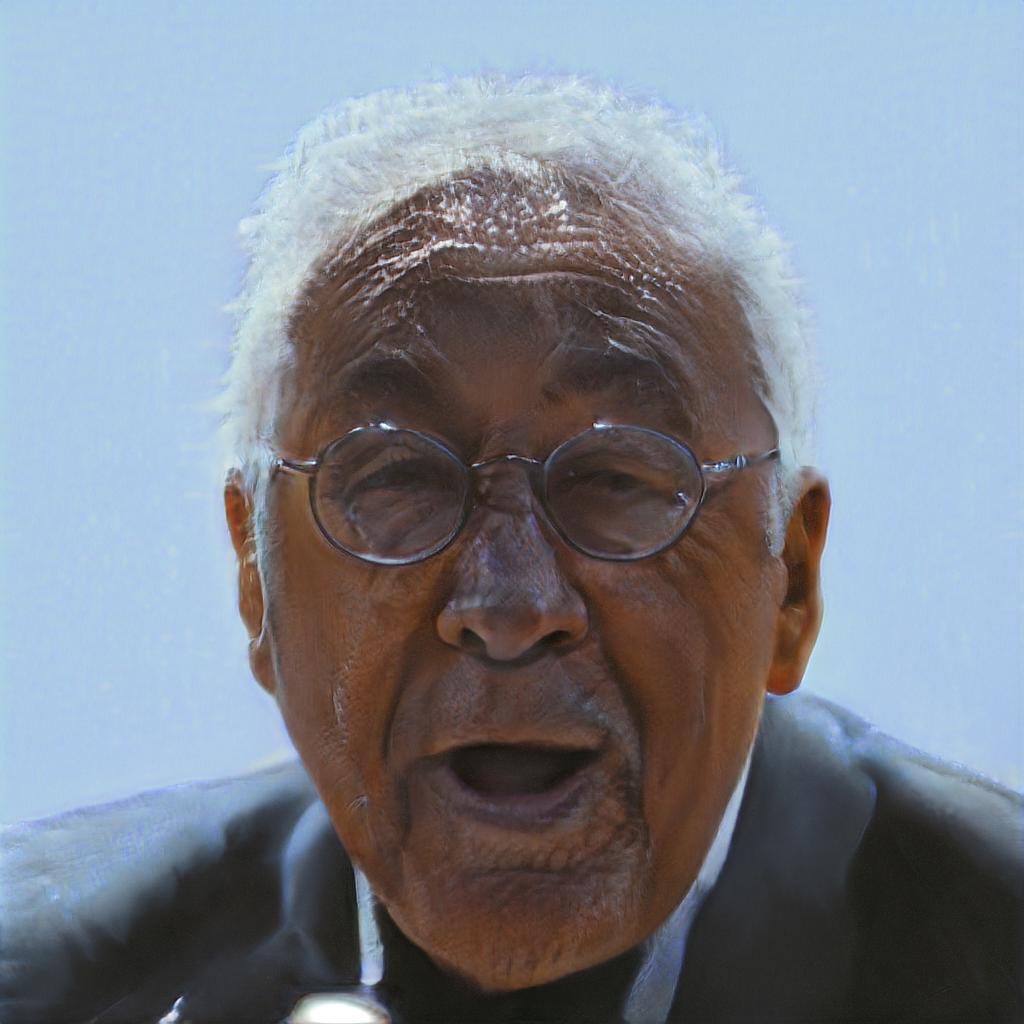} \\
\end{tabular}
\caption{Master face images with the highest MSC score, generated by (a) Random Search. (b) DE. (c) LSHADE-RSP. (d) IMODE. (e) CMA-ESA. (f) NGOpt. (g) lq-CMA-ES. (h) ACM-ES. (i) LM-MA-ES. (j) LM-MA-ES + LQ-Filter. (k) LM-MA-ES + Success Predictor.}
\label{fig:gen_faces}
\end{figure}

In general, algorithms from the CMA-ES family perform better than other algorithms, like the DE family. In particular, the original implementation of the LM-MA-ES\cite{loshchilov2017limited} performs better than the CMA-ES as expected on a high-dimensional optimization problem. In fact, LM-MA-ES performs the best among the original variants with no additional assisting models. 
It is worth mentioning that even though the lq-CMA-ES achieved worse results than the original CMA-ES, as might be expected from its usage of predicted fitness values instead of the actual ones, its training process is faster. 
In Fig.~\ref{fig:gen_faces}, we present the generated master face images with the highest MSC score for each of the face descriptors. 

\begin{table}[t]
\caption{Generated master faces MSC score (\%) for different optimization methods.}
\centering
\begin{tabular}{@{}l@{~}c@{~}c@{~}c@{~}c@{~}c@{~}c@{}}
\toprule& 
\multicolumn{2}{c}{Dlib} &
\multicolumn{2}{c}{FaceNet} & 
\multicolumn{2}{c}{SphereFace}
\\
\cmidrule(lr){2-3}
\cmidrule(lr){4-5}
\cmidrule(lr){6-7}

Optimization Algorithm & Train & Test & Train & Test 
& Train & Test\\
\midrule
RandomSearch 
& $5.78$ & $4.73$
& $9.69$ & $9.94$
& $7.60$ & $6.78$ \\
DE 
& $7.08$ & $5.67$
& $12.29$ & $11.87$  
& $14.12$ & $13.68$ \\
LSHADE-RSP
& $10.09$ & $8.30$
& $15.90$ & $15.20$ 
& $15.73$ & $11.52$ \\
IMODE
& $7.70$ & $6.06$
& $12.36$ & $12.04$
& $11.64$ & $11.52$ \\
CMA-ES 
& $18.17$ & $16.32$
& $16.75$ & $16.43$ 
& $16.84$ & $15.49$ \\
NGOpt 
& $14.04$ & $12.71$ 
& $16.62$ & $15.20$ 
& $17.24$ & $16.37$\\
lq-CMA-ES
& $7.38$ & $6.12$
& $13.10$ & $12.57$ 
& $9.31$ & $9.82$ \\
ACM-ES
& $18.09$ & $17.61$
& $16.00$ & $14.50$ 
& $16.67$ & $16.84$\\
LM-MA-ES (LME)
& $18.27$ & $17.97$ 
& $17.12$ & $\mathbf{16.96}$ 
& $17.39$ & $16.96$\\
LME + LQ-Filter
& $18.87$ & $16.67$ 
& $16.60$ & $14.91$ 
& $\mathbf{17.79}$ & $17.42$\\
LME + Success Pred.
& $\mathbf{22.43}$ & $\mathbf{21.56}$ 
& $\mathbf{17.36}$ & $16.60$ 
& $17.51$ & $\mathbf{17.60}$\\ 
\bottomrule
\end{tabular}
\label{table:ea_comp}
\end{table}

\begin{table}
\caption{Percentage of dataset covered by nine generated images}
\label{table:ds_cover}
\begin{center}
\begin{tabular}{lccc}
\toprule
FD & Dlib & FaceNet & SphereFace  \\
\midrule
\makecell[l]{Coverage Search LM-MA-ES\\on clustered data}& $51.43\%$ &  $40.17\%$ & $39.78\%$\\ 
\hline
\makecell[l]{$greedy$-Coverage Search\\ LM-MA-ES}& $63.22\%$ &  $42.08\%$ & $43.14\%$\\ 
\hline
\makecell[l]{$greedy$-Coverage Search\\ LM-MA-ES+Success Predictor}& $\mathbf{63.92}\%$&  $\mathbf{43.82}\%$ & $\mathbf{44.15}\%$ \\ 
\bottomrule
\end{tabular}
\end{center}
\end{table}

\begin{figure*}
\vspace{-1mm}
\centering
\begin{tabular}{@{~}l@{~}c@{~}c@{~}c@{~}c@{~}c@{~}c@{~}c@{~}c@{~}c@{~}}
   (a) &
    \includegraphics[align=c,width=1.5cm]{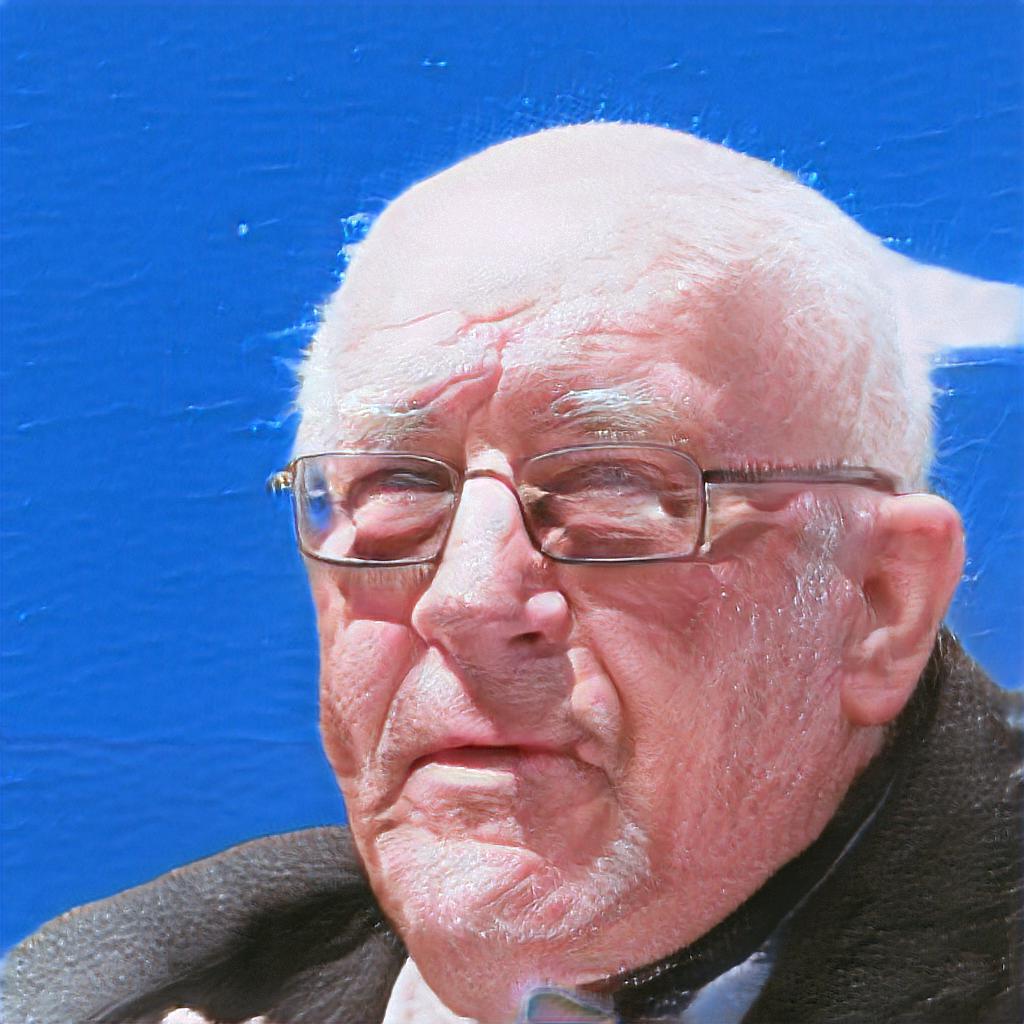} & \includegraphics[align=c,width=1.5cm]{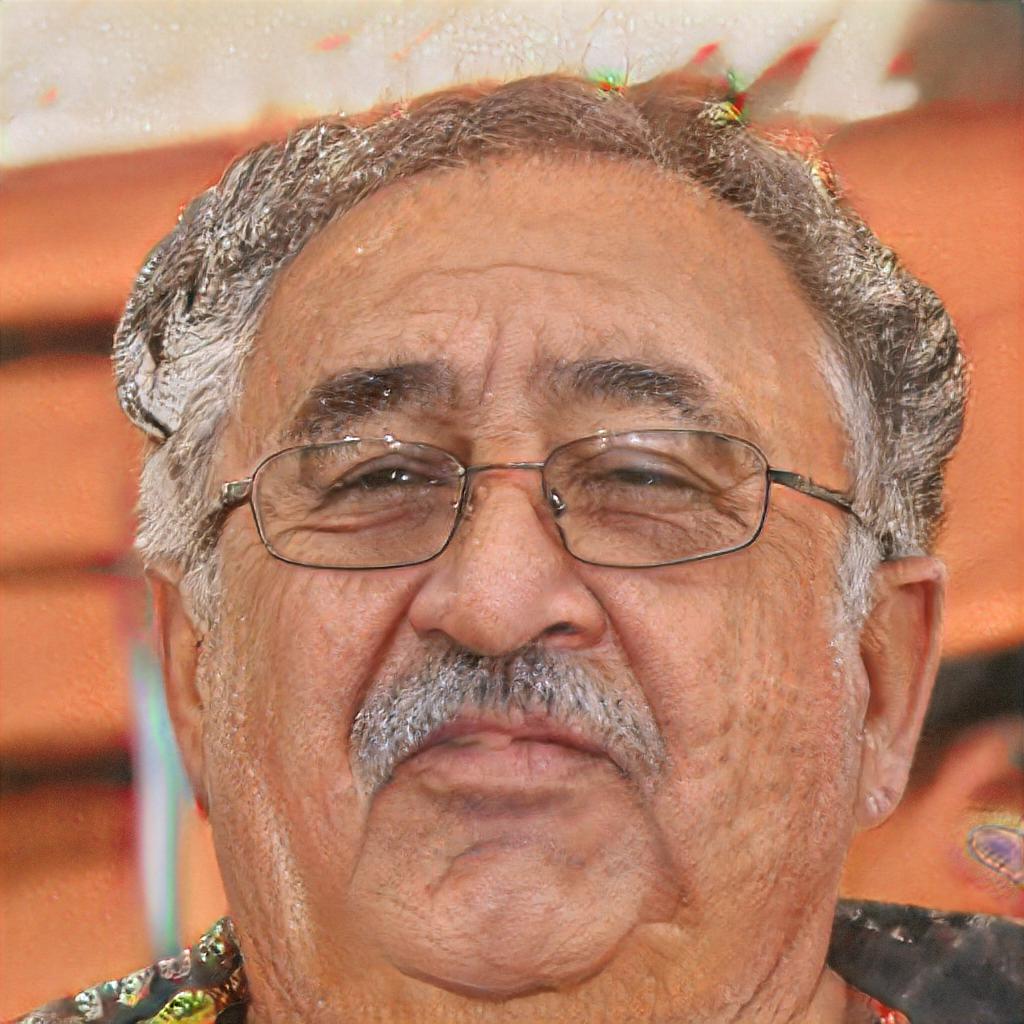} & \includegraphics[align=c,width=1.5cm]{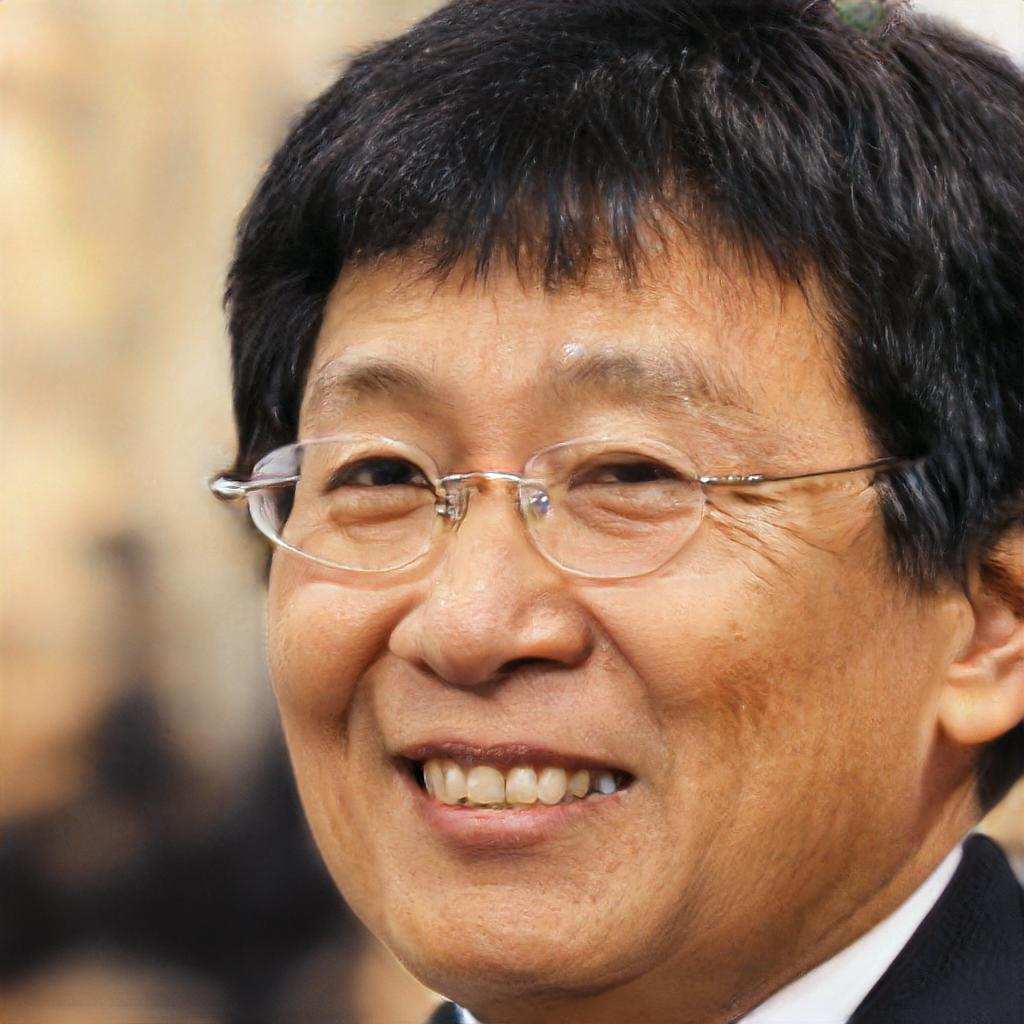} &  \includegraphics[align=c,width=1.5cm]{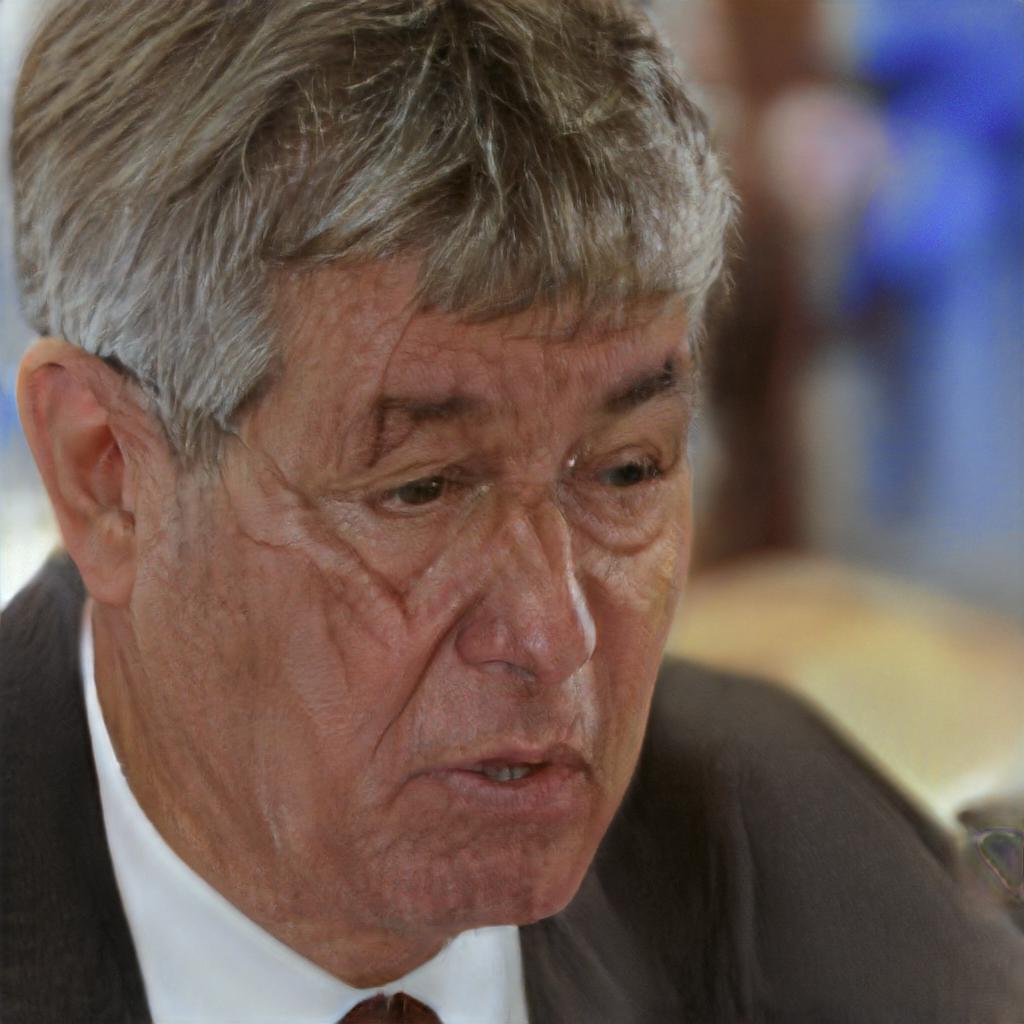} &
    \includegraphics[align=c,width=1.5cm]{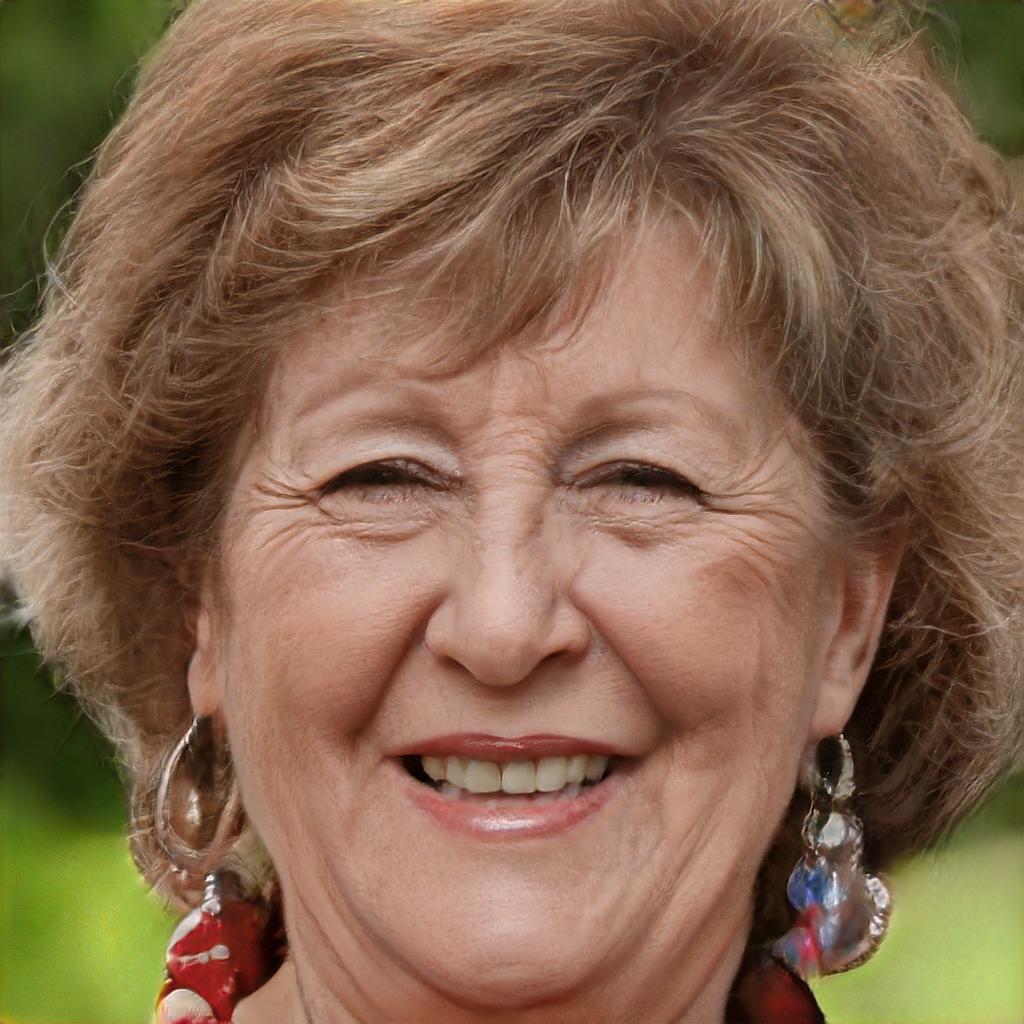} & \includegraphics[align=c,width=1.5cm]{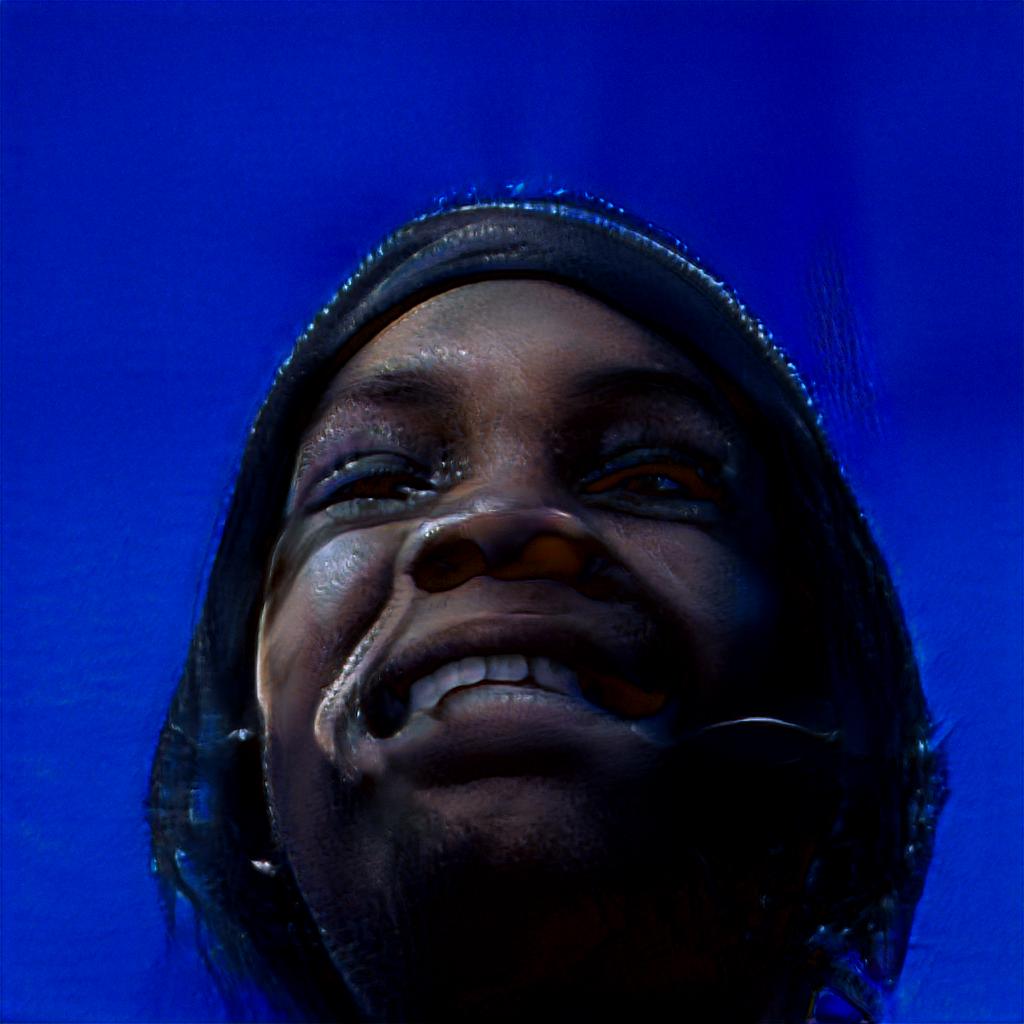} & \includegraphics[align=c,width=1.5cm]{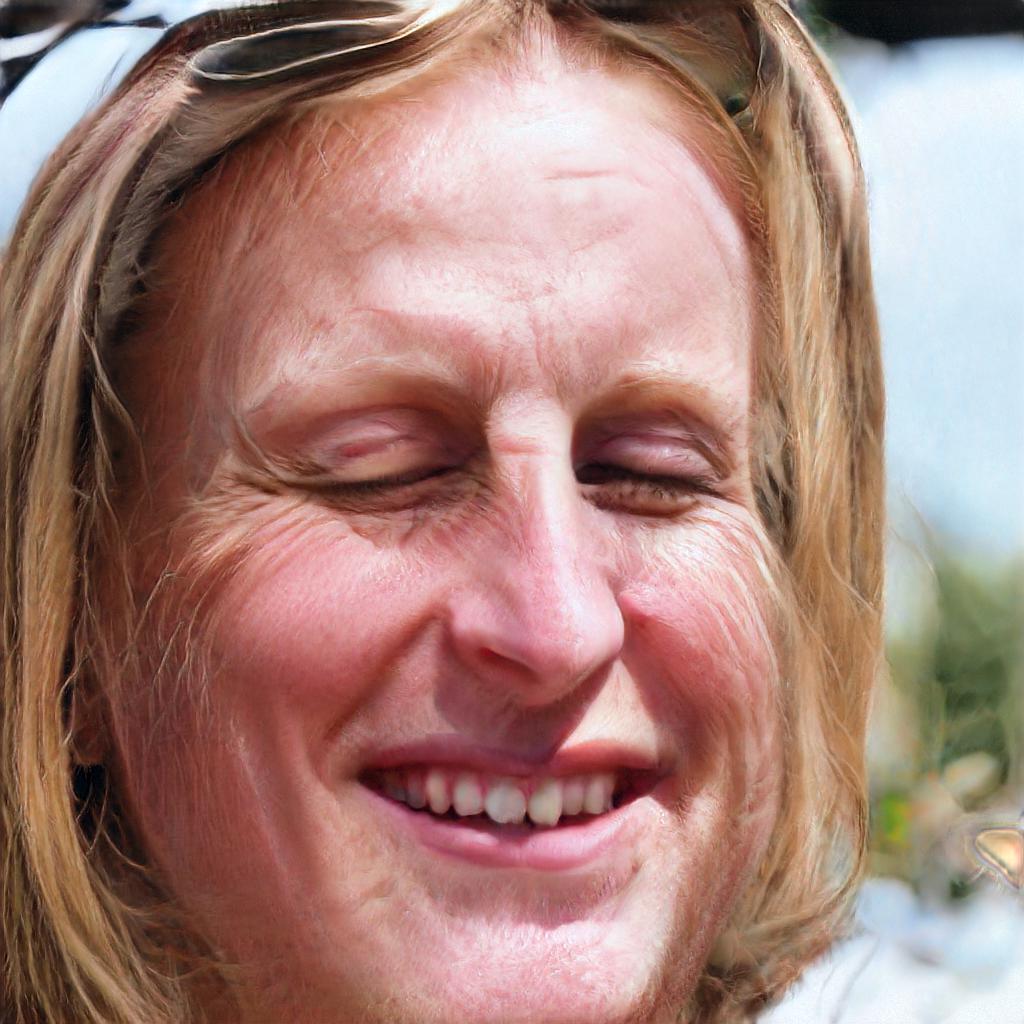} &  \includegraphics[align=c,width=1.5cm]{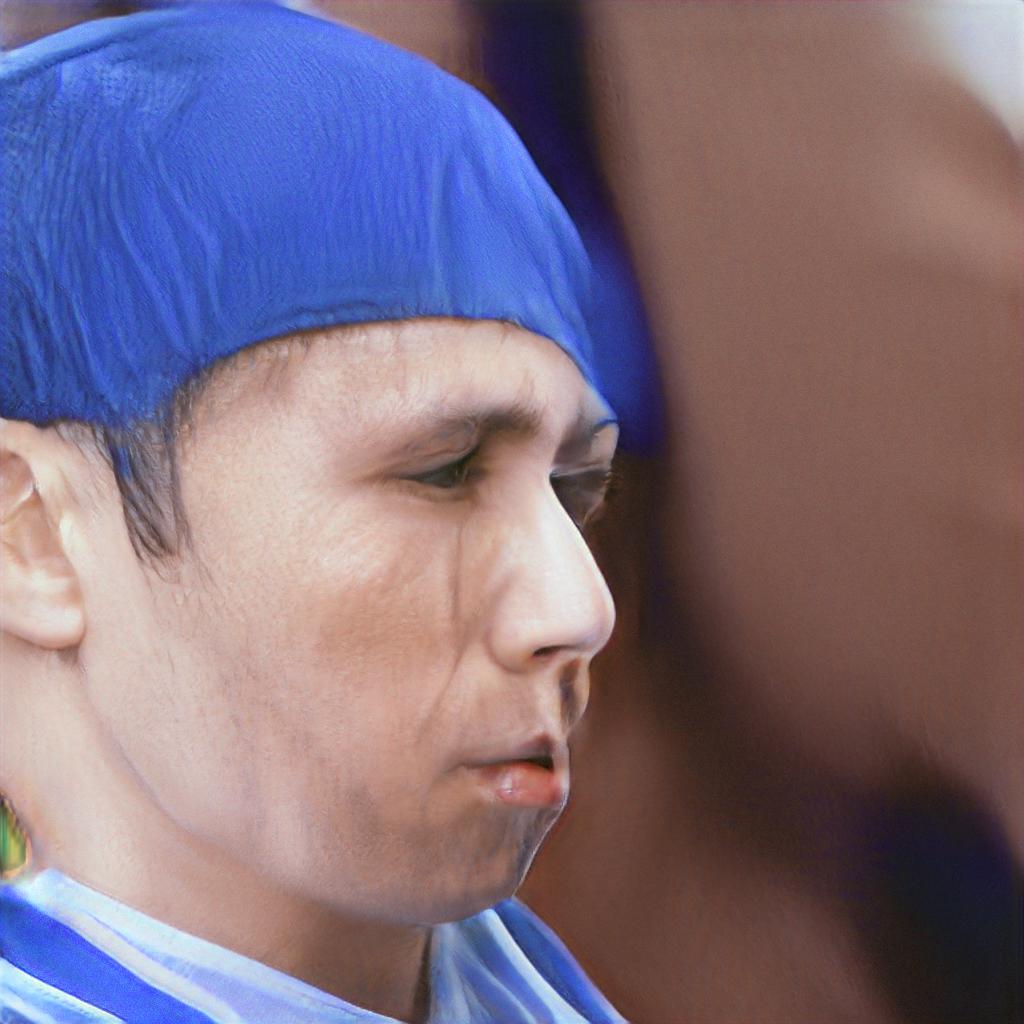} &
    \includegraphics[align=c,width=1.5cm]{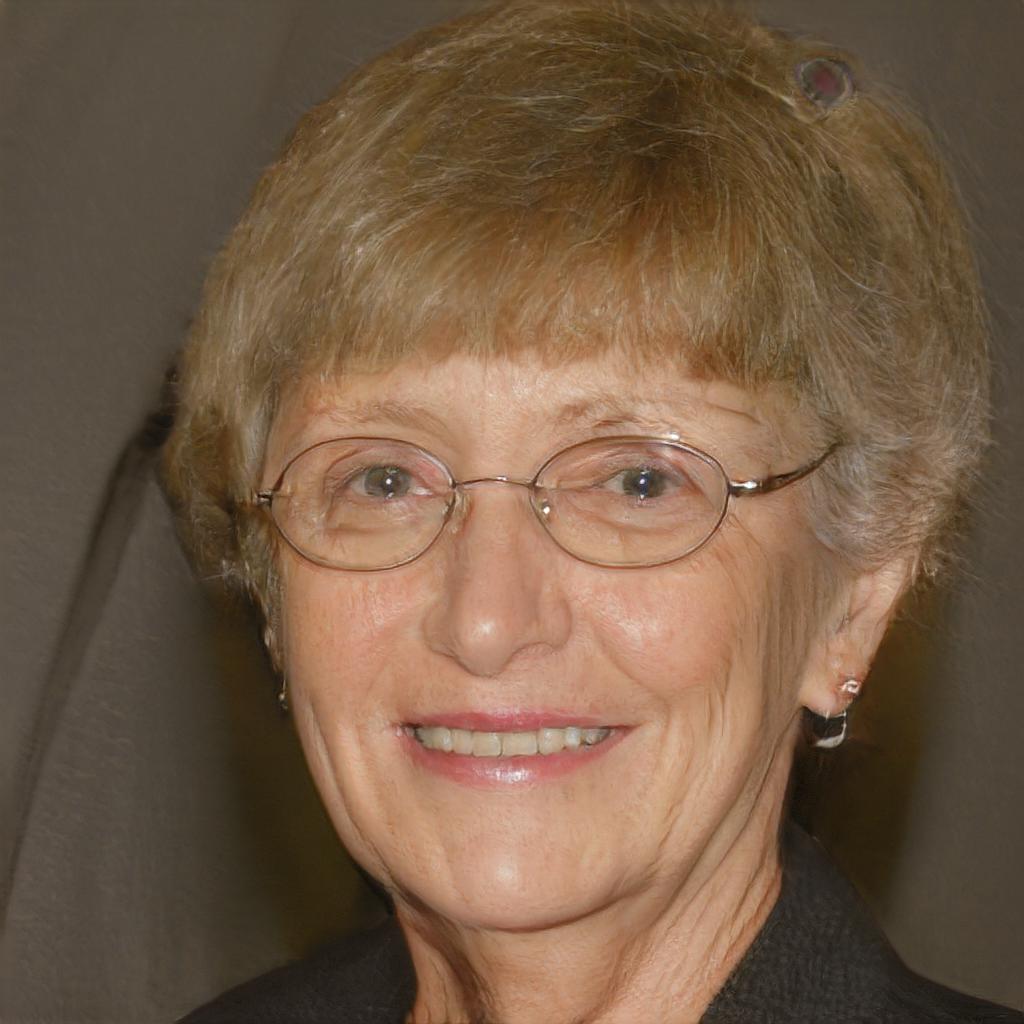} \\
    \small &$9.39\%$ & $7.00\%$ & $5.53\%$ & $4.94\%$ & $4.71\%$ & $3.62\%$ & $2.04\%$ & $1.70\%$ & $0.85\%$ \\
    (b) &
    \includegraphics[align=c,width=1.5cm]{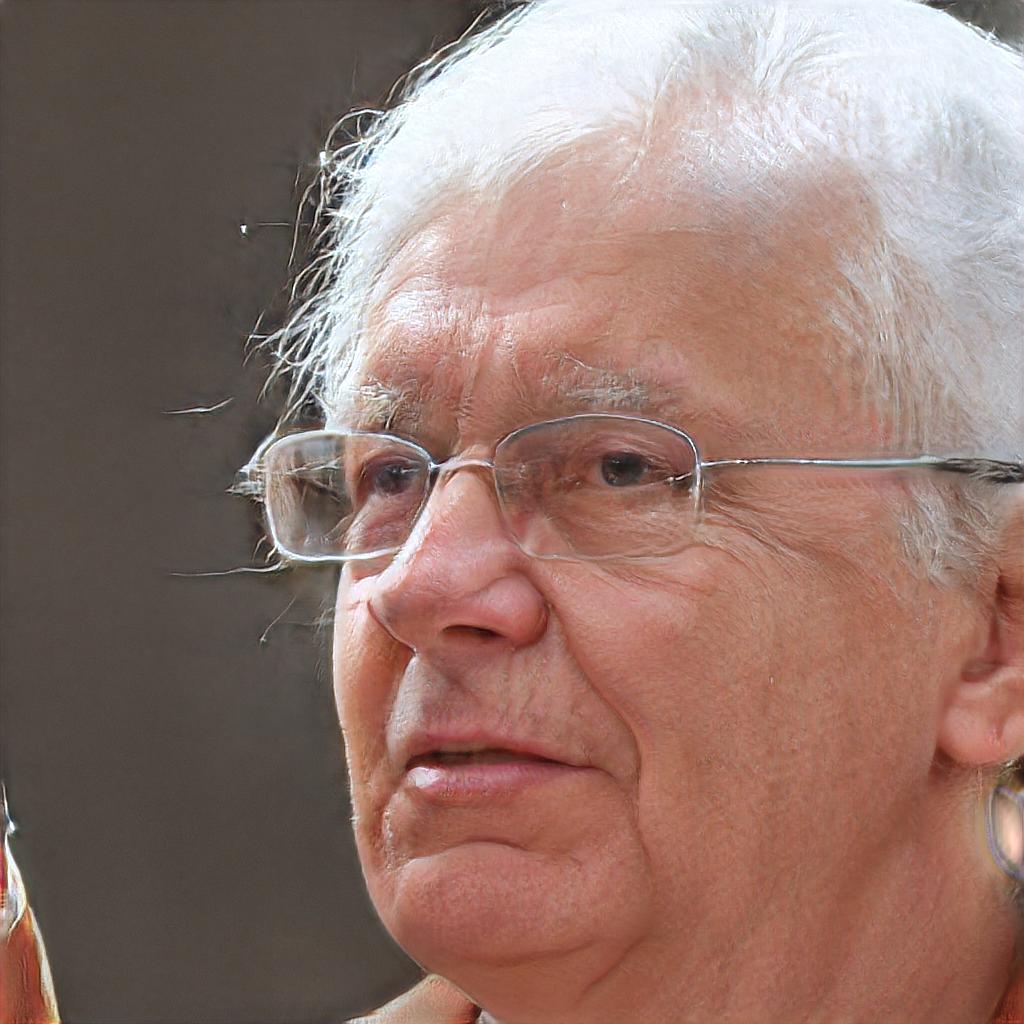} & \includegraphics[align=c,width=1.5cm]{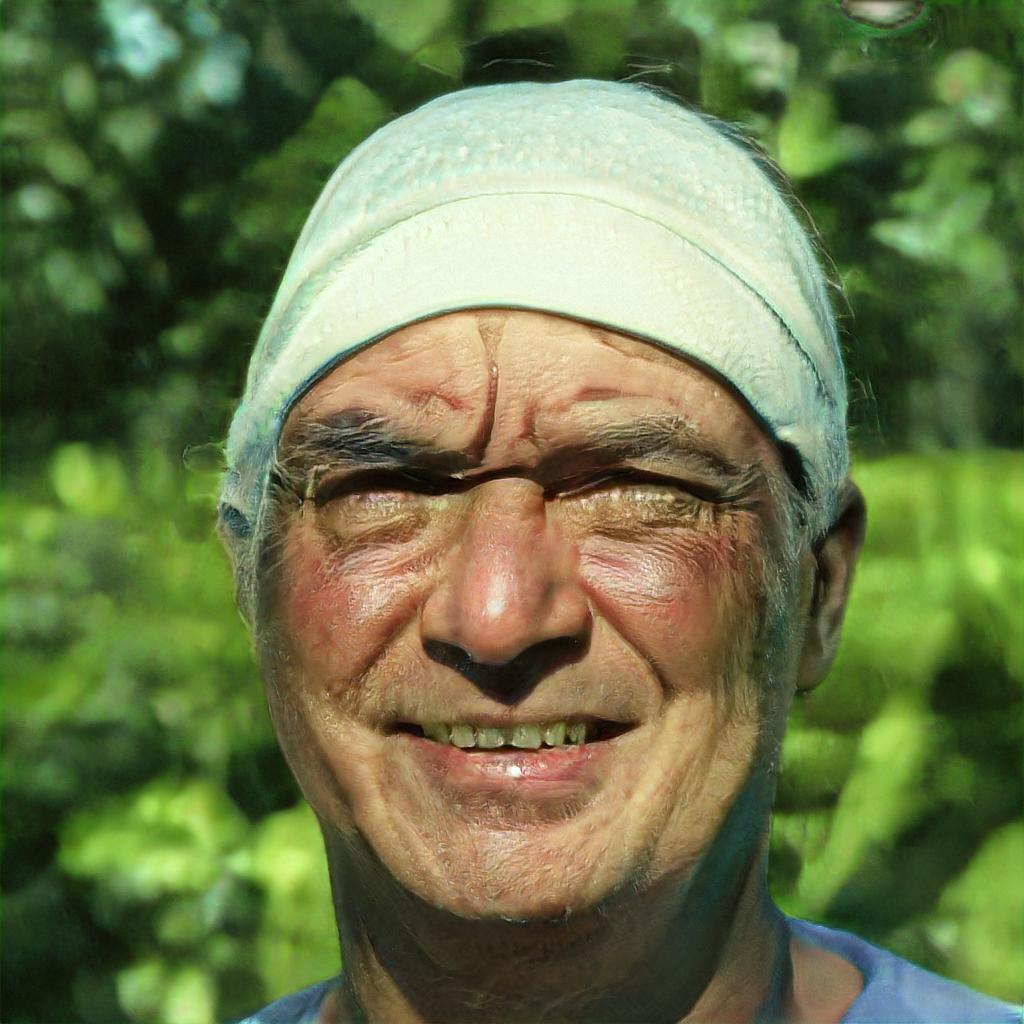} & \includegraphics[align=c,width=1.5cm]{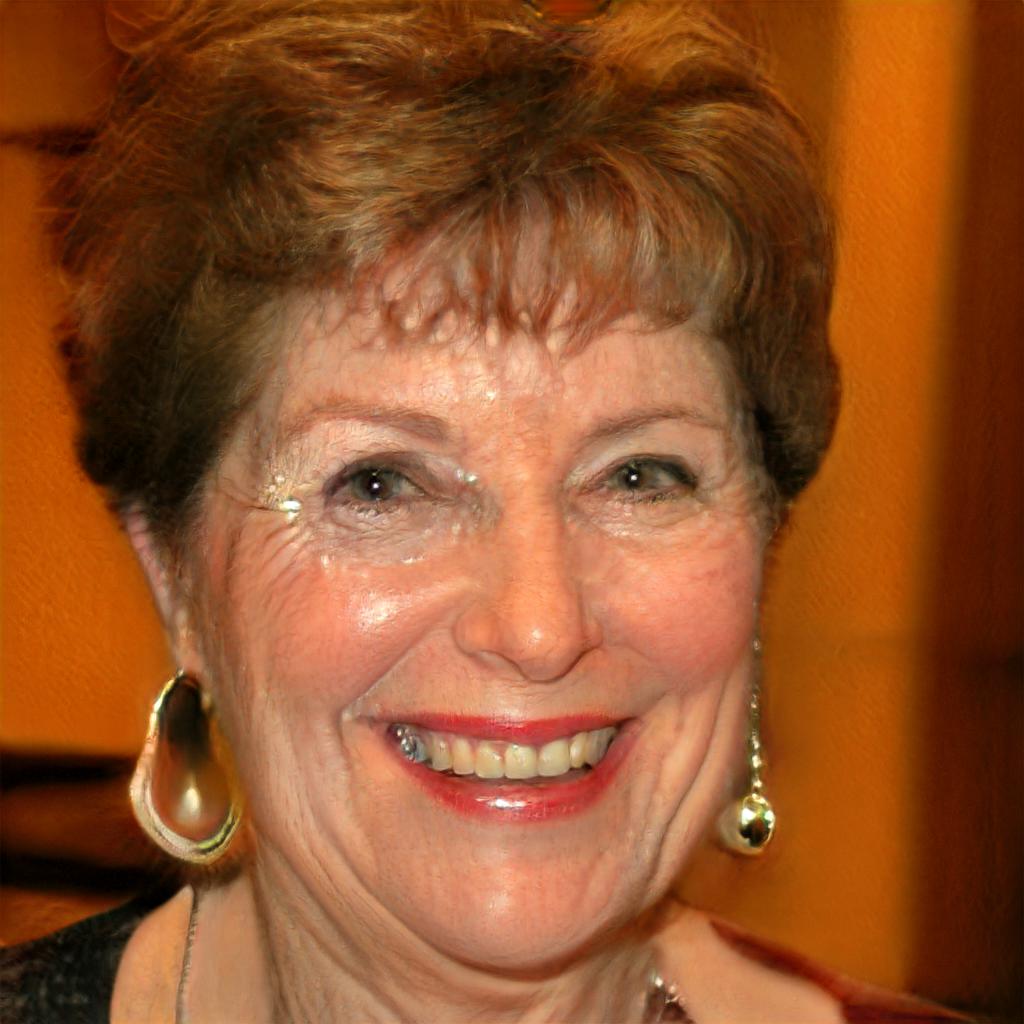} &      \includegraphics[align=c,width=1.5cm]{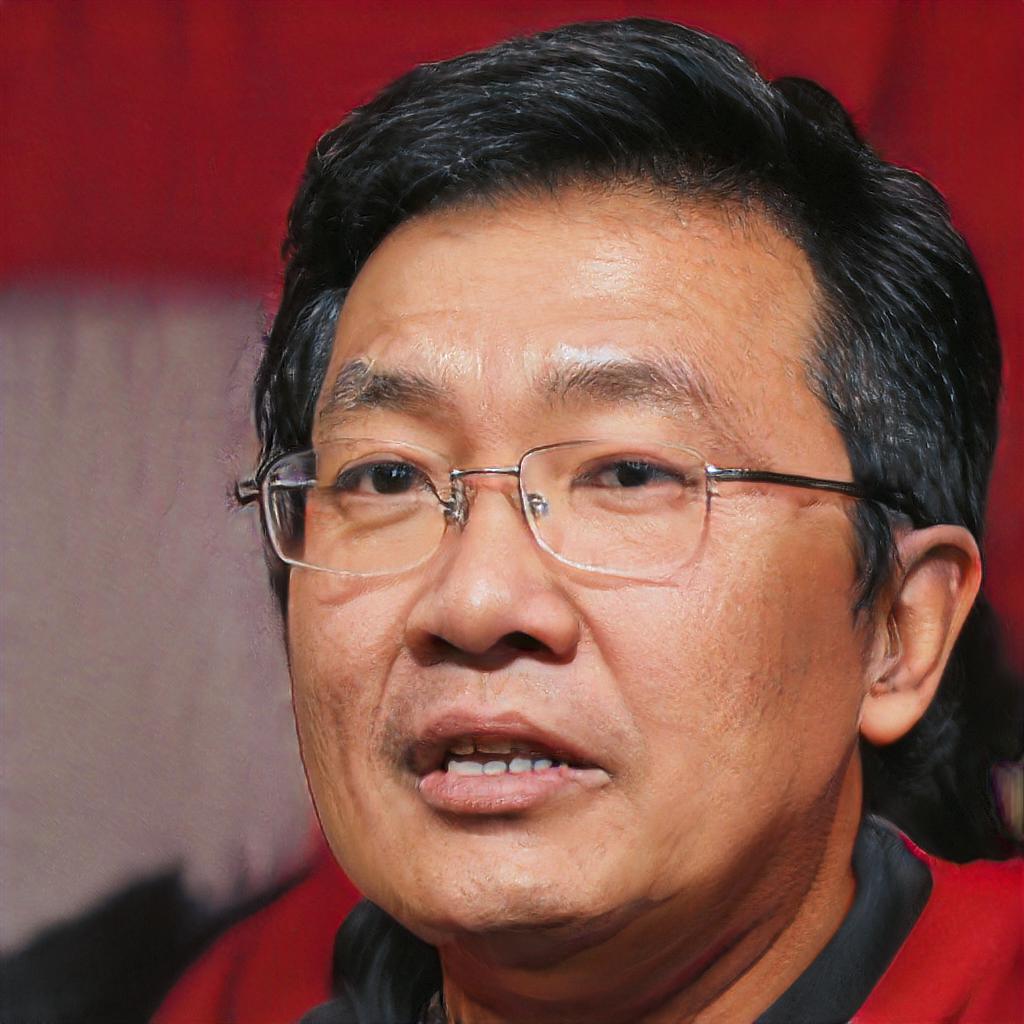} &   
    \includegraphics[align=c,width=1.5cm]{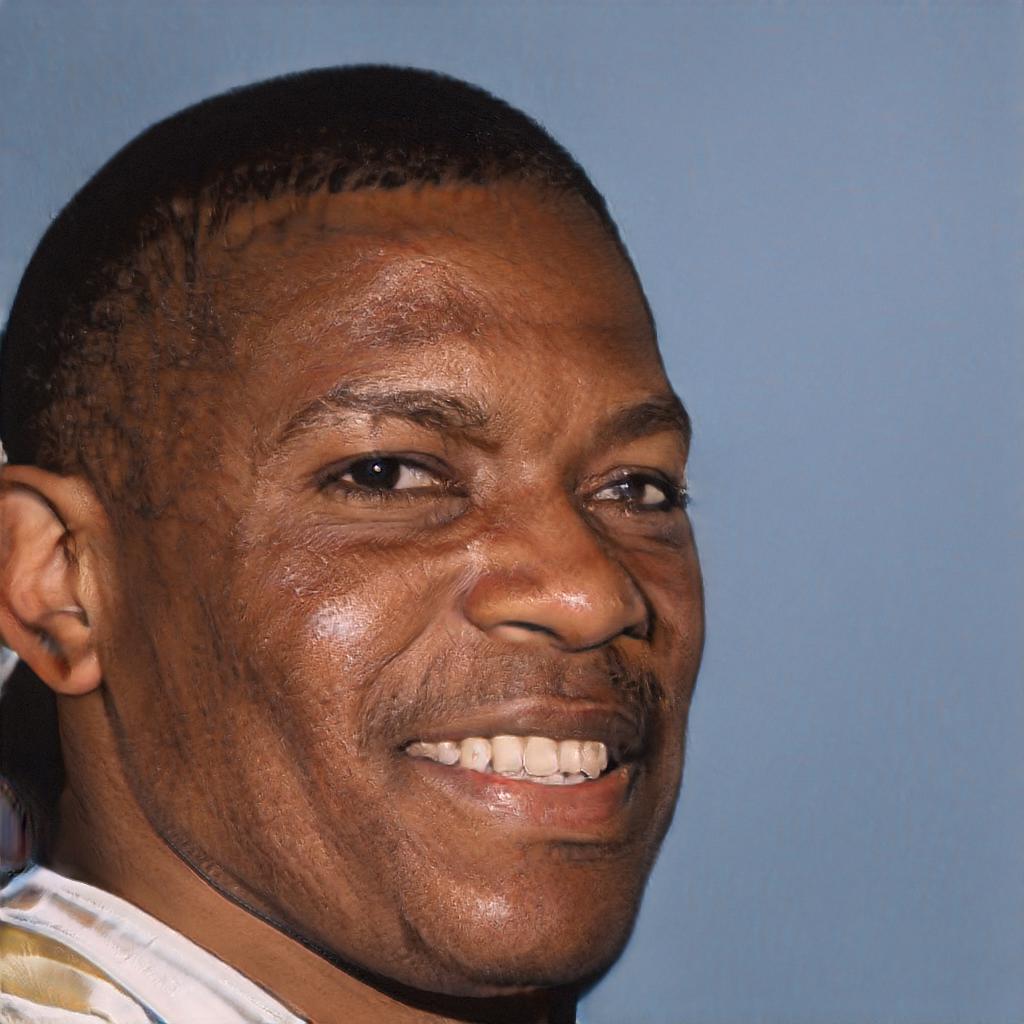} & \includegraphics[align=c,width=1.5cm]{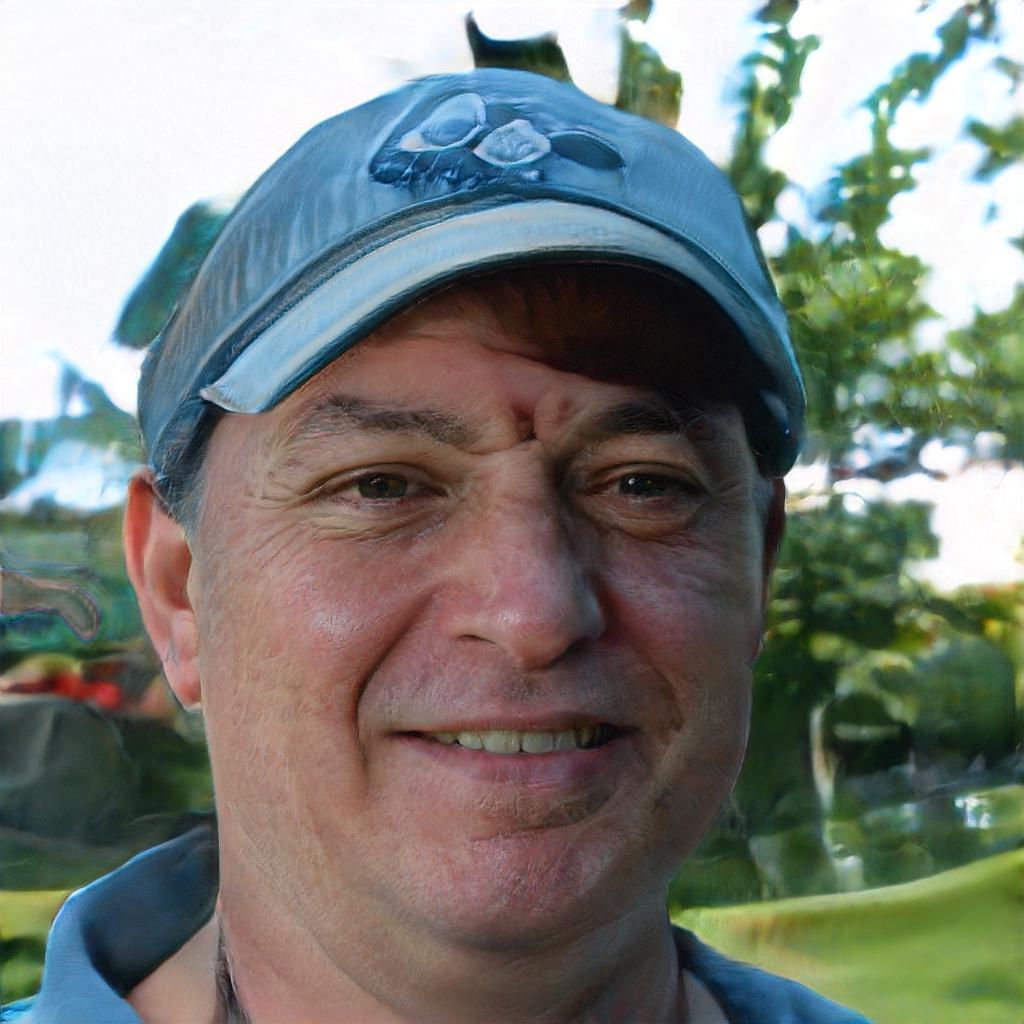} & \includegraphics[align=c,width=1.5cm]{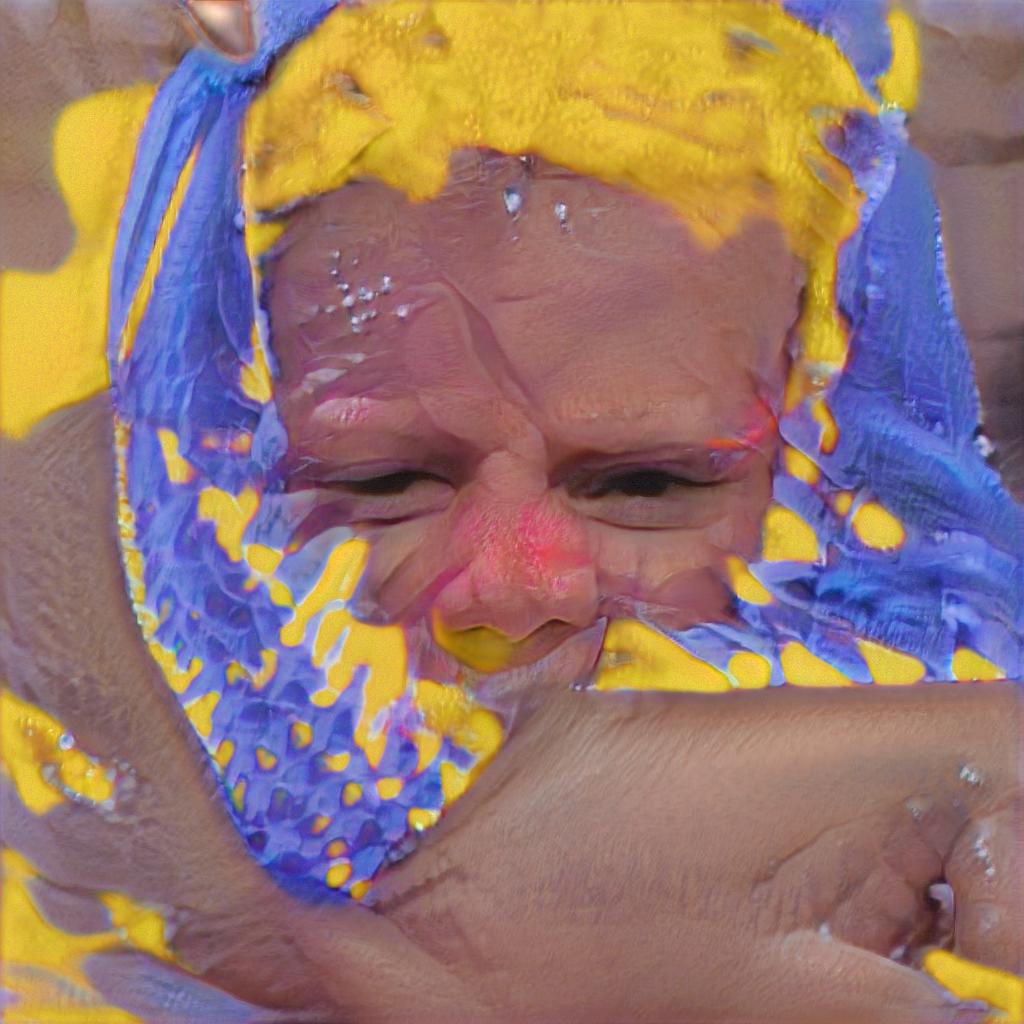} &     \includegraphics[align=c,width=1.5cm]{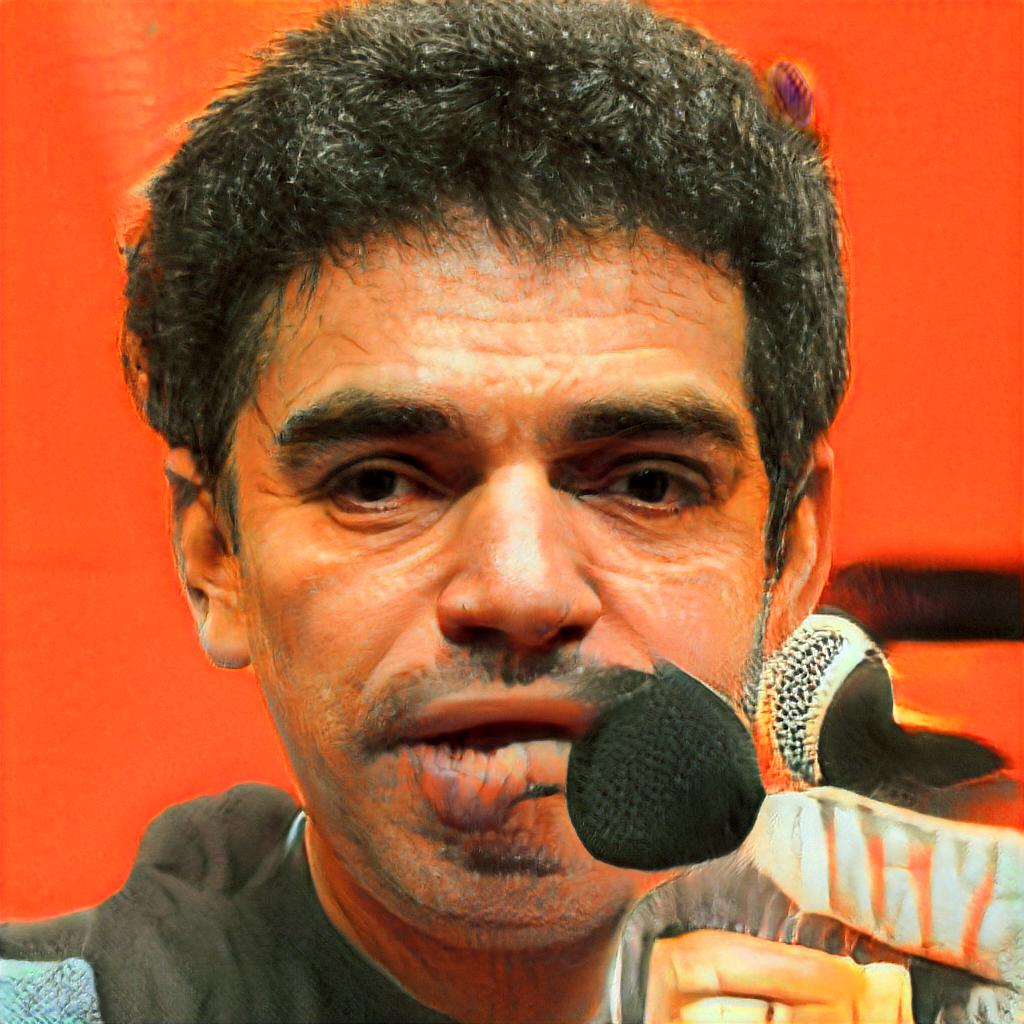} &   
    \includegraphics[align=c,width=1.5cm]{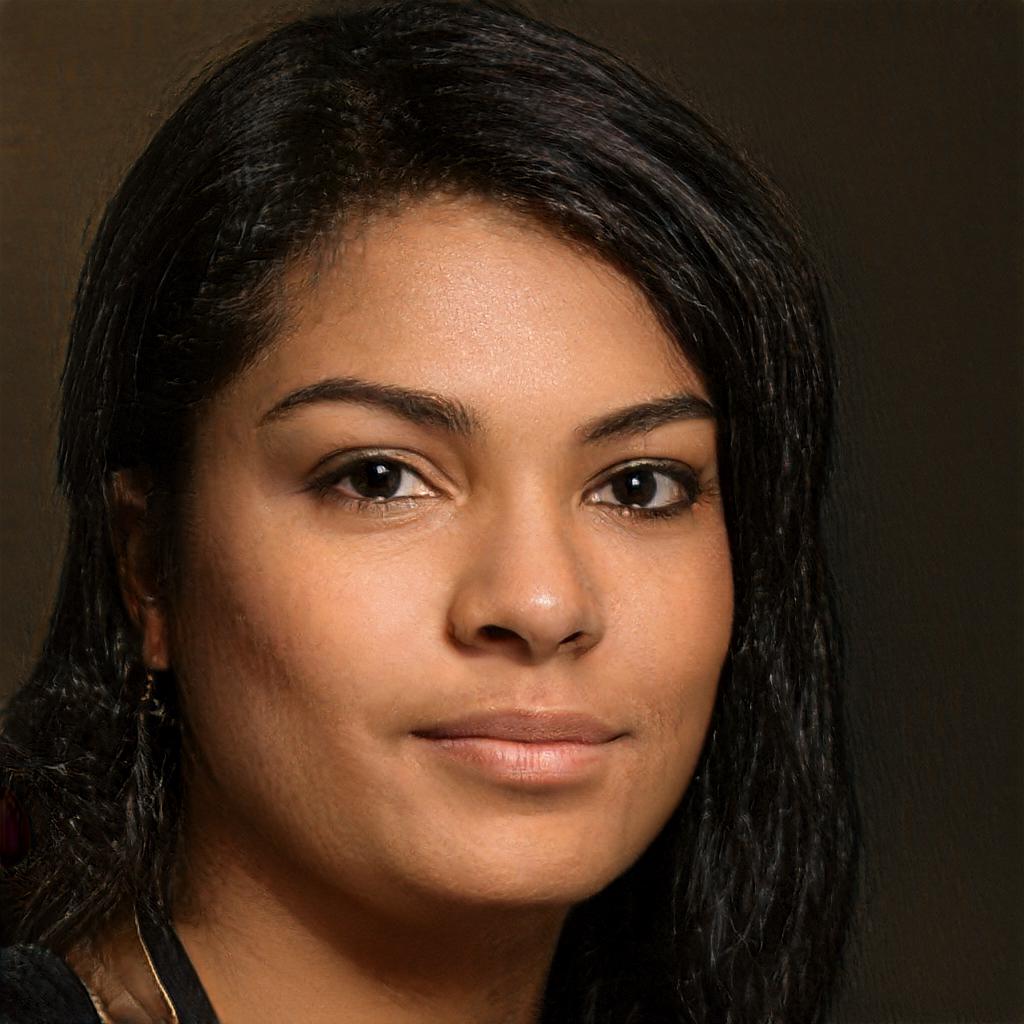} \\
      \small &$11.39\%$ & $6.45\%$ & $4.84\%$ & $4.82\%$ & $3.81\%$ & $3.32\%$ & $2.89\%$ & $1.67\%$ & $0.97\%$ \\
      (c) &
    \includegraphics[align=c,width=1.5cm]{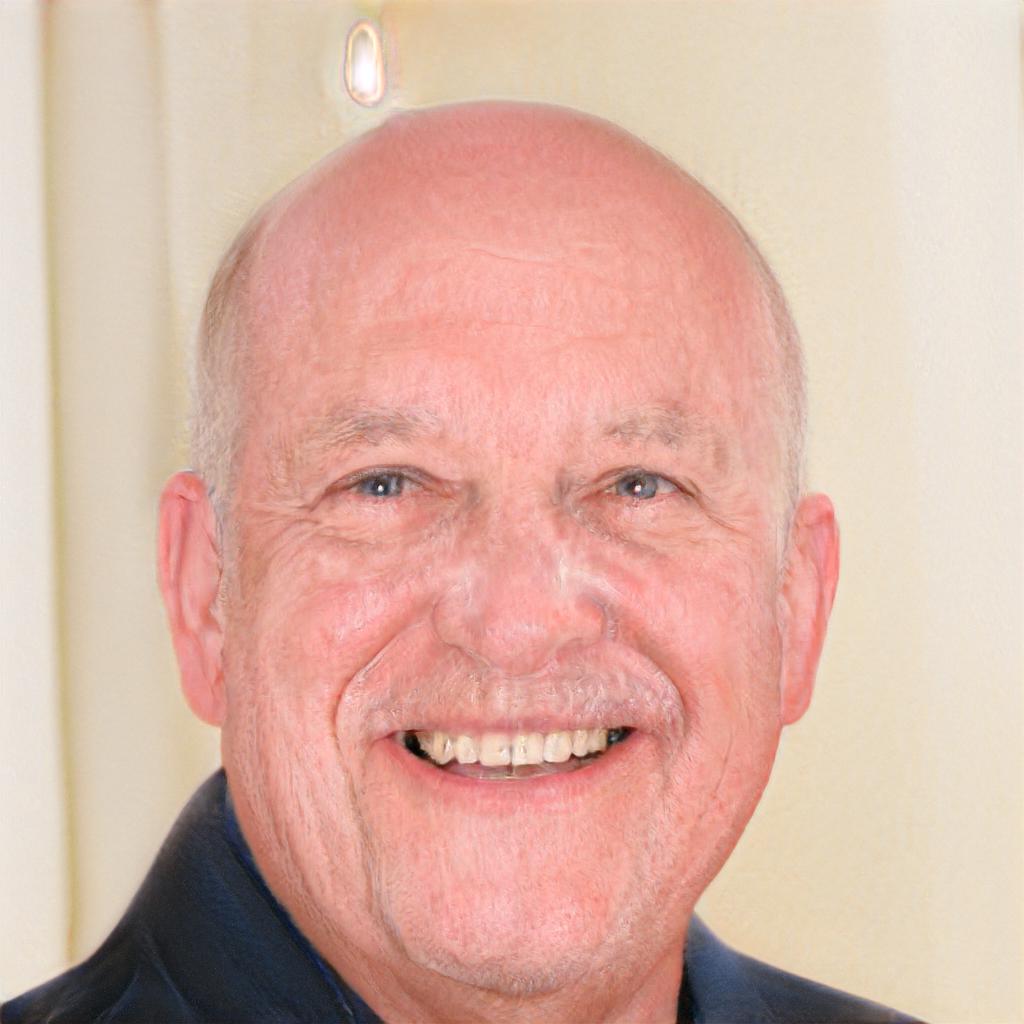} & \includegraphics[align=c,width=1.5cm]{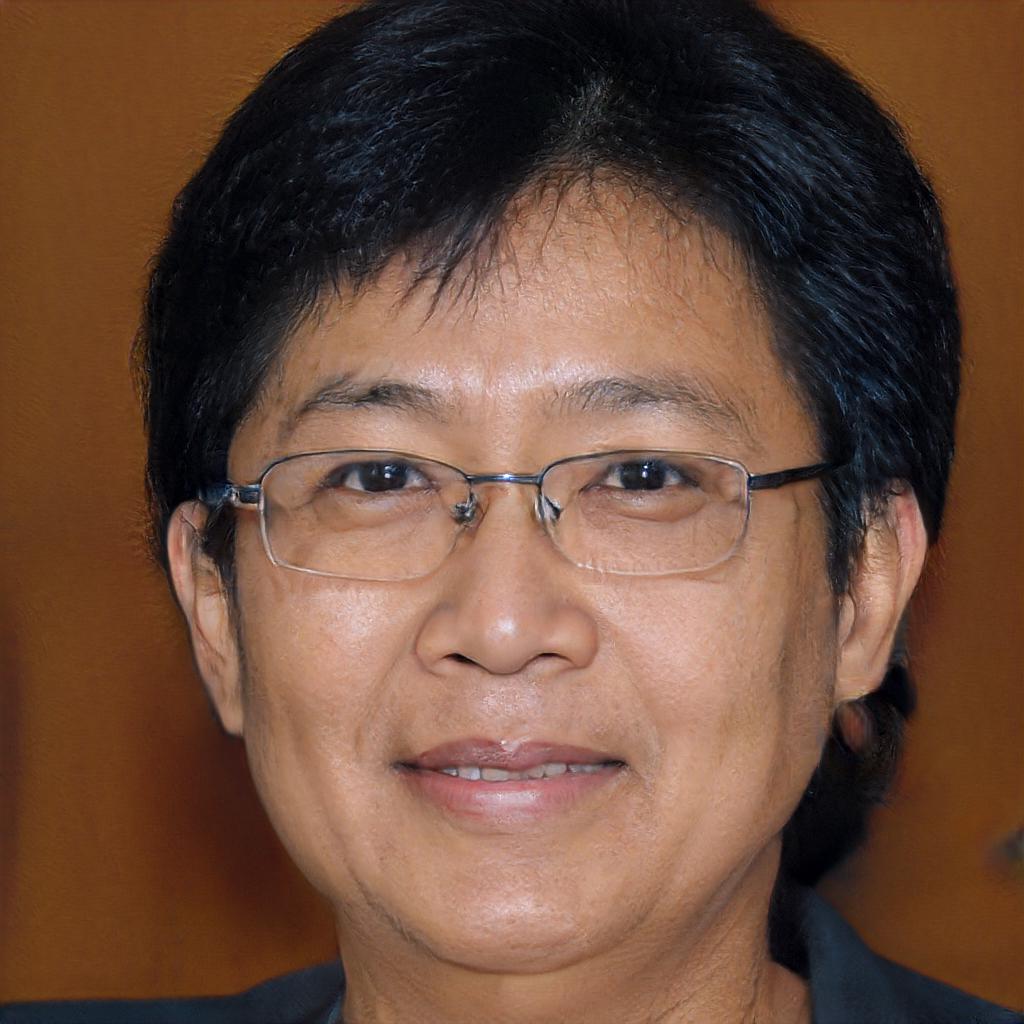} & \includegraphics[align=c,width=1.5cm]{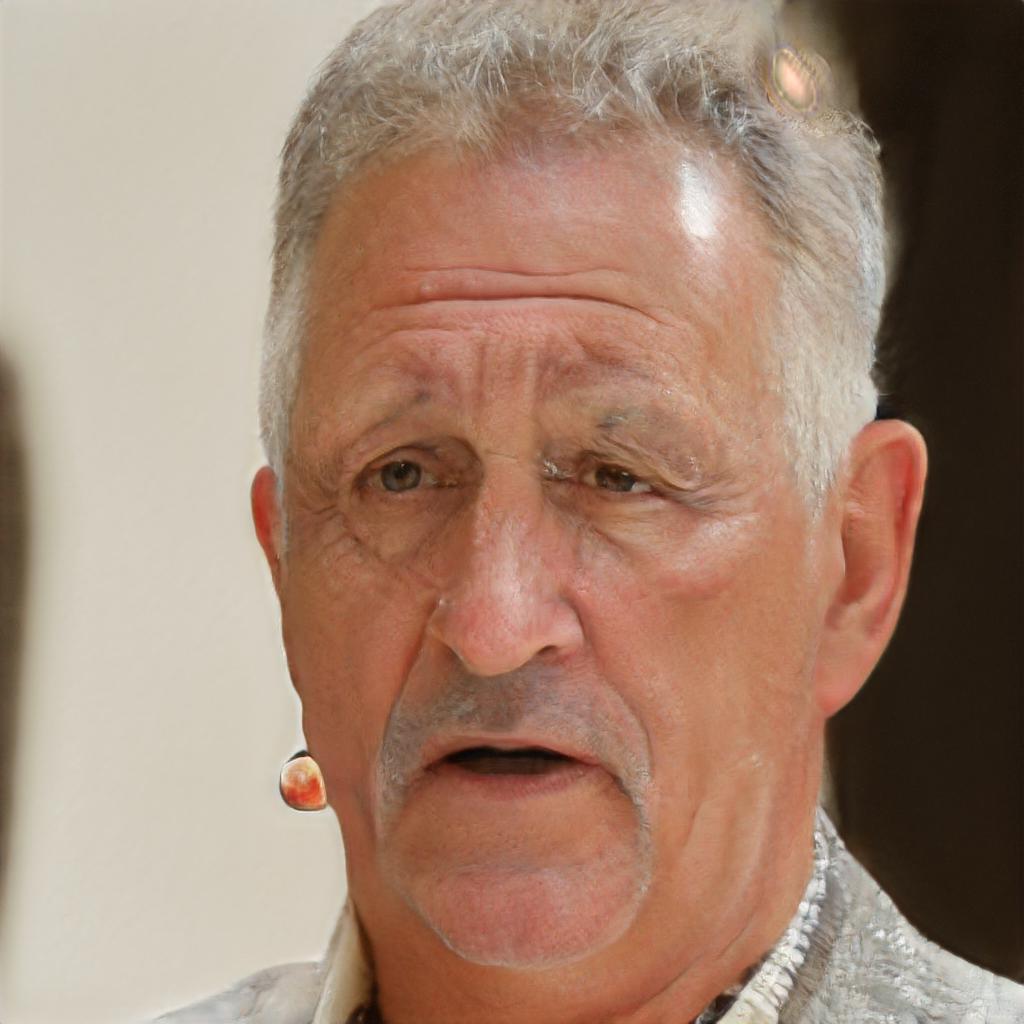} &      \includegraphics[align=c,width=1.5cm]{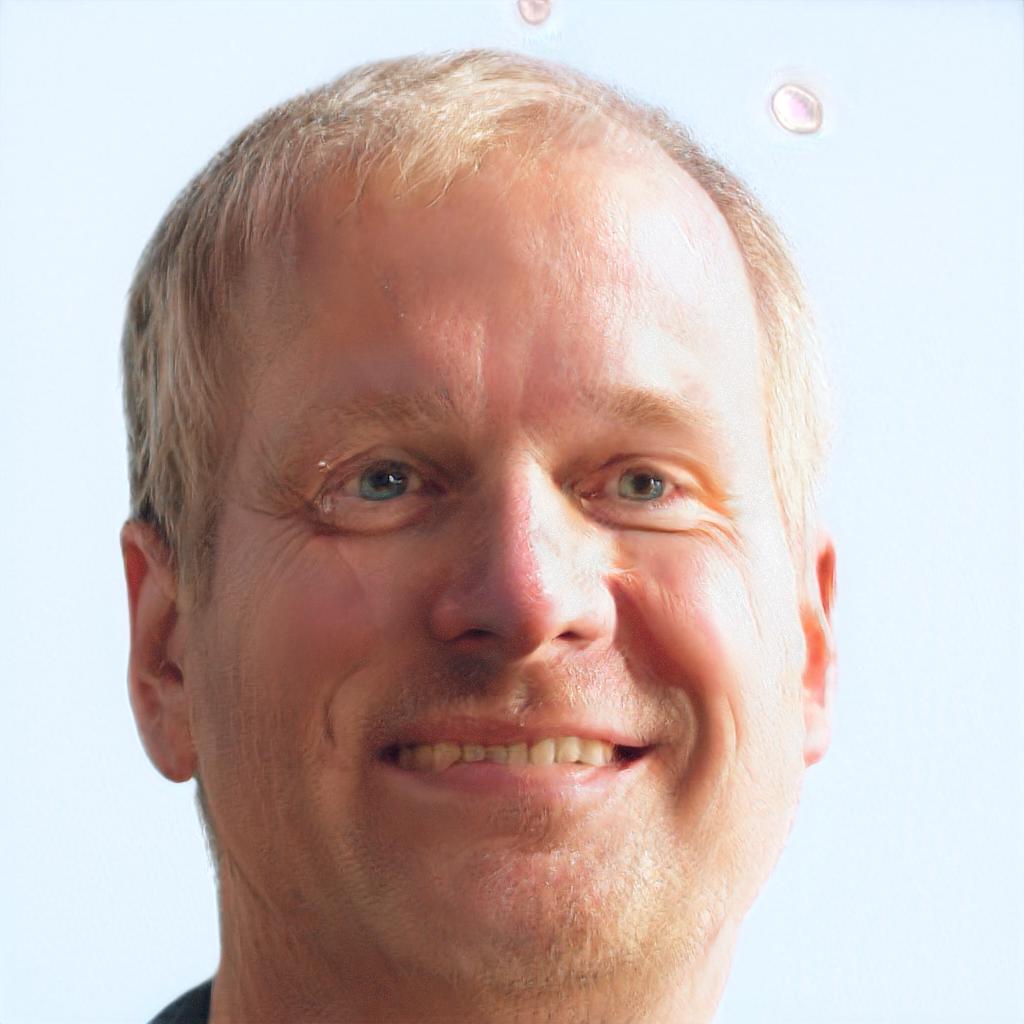} &   
    \includegraphics[align=c,width=1.5cm]{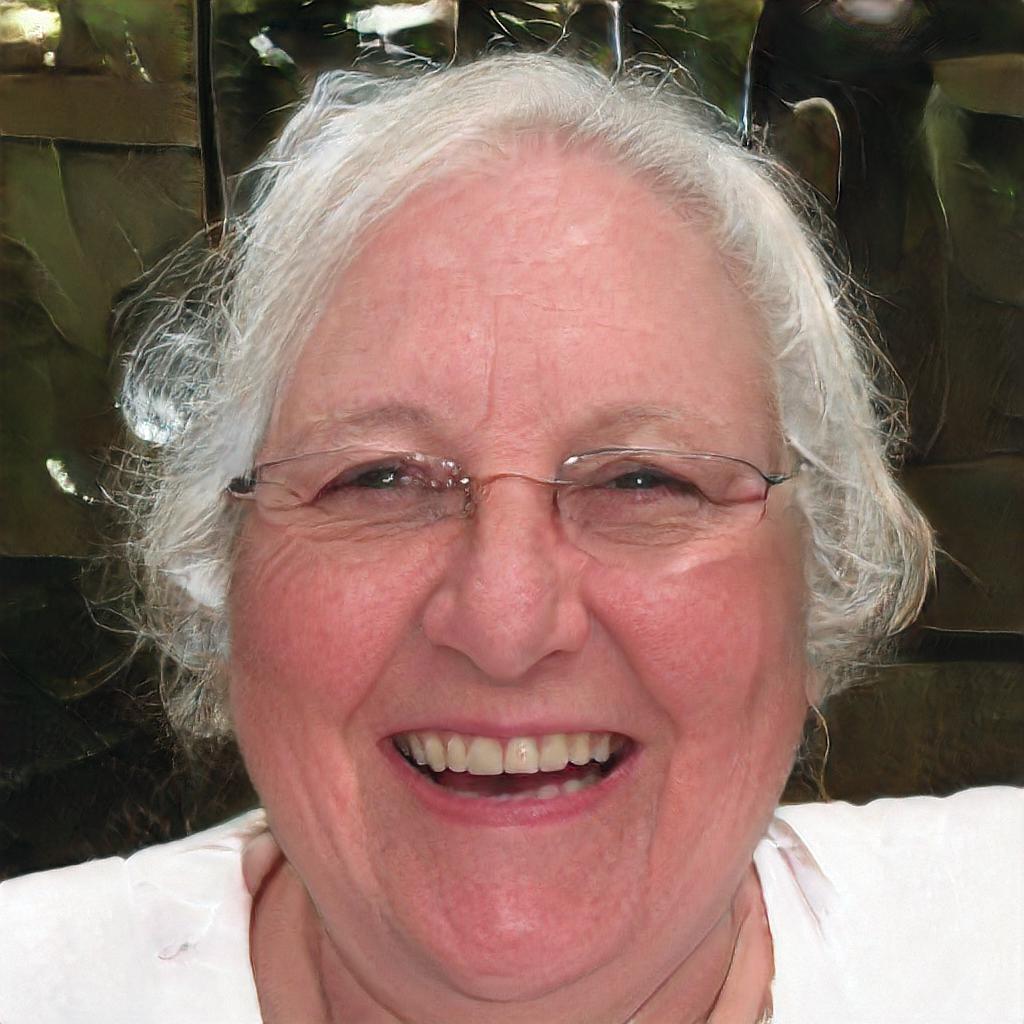} & \includegraphics[align=c,width=1.5cm]{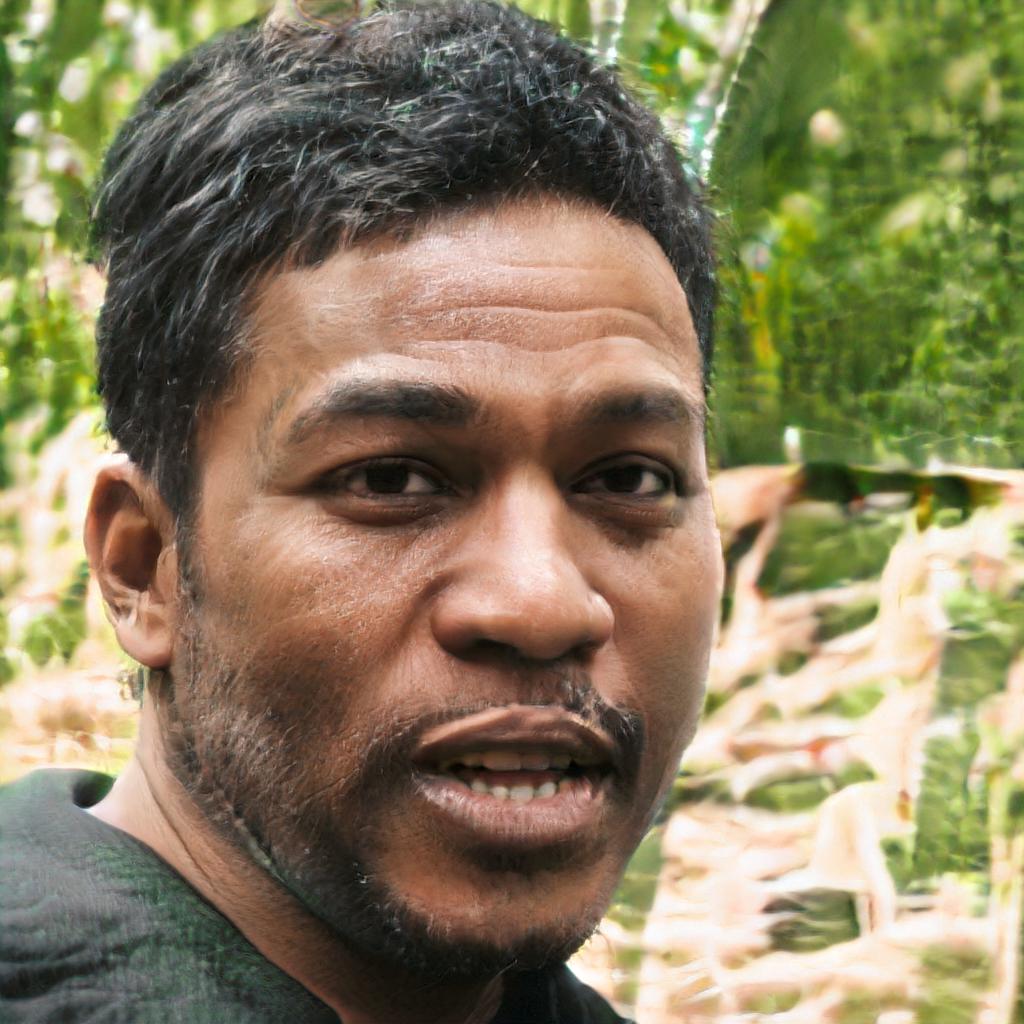} & \includegraphics[align=c,width=1.5cm]{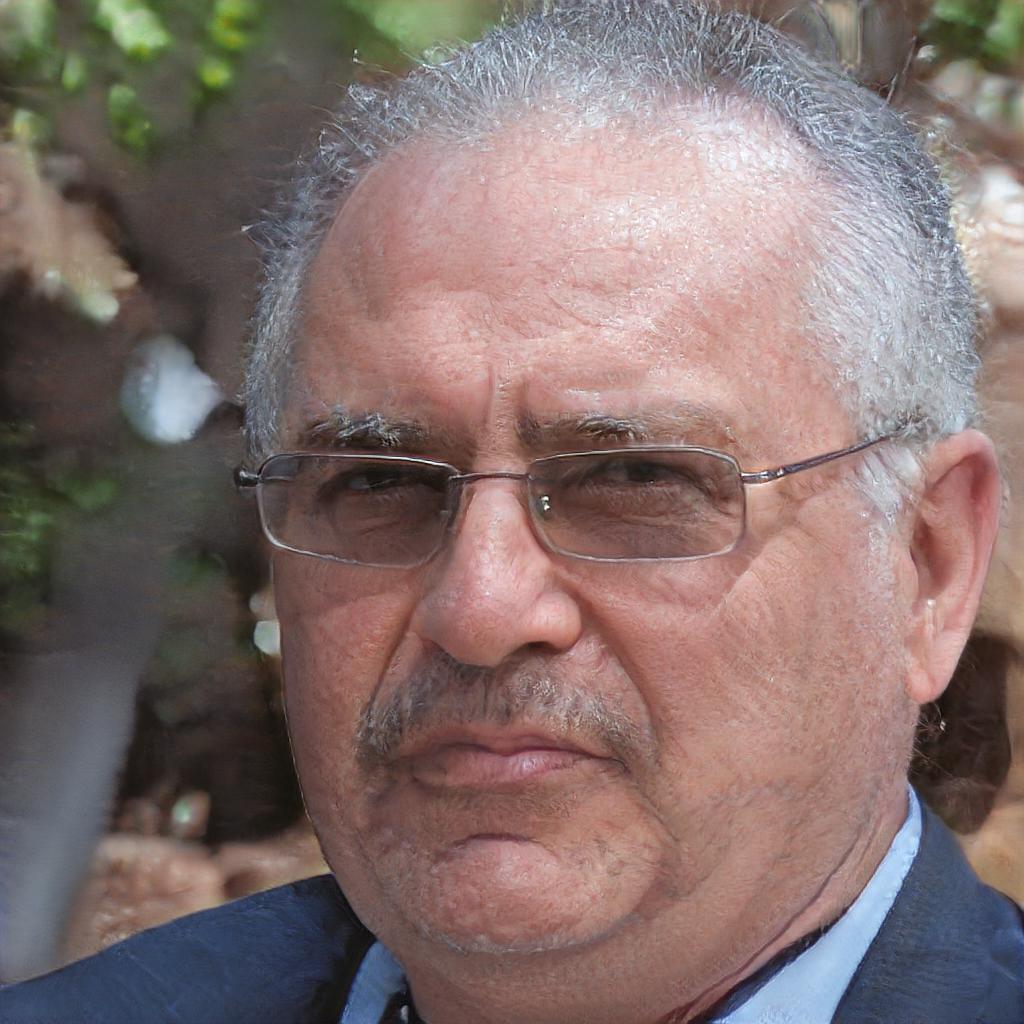} &      \includegraphics[align=c,width=1.5cm]{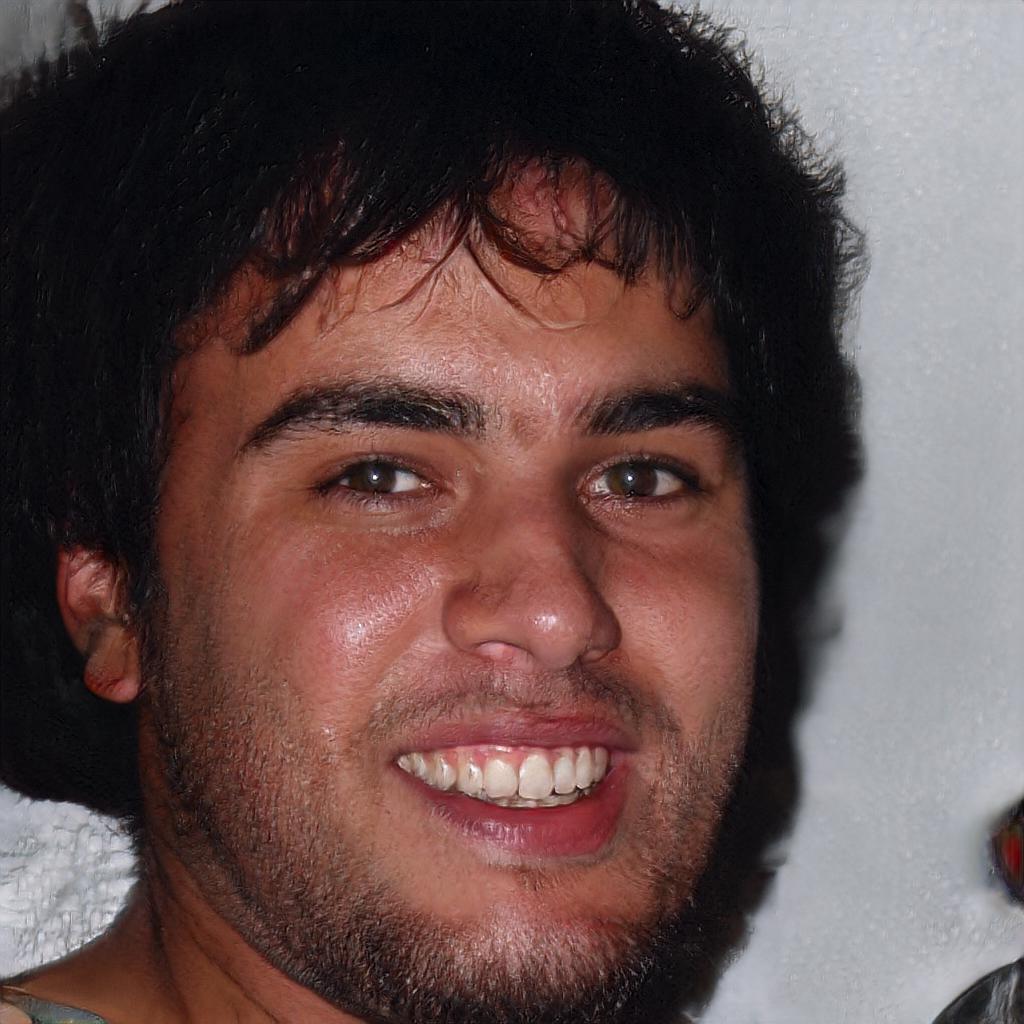} &   
    \includegraphics[align=c,width=1.5cm]{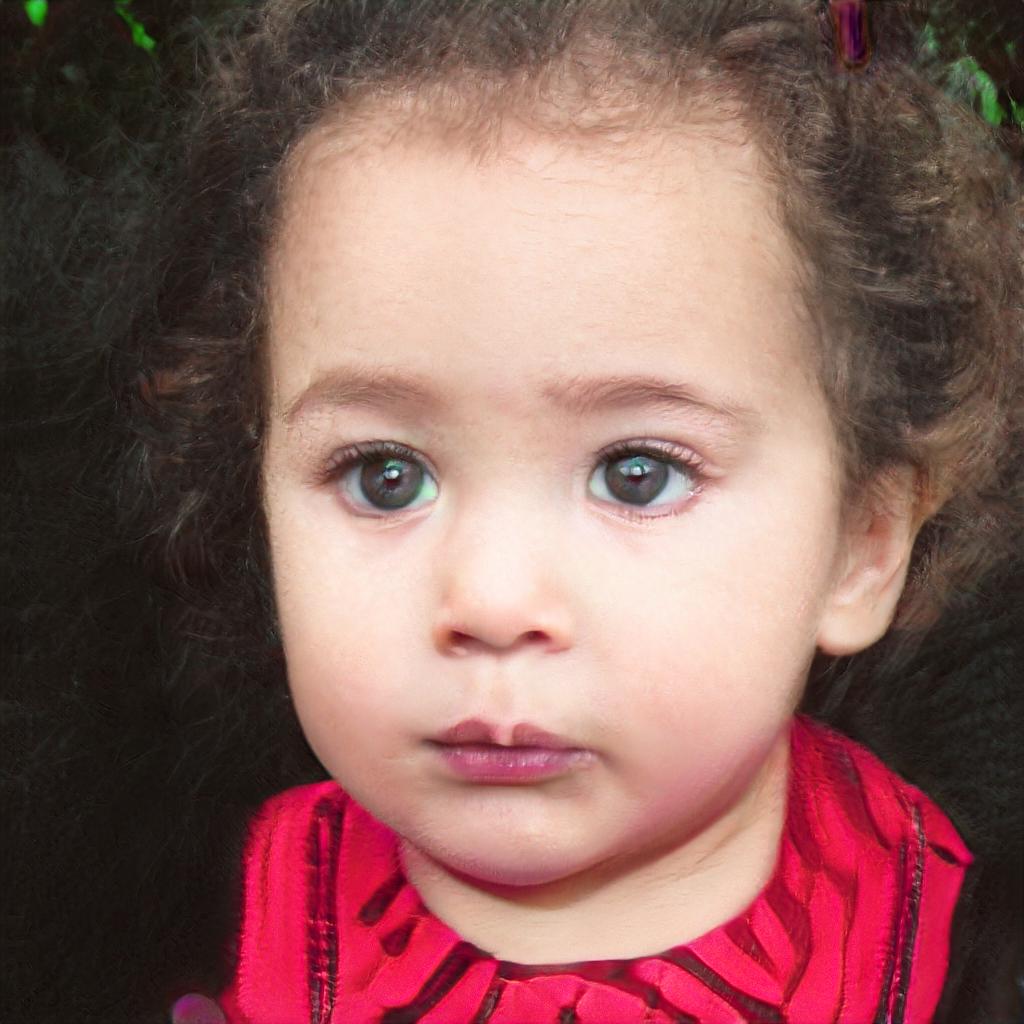} \\
      \small & $9.22\%$ & $6.98\%$ & $6.94\%$ & $6.60\%$ & $6.57\%$ & $6.14\%$ & $3.88\%$ & $3.25\%$ & $1.84\%$ \\
      (d) &
    \includegraphics[align=c,width=1.5cm]{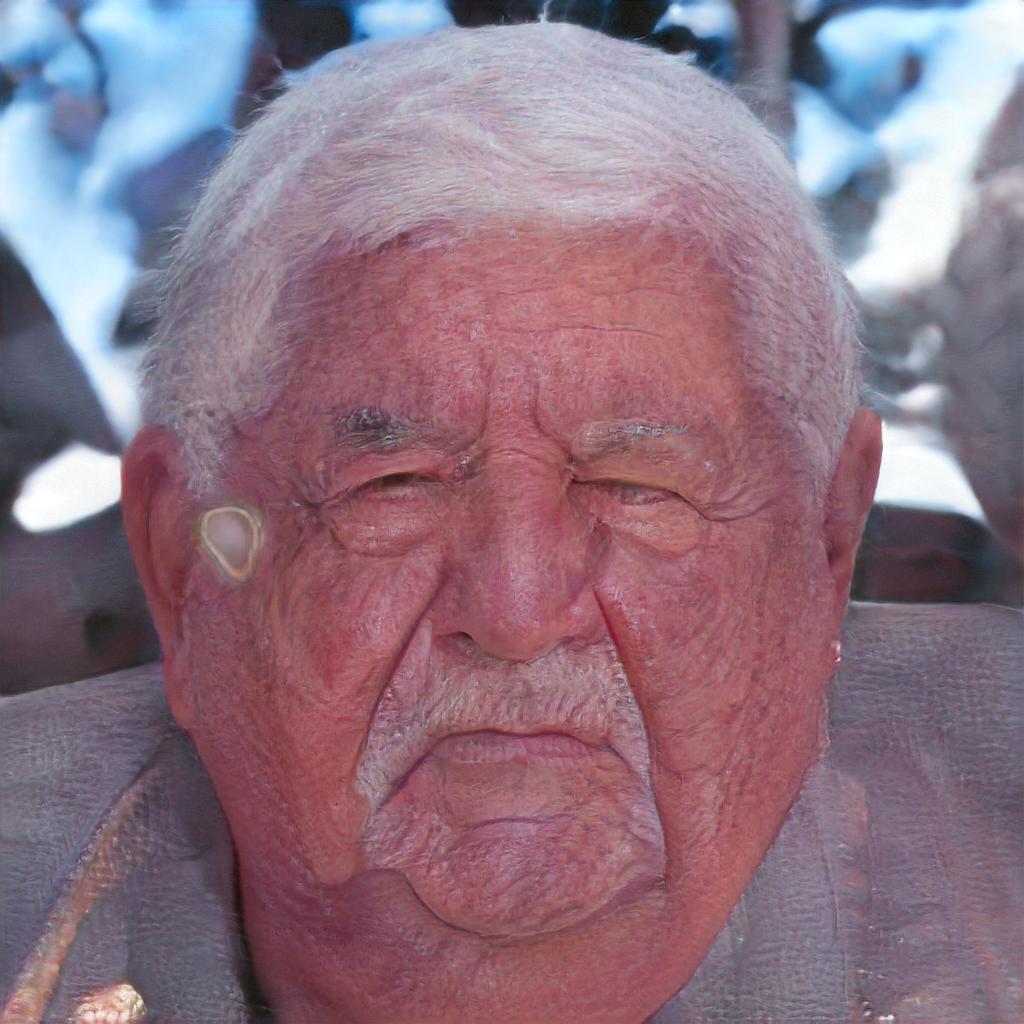} & \includegraphics[align=c,width=1.5cm]{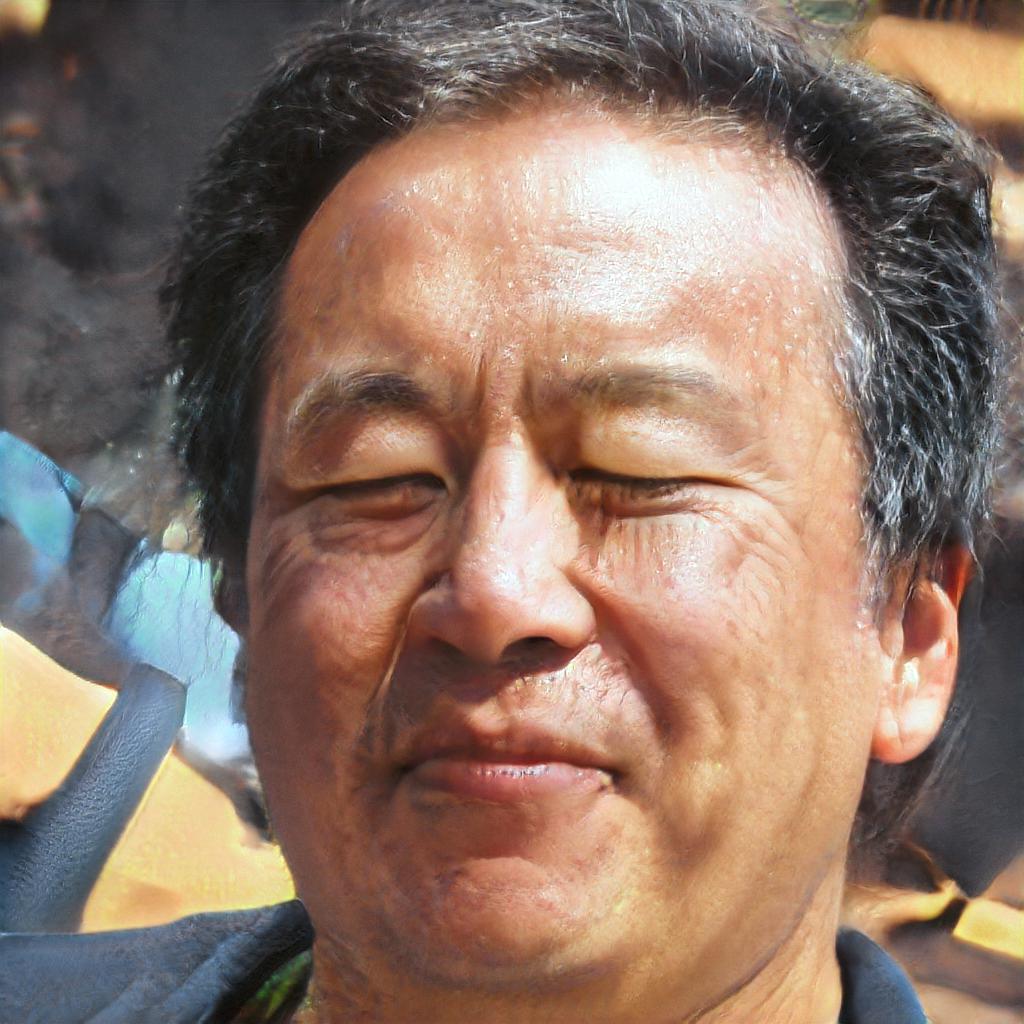} & \includegraphics[align=c,width=1.5cm]{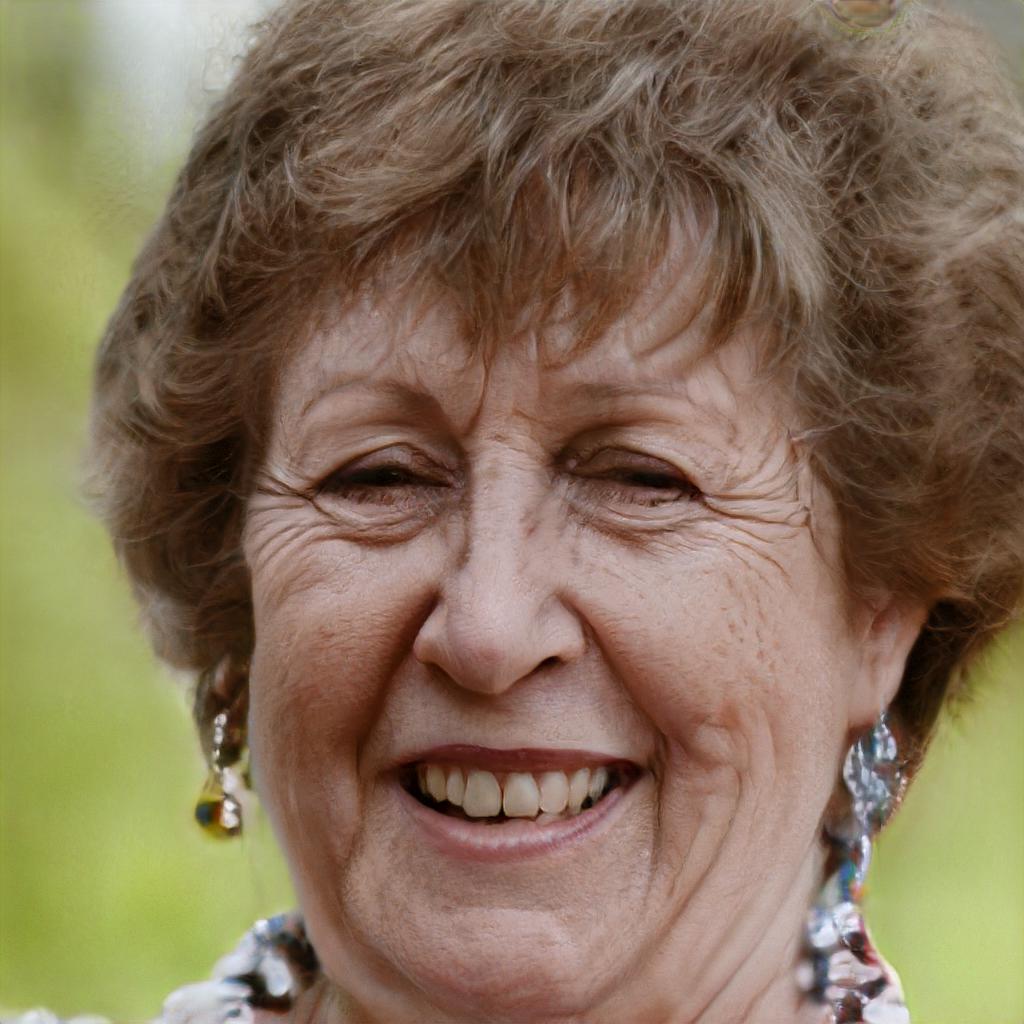} &  \includegraphics[align=c,width=1.5cm]{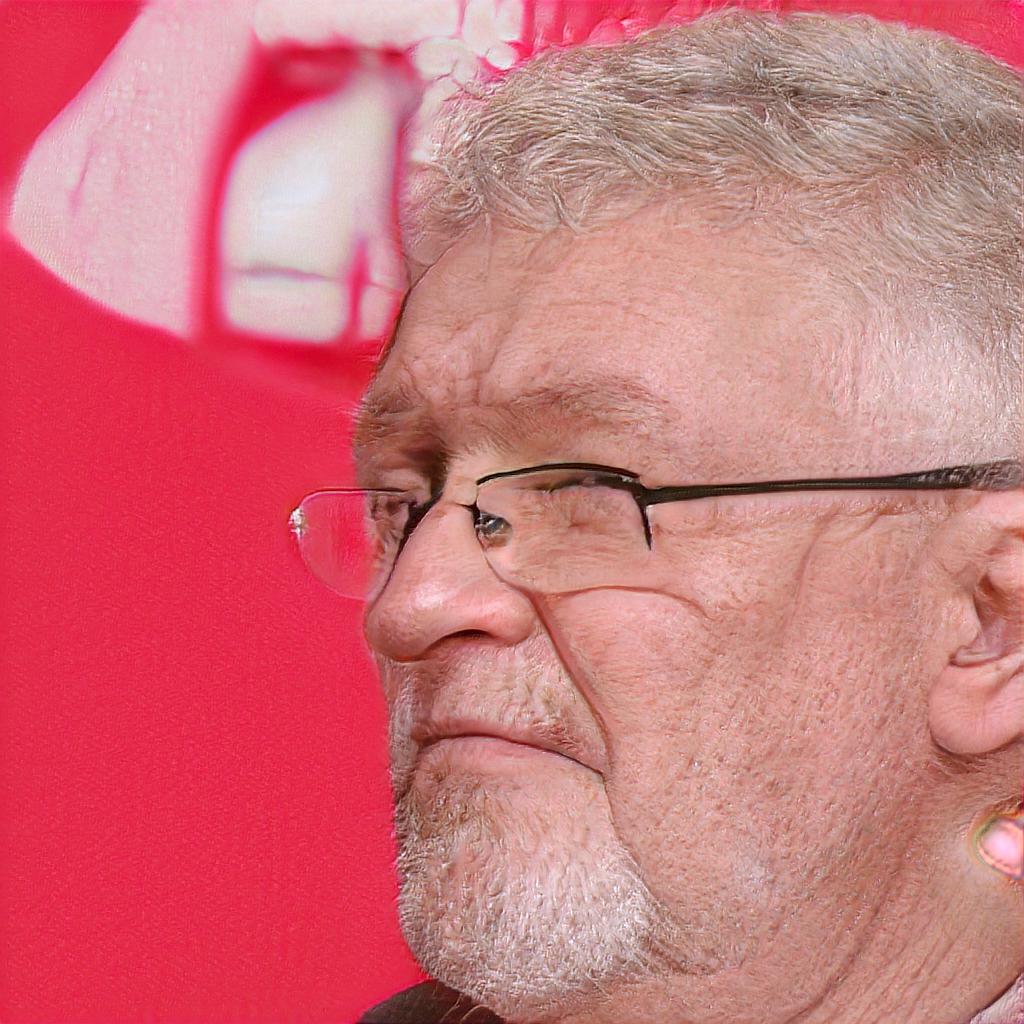} &
    \includegraphics[align=c,width=1.5cm]{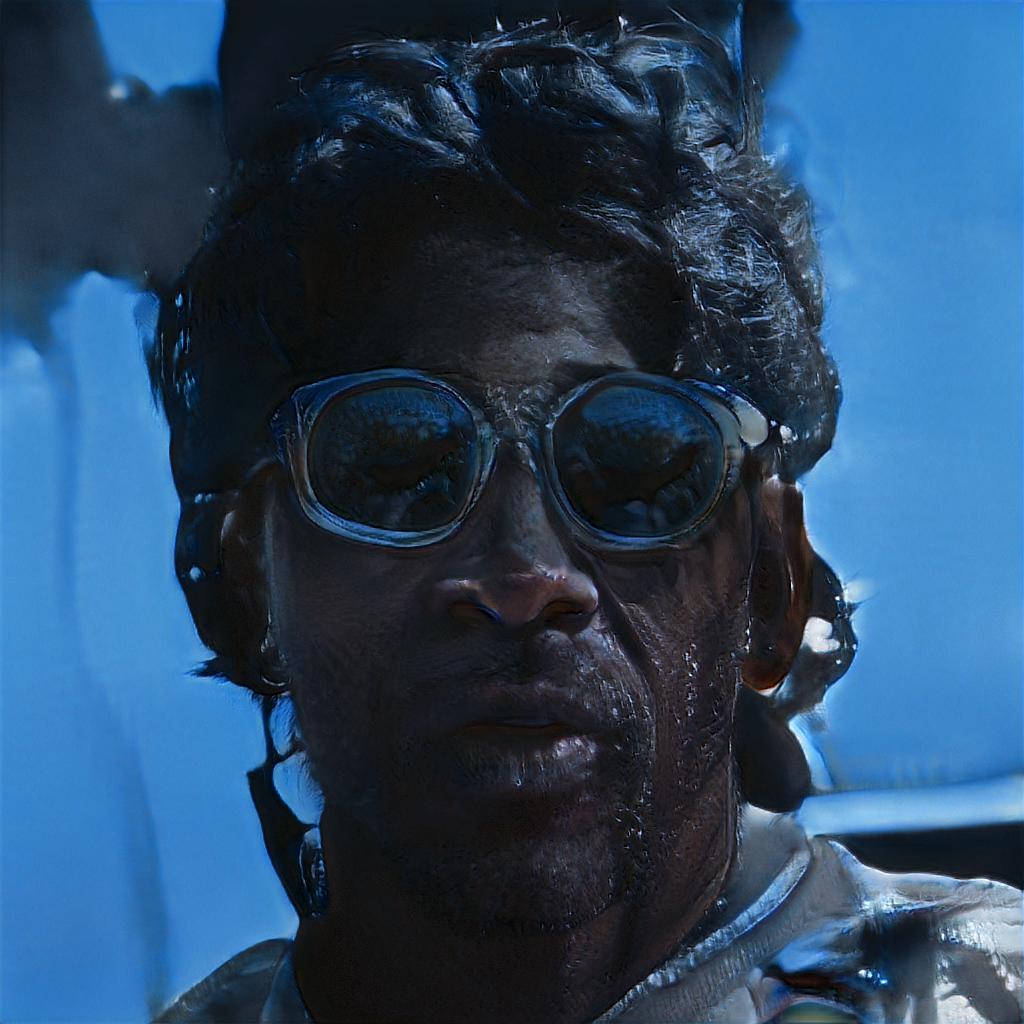} & \includegraphics[align=c,width=1.5cm]{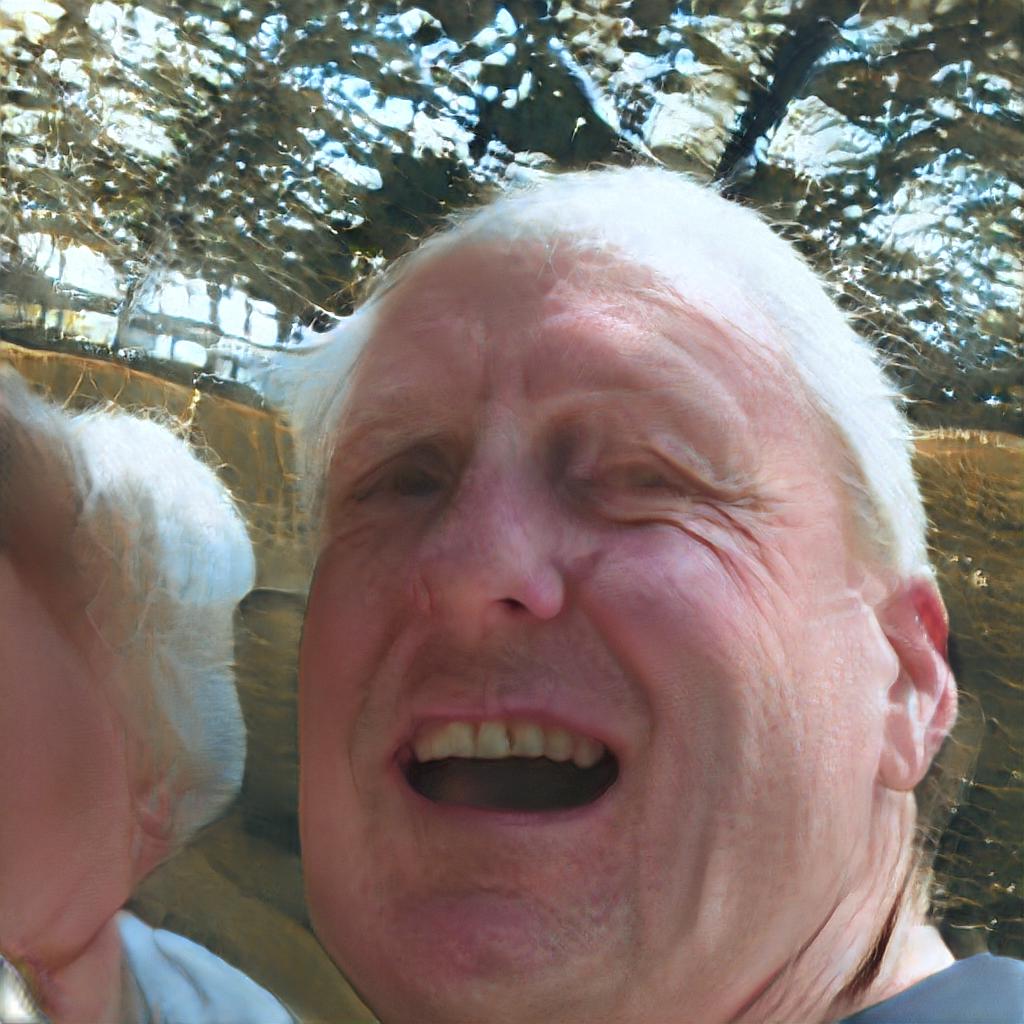} & \includegraphics[align=c,width=1.5cm]{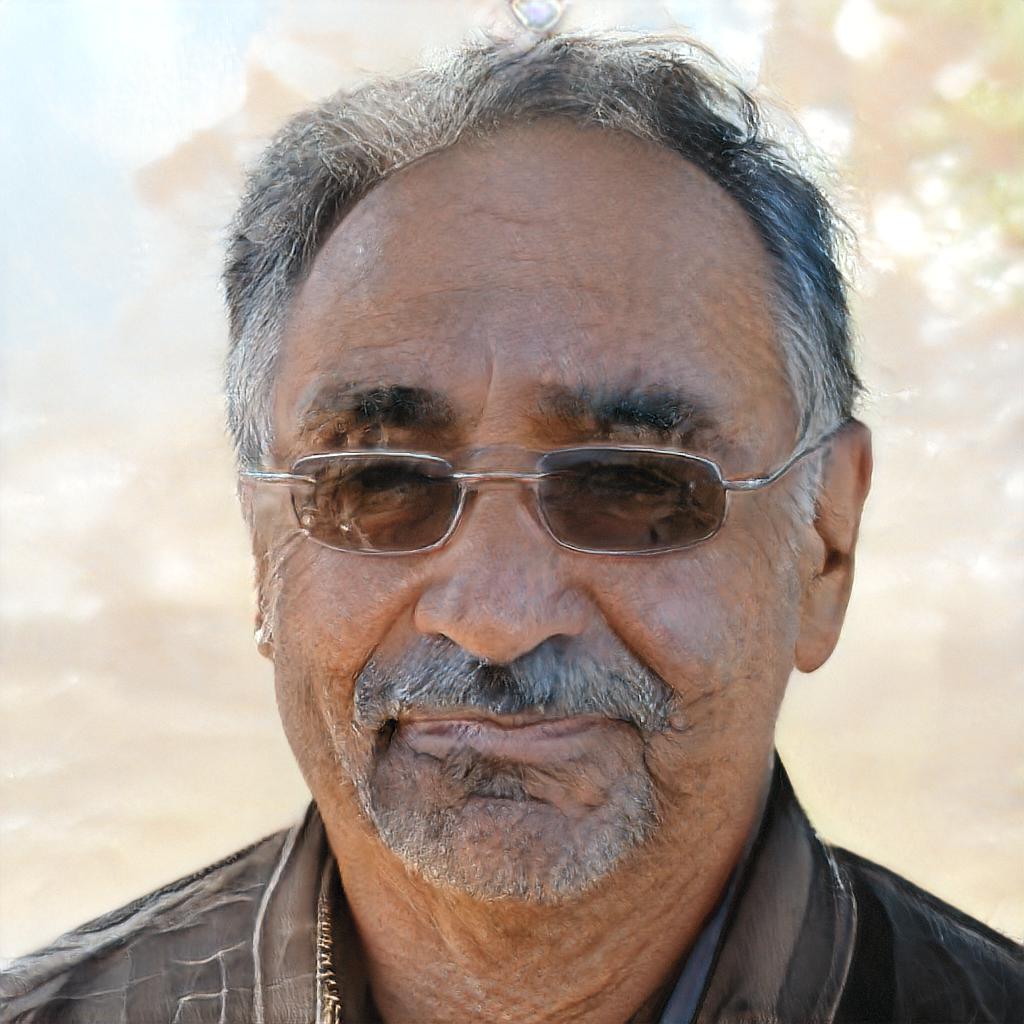} &  \includegraphics[align=c,width=1.5cm]{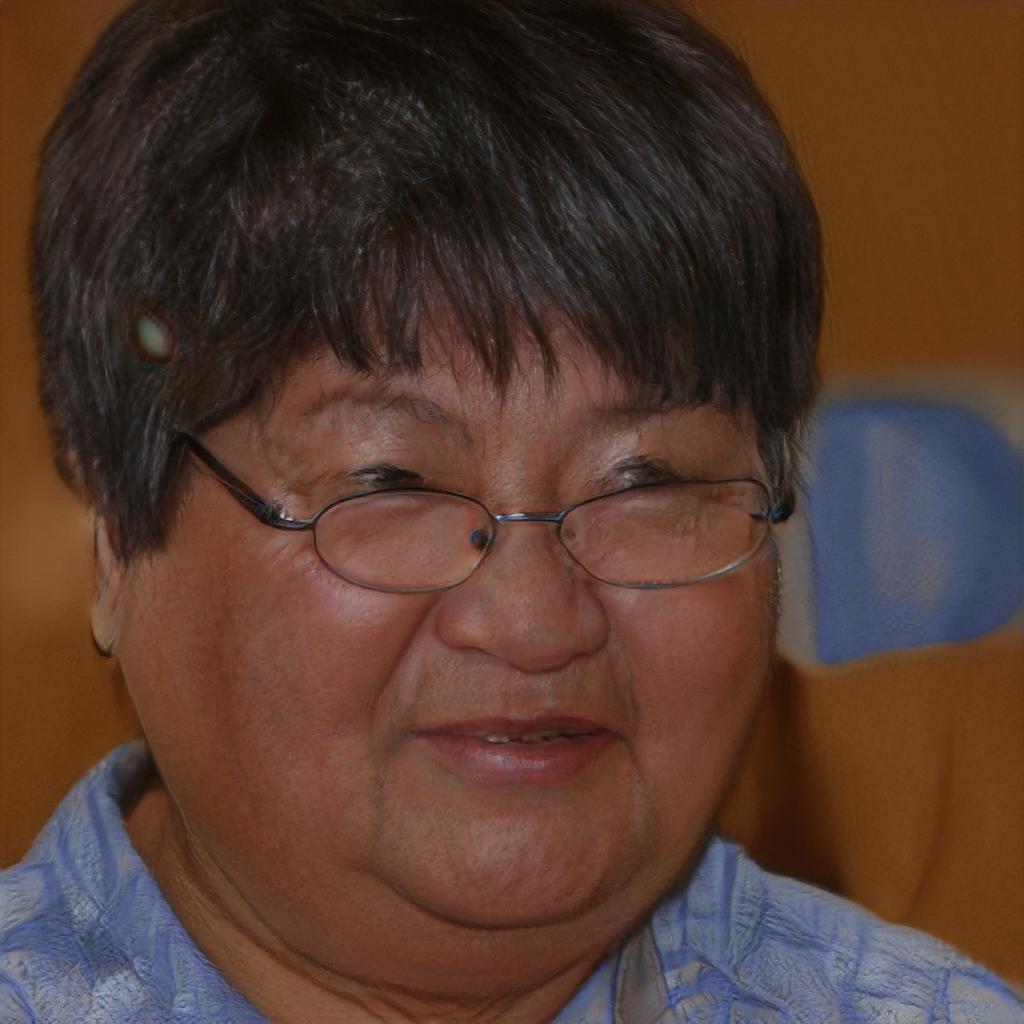} &
    \includegraphics[align=c,width=1.5cm]{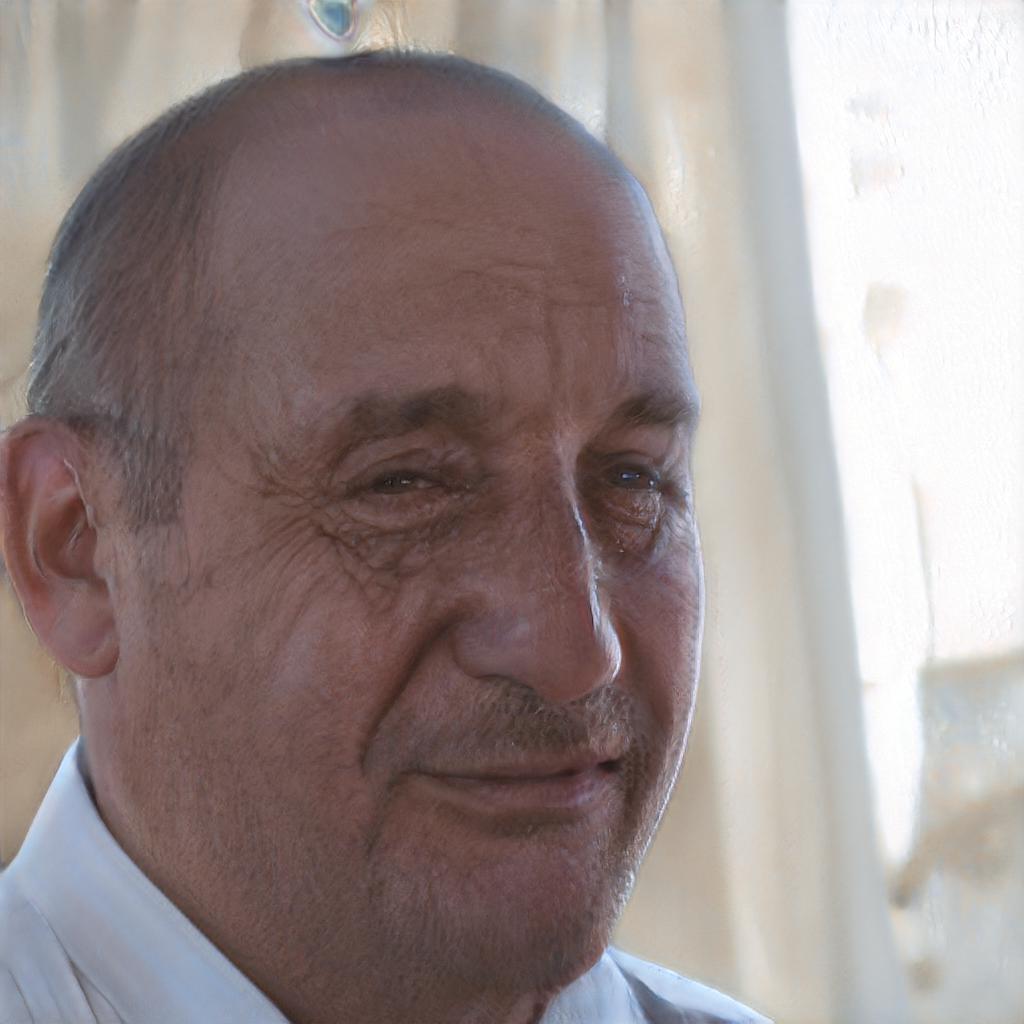} \\
    \small & $16.81\%$ & $5.50\%$ & $4.83\%$ & $3.42\%$ & $3.40\%$ & $3.09\%$ & $3.00\%$ & $1.60\%$ & $1.49\%$ \\
      (e) &
    \includegraphics[align=c,width=1.5cm]{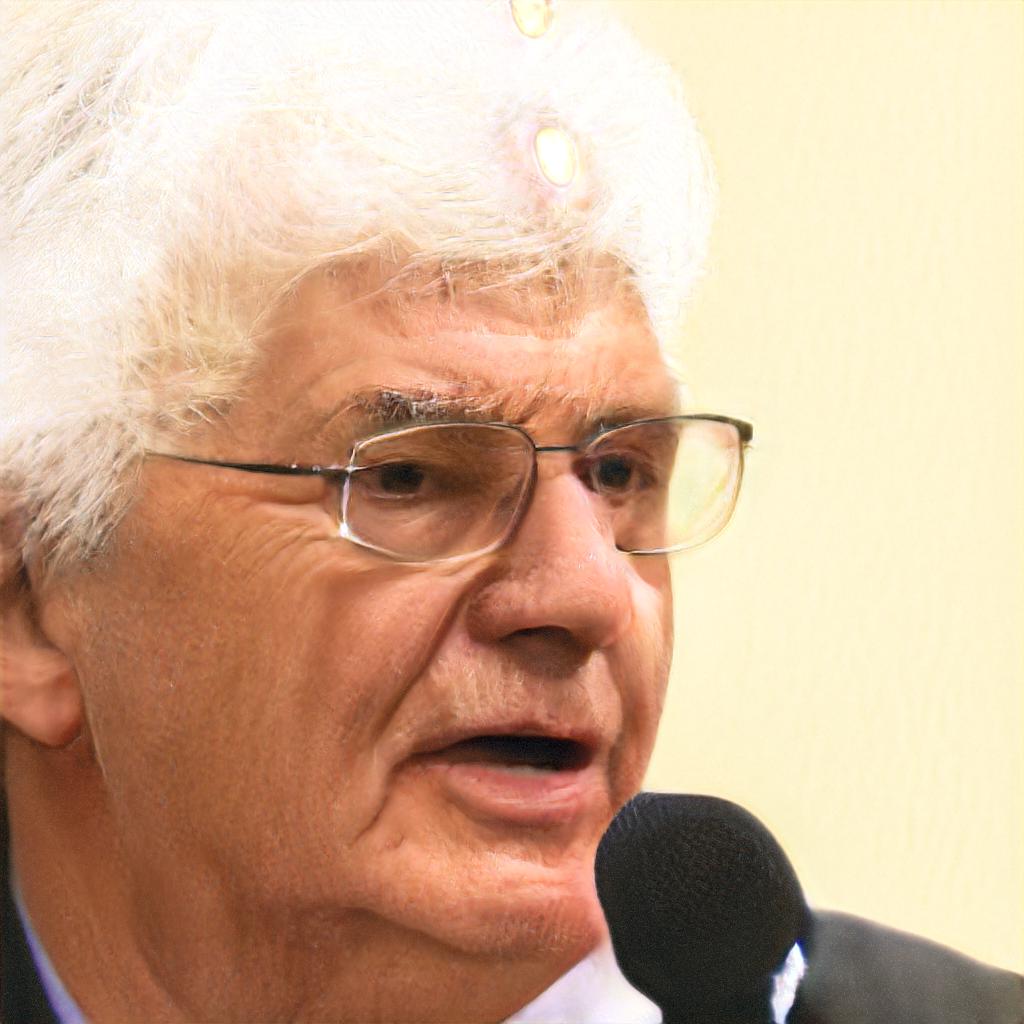} & \includegraphics[align=c,width=1.5cm]{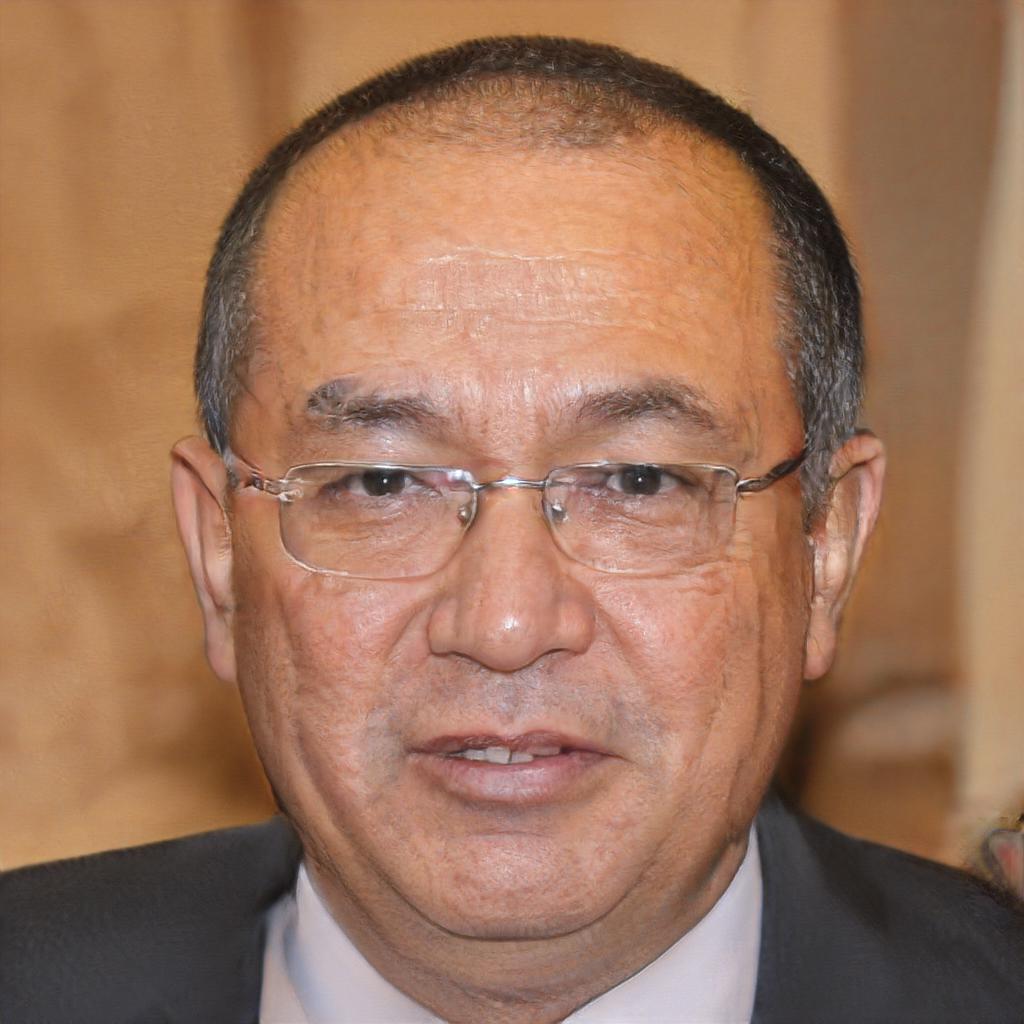} & \includegraphics[align=c,width=1.5cm]{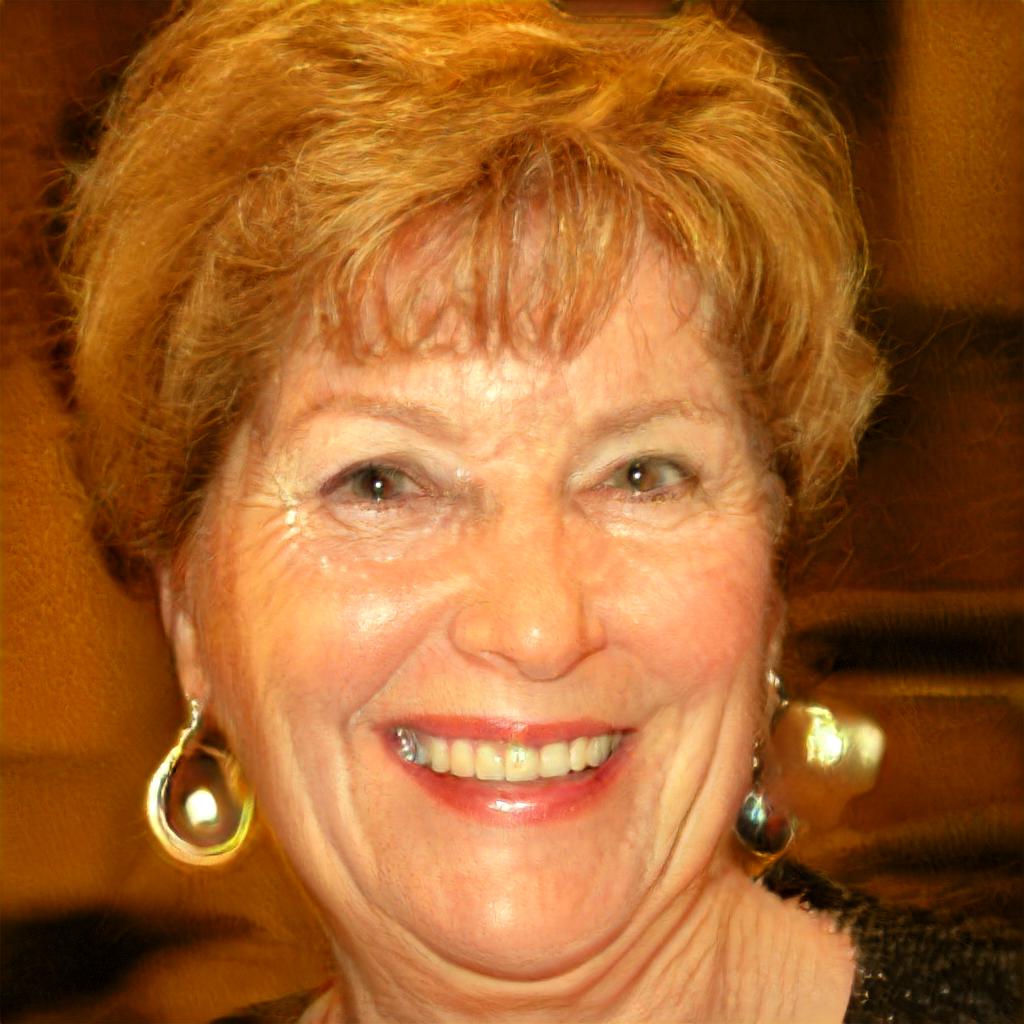} &      \includegraphics[align=c,width=1.5cm]{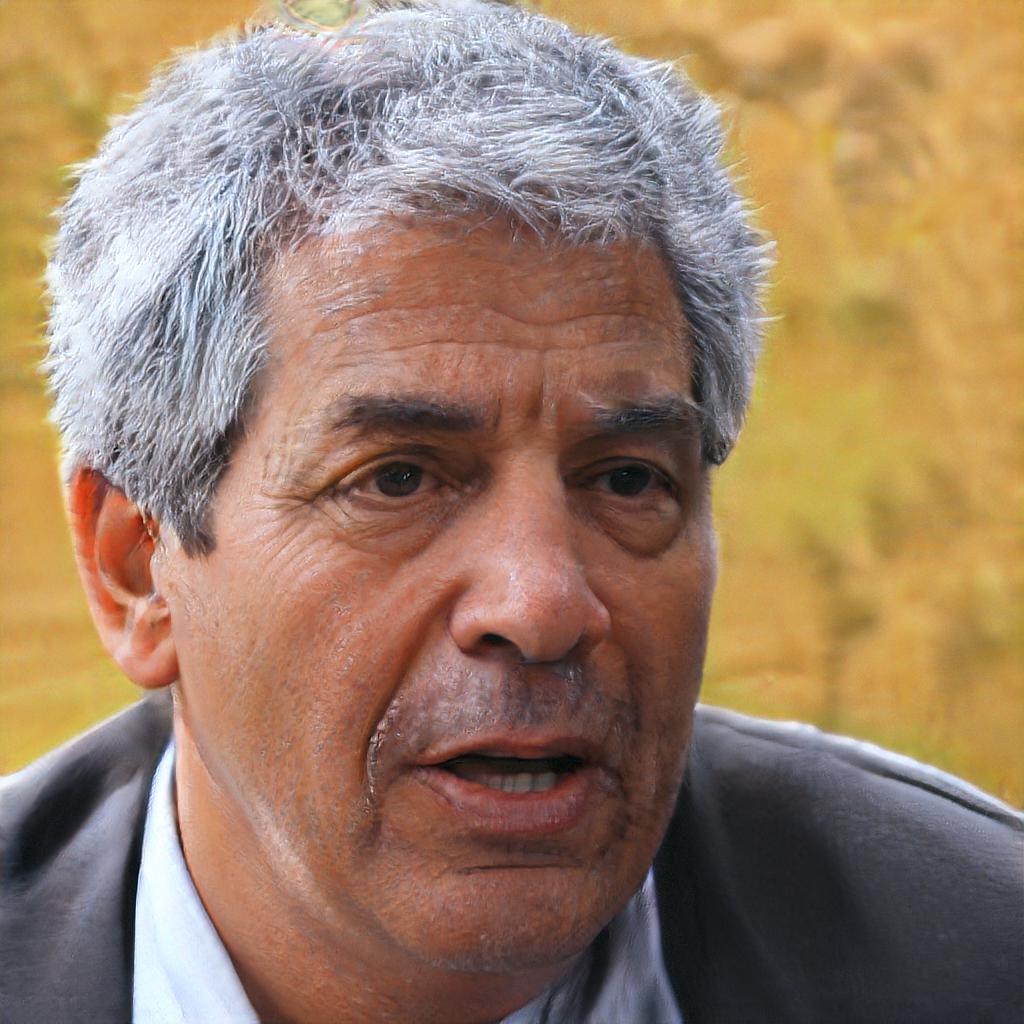} &   
    \includegraphics[align=c,width=1.5cm]{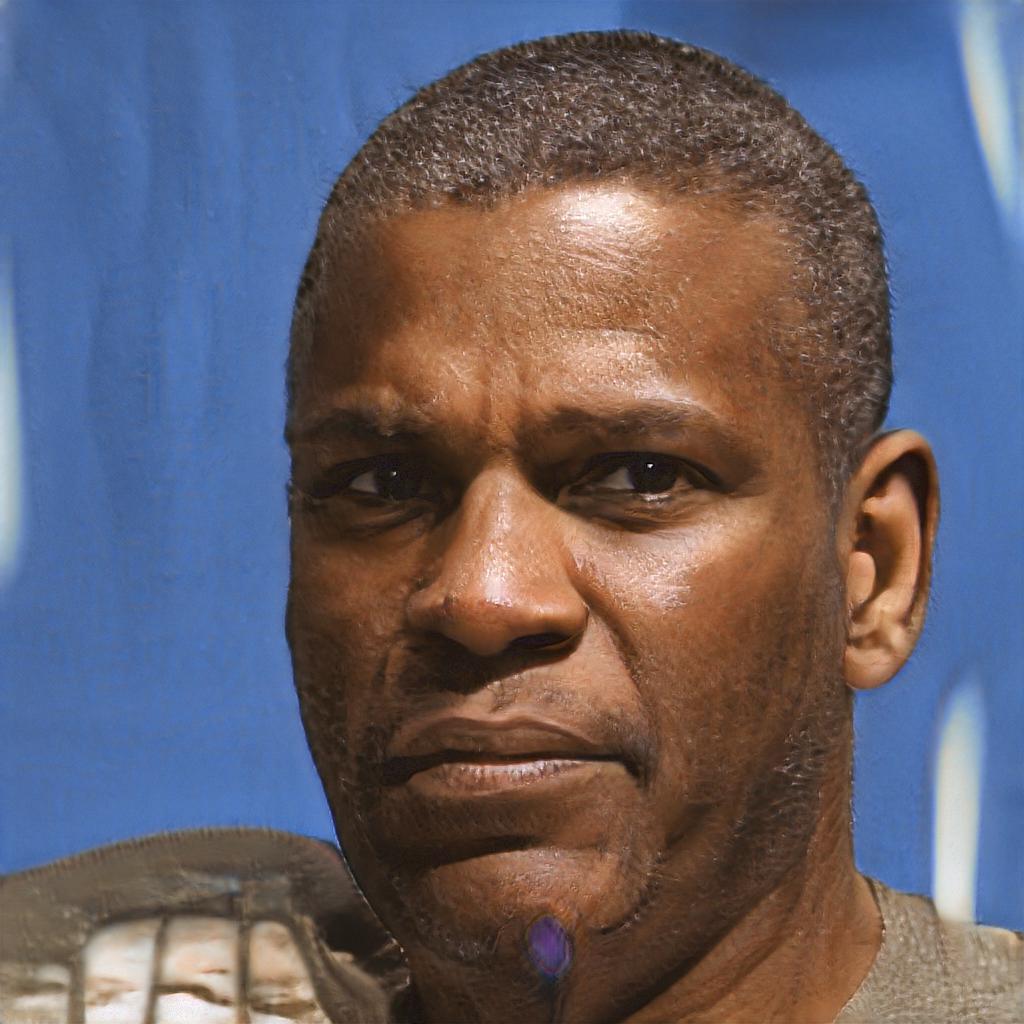} & \includegraphics[align=c,width=1.5cm]{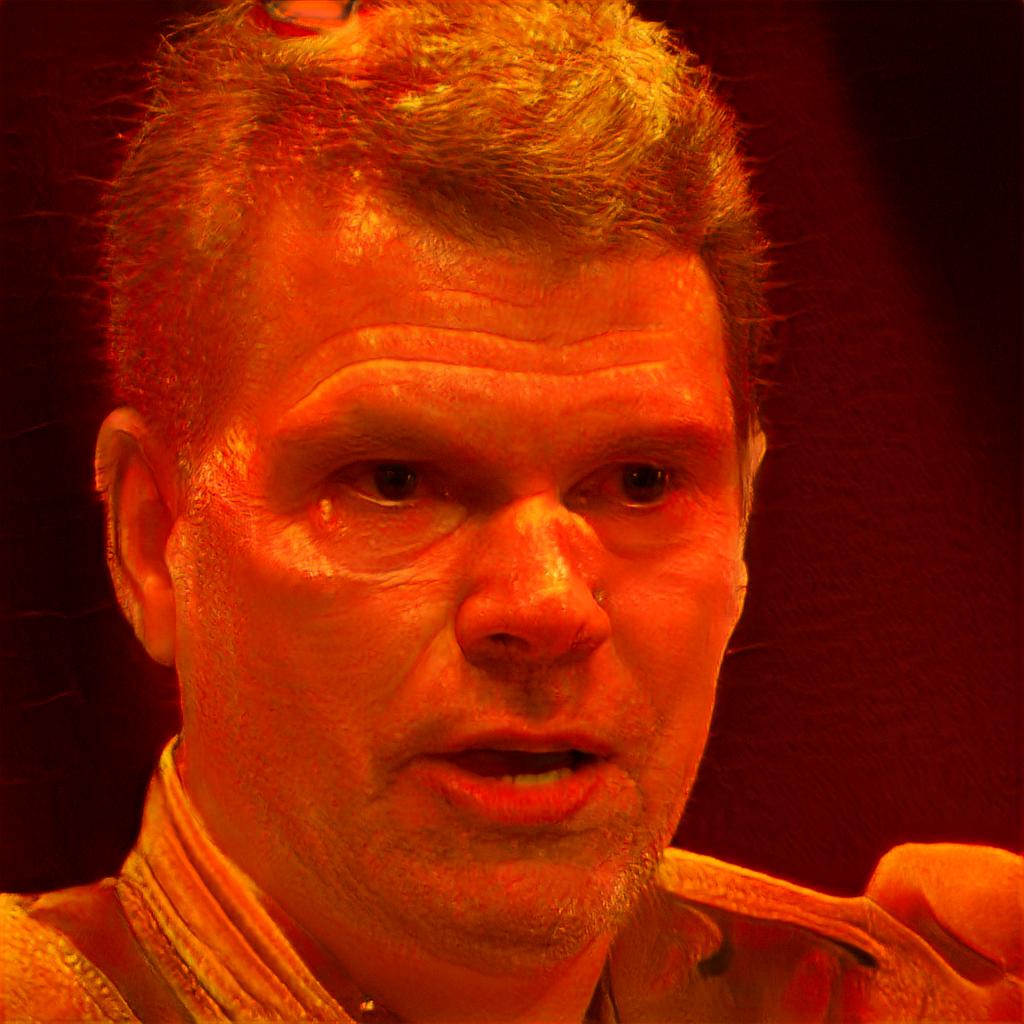} & \includegraphics[align=c,width=1.5cm]{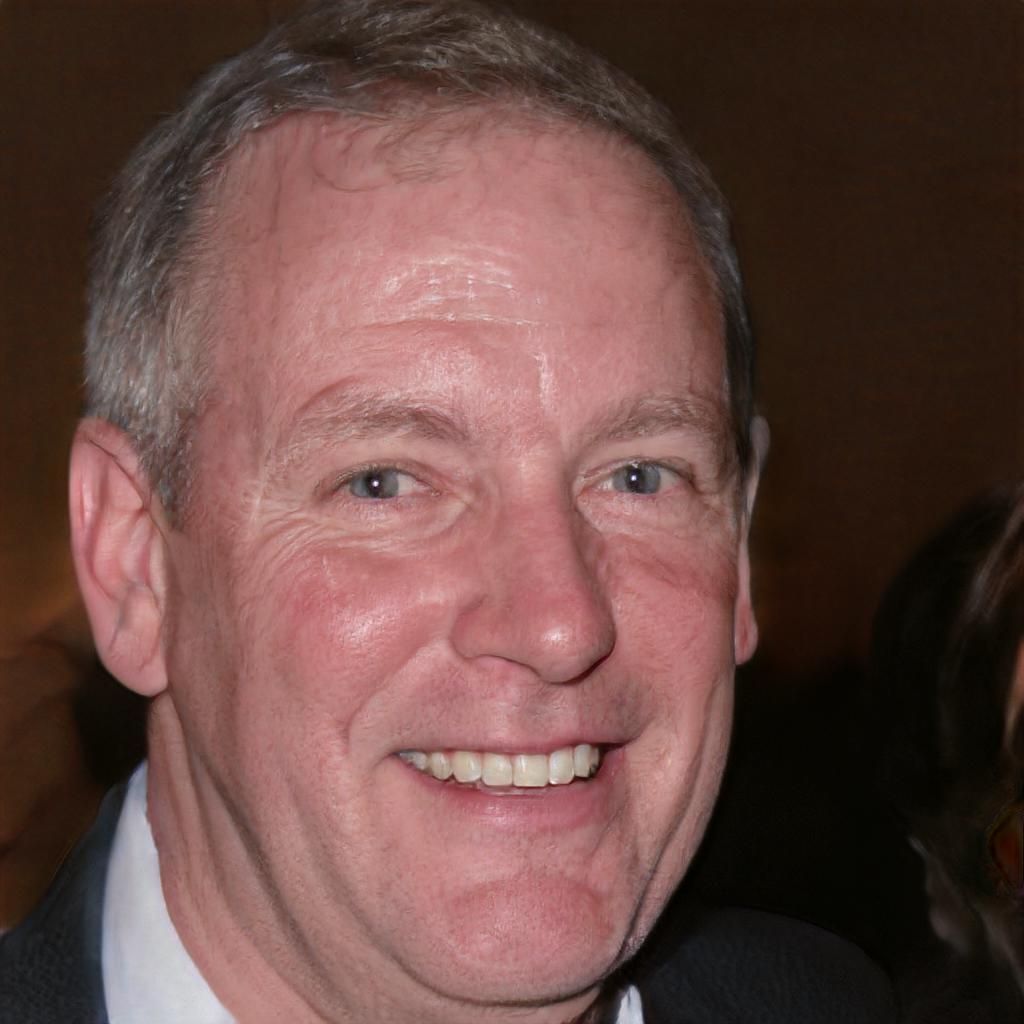} &     \includegraphics[align=c,width=1.5cm]{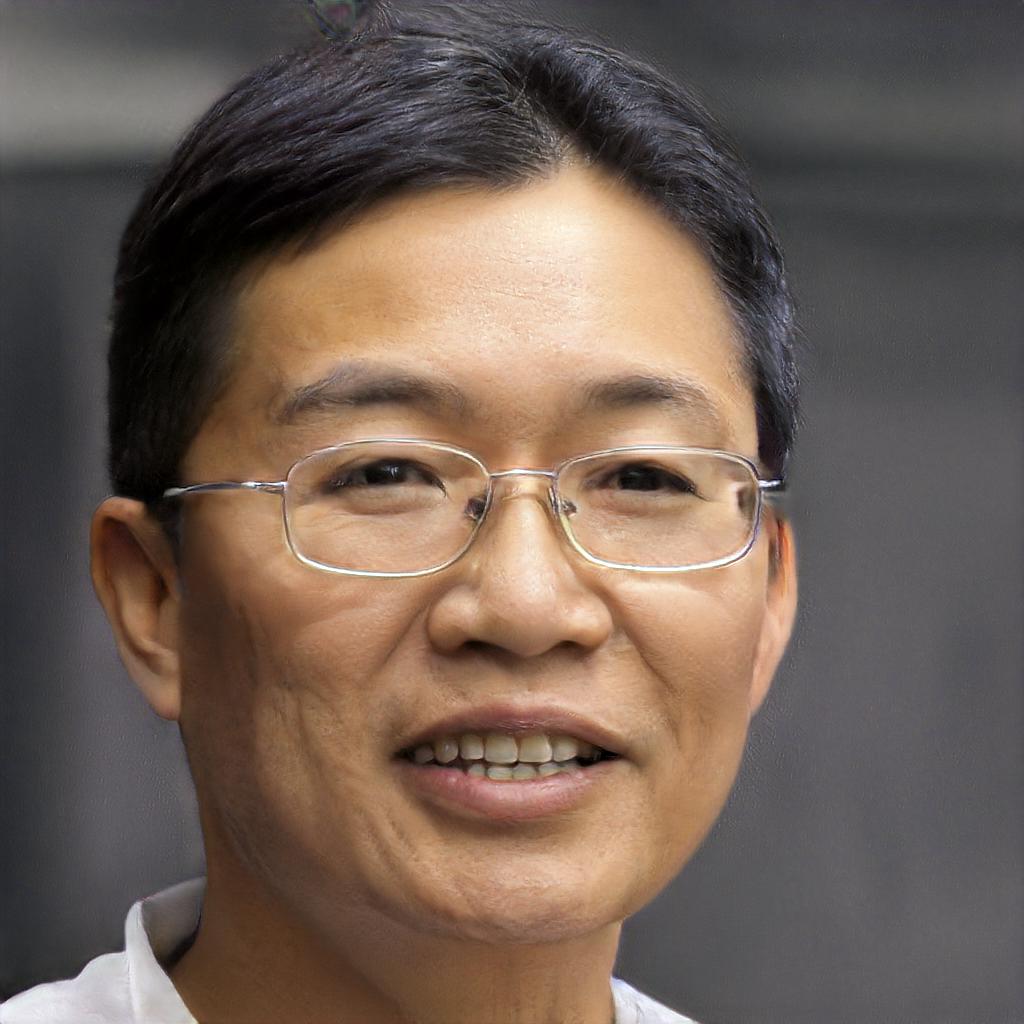} &   
    \includegraphics[align=c,width=1.5cm]{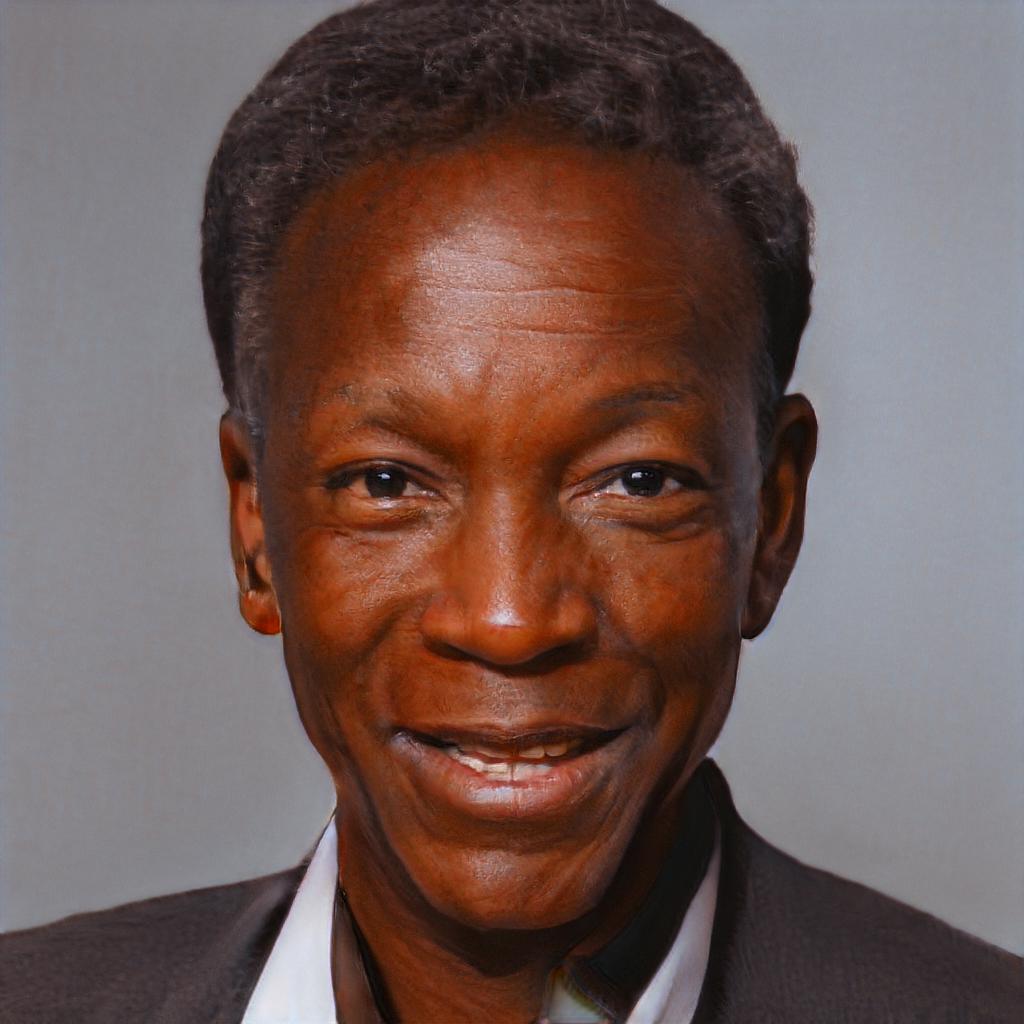} \\
      \small & $16.28\%$ & $5.70\%$ & $4.40\%$ & $3.56\%$ & $3.30\%$ & $2.90\%$ & $2.61\%$ & $1.91\%$ & $1.39\%$ \\
      (f) &
    \includegraphics[align=c,width=1.5cm]{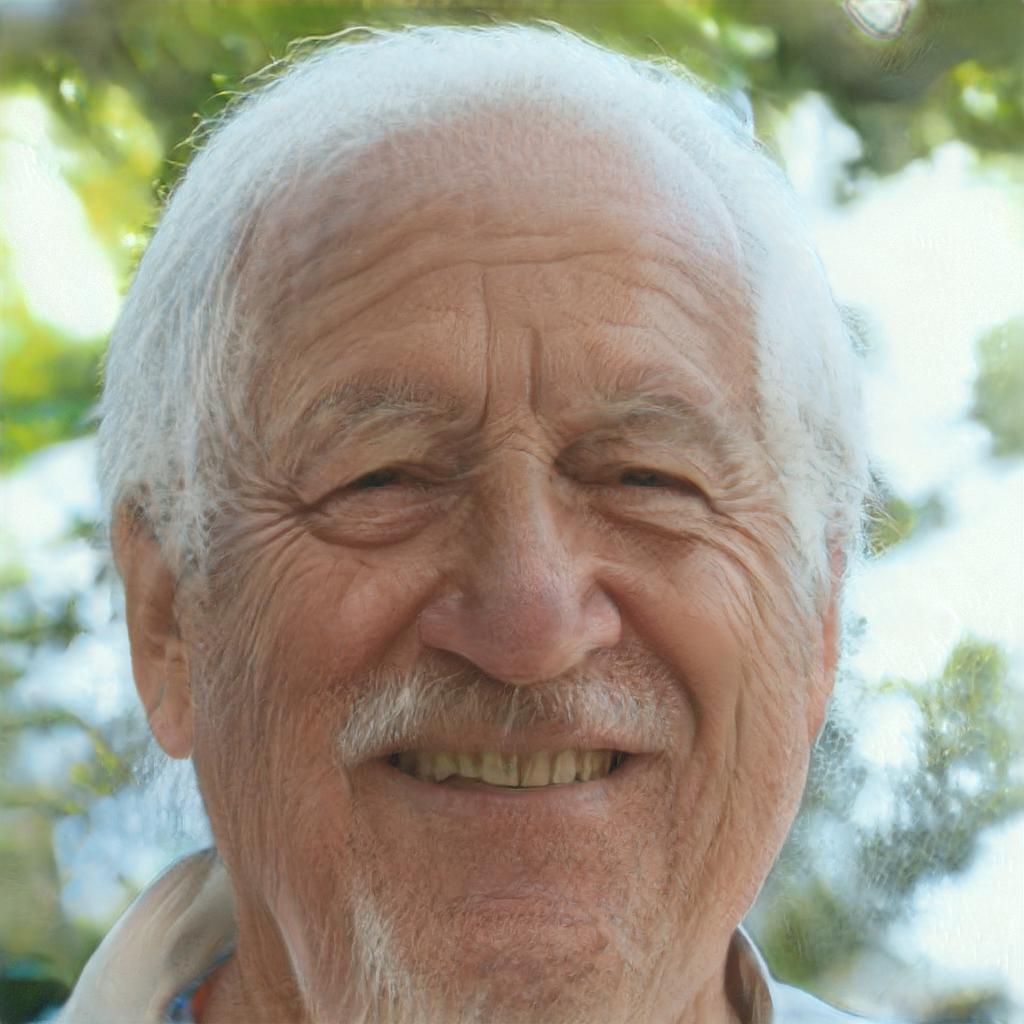} & \includegraphics[align=c,width=1.5cm]{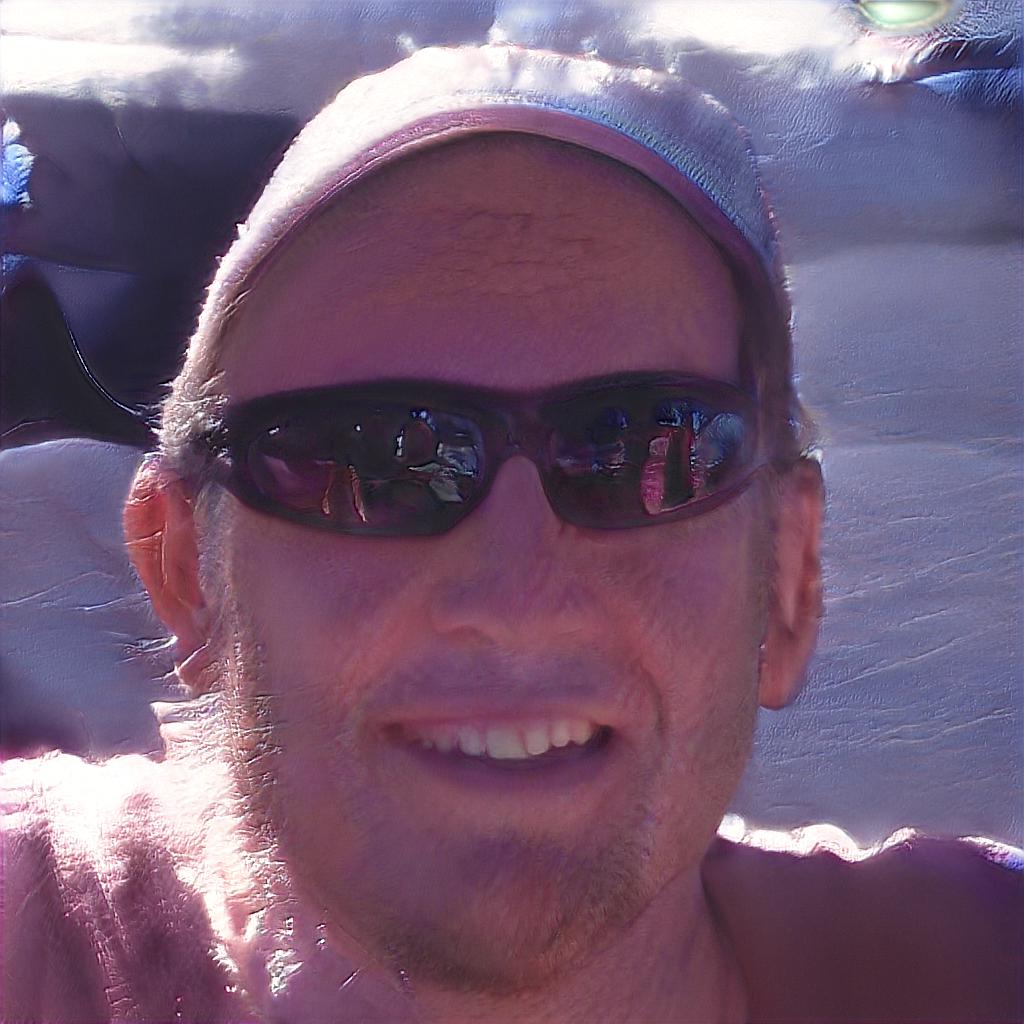} & \includegraphics[align=c,width=1.5cm]{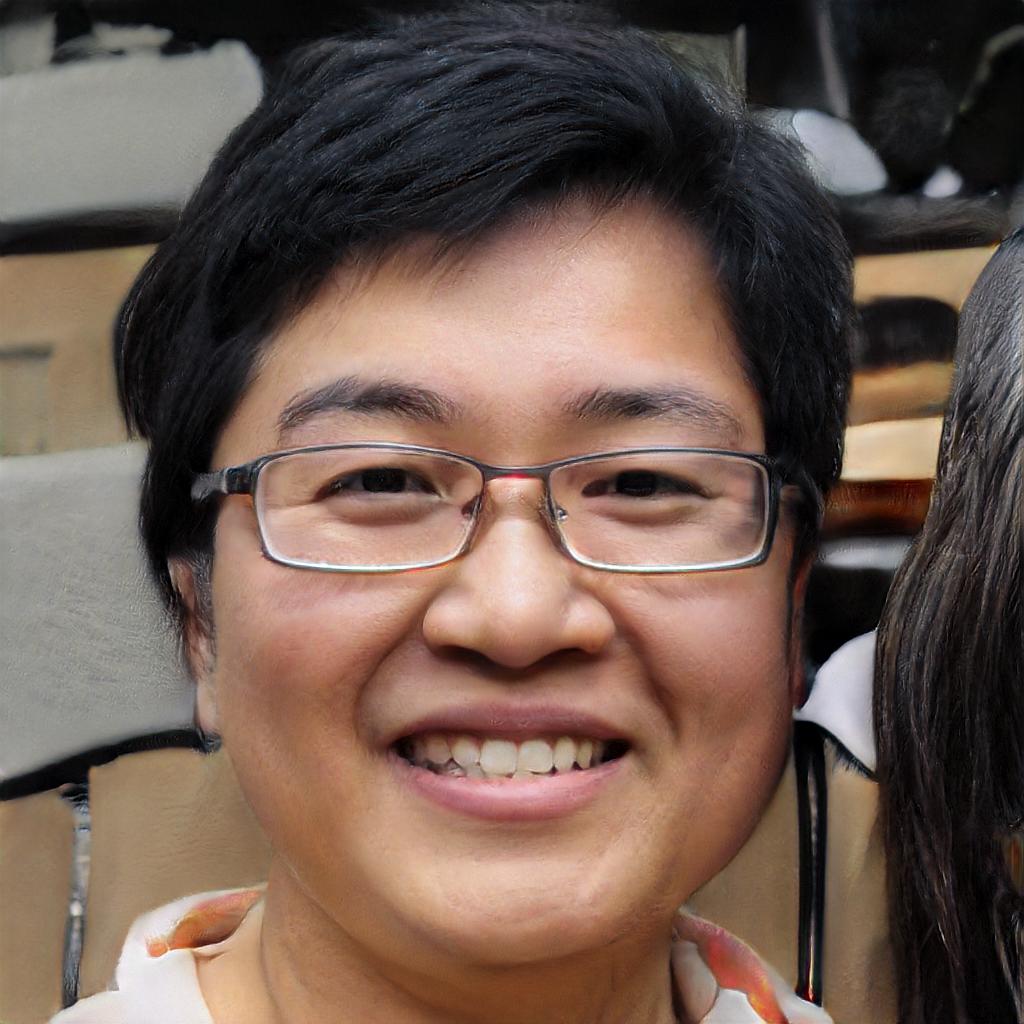} &      \includegraphics[align=c,width=1.5cm]{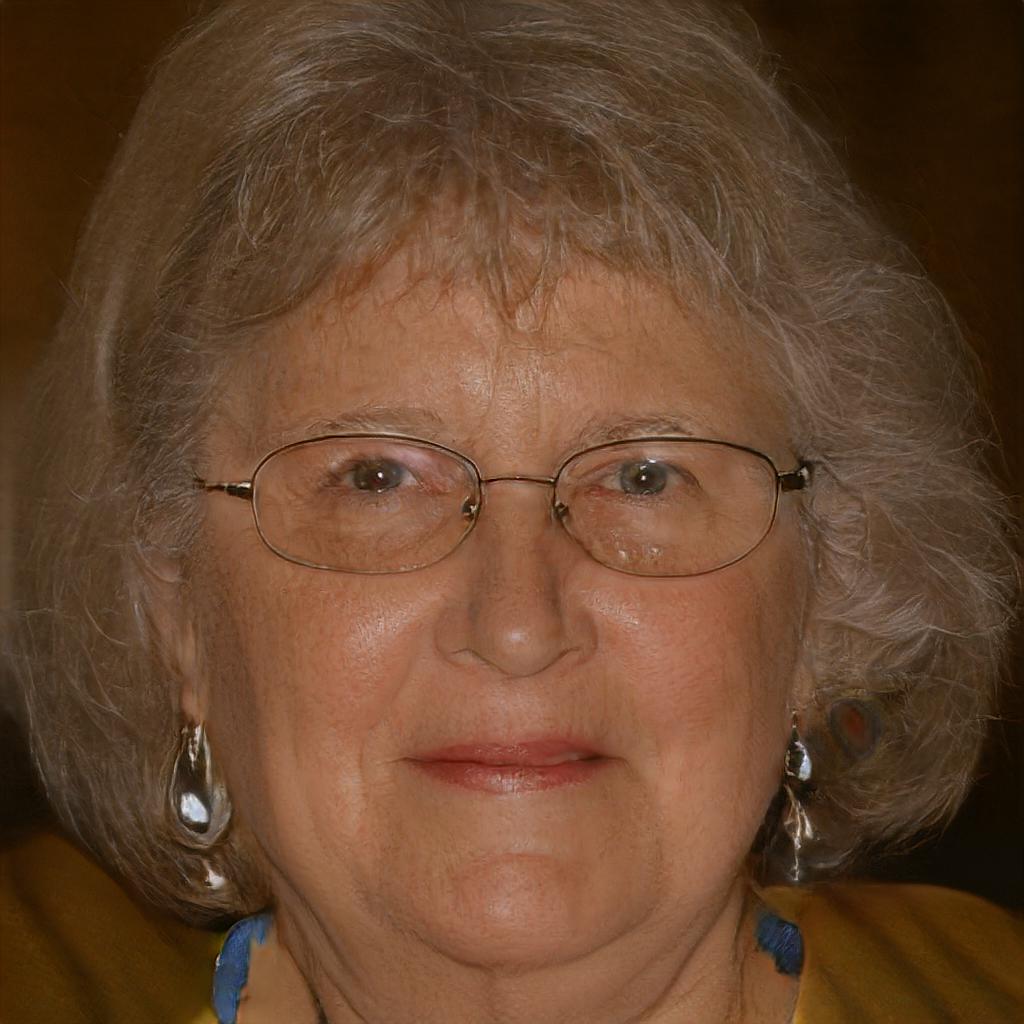} &   
    \includegraphics[align=c,width=1.5cm]{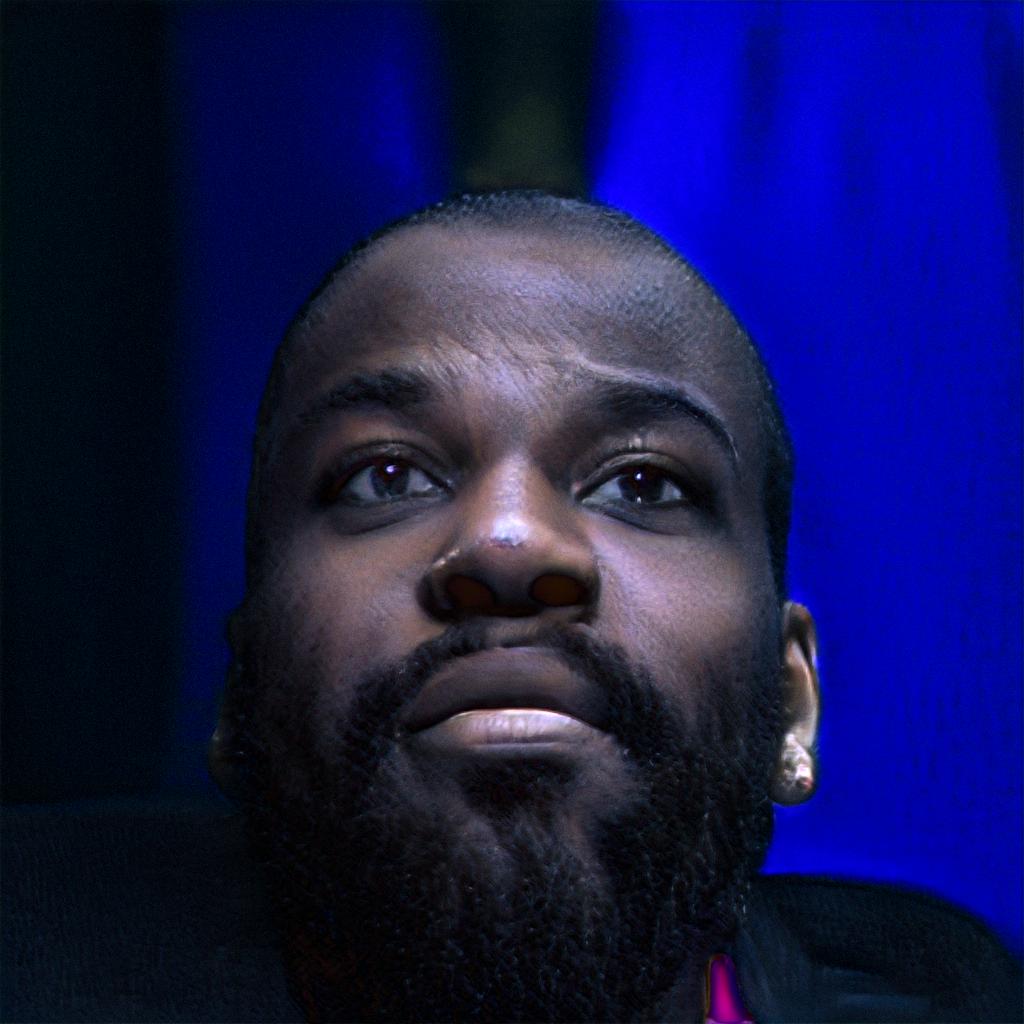} & \includegraphics[align=c,width=1.5cm]{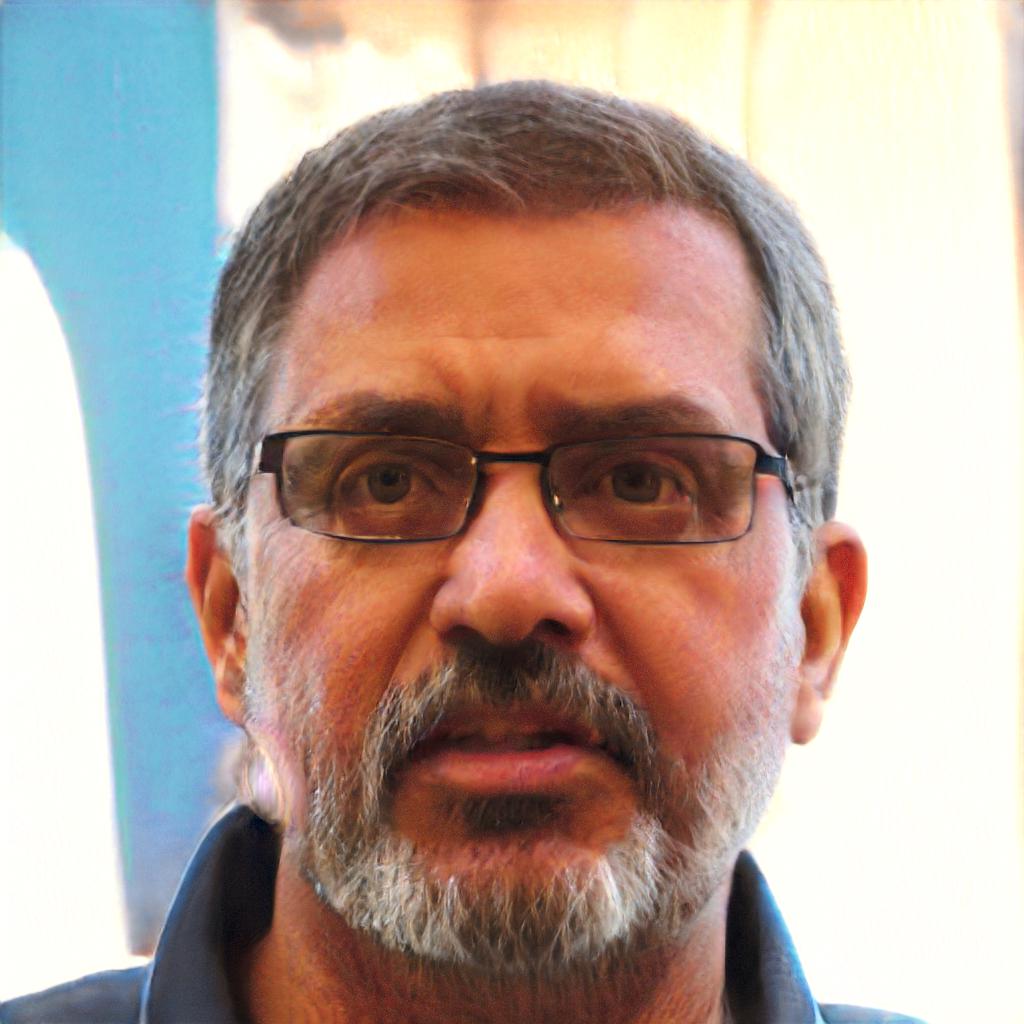} & \includegraphics[align=c,width=1.5cm]{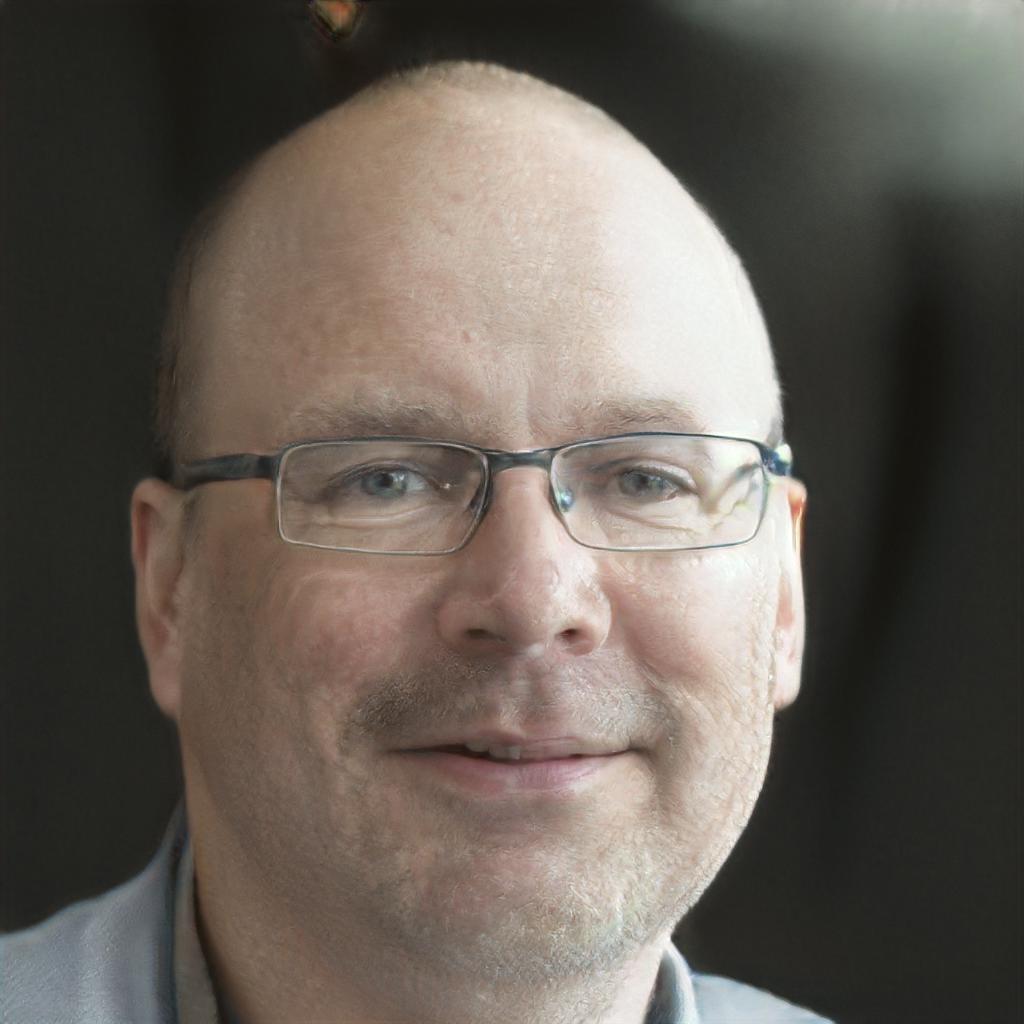} &      \includegraphics[align=c,width=1.5cm]{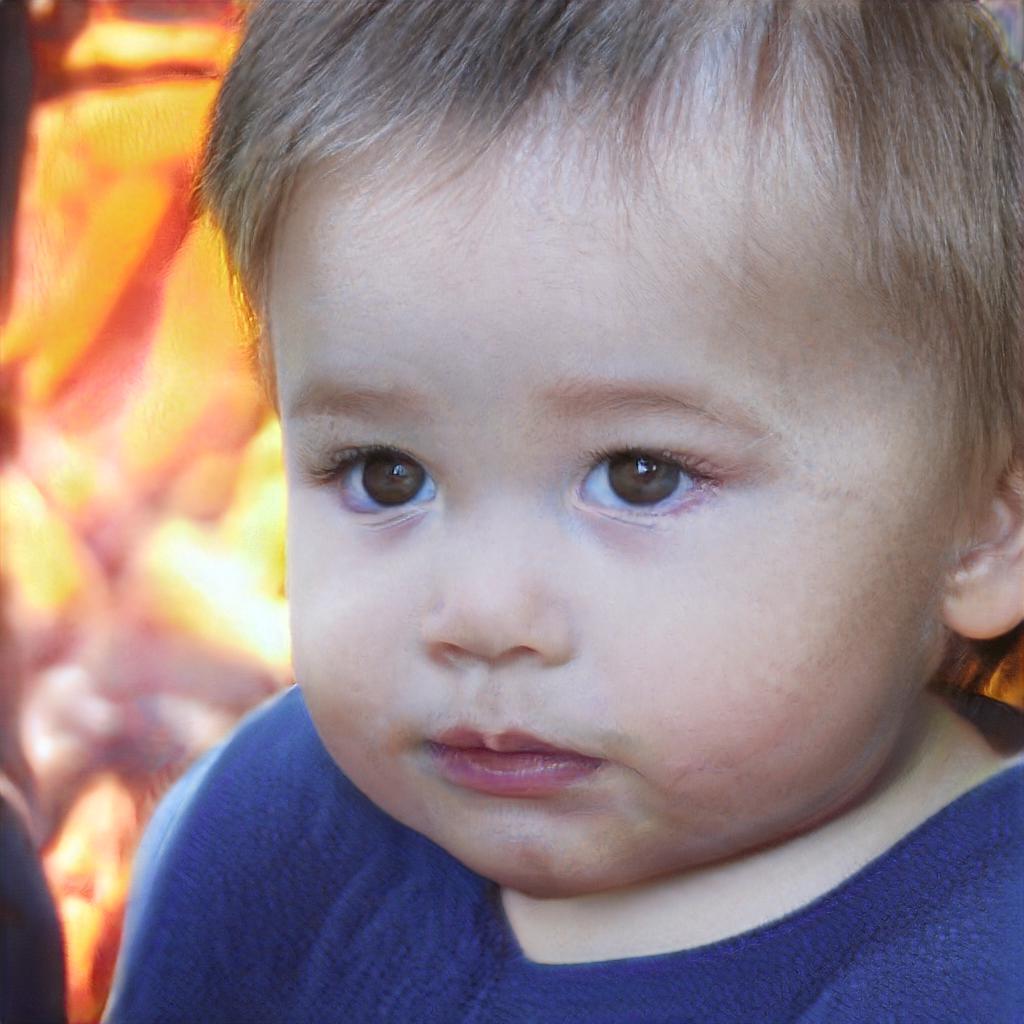} &   
    \includegraphics[align=c,width=1.5cm]{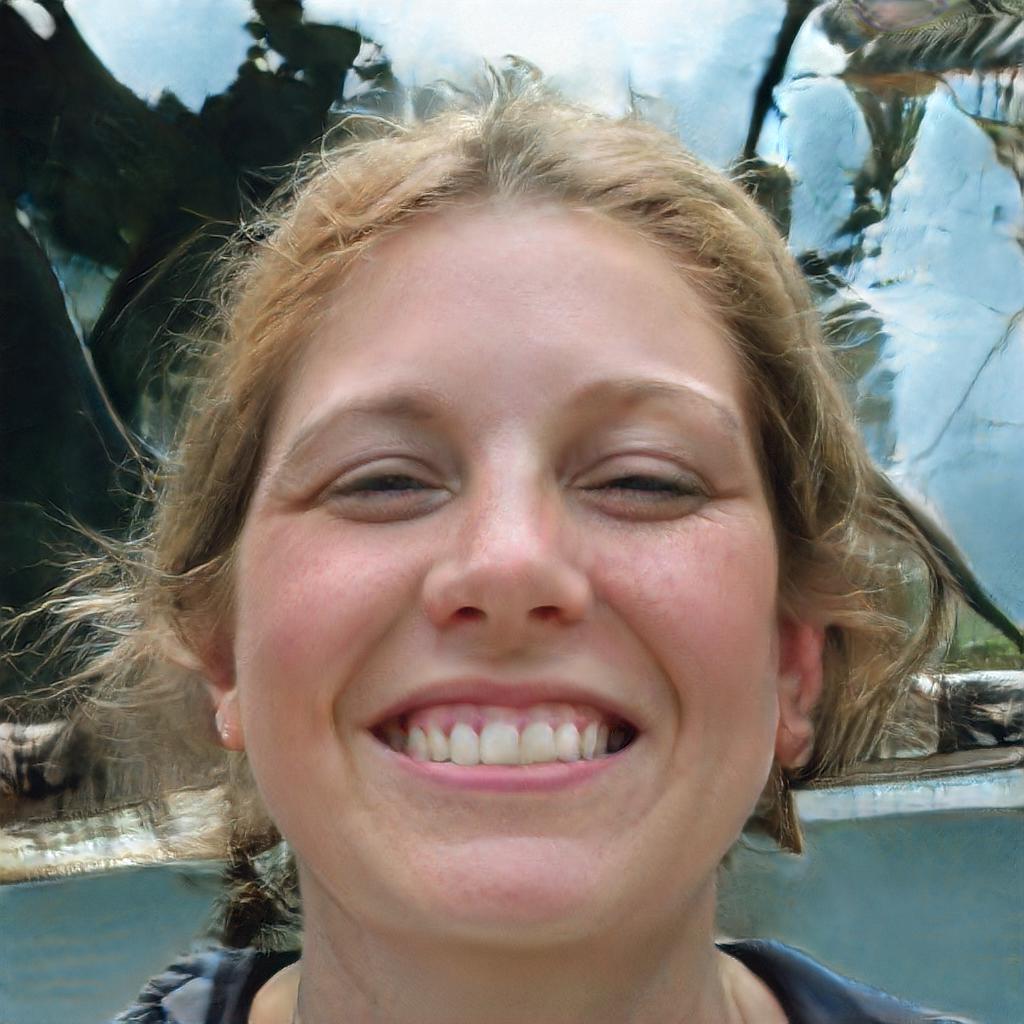} \\
      \small &$17.30\%$ & $9.84\%$ & $7.70\%$ & $6.39\%$ & $6.11\%$ & $5.29\%$ & $4.85\%$ & $3.45\%$ & $2.29\%$ \\
      (g) &
    \includegraphics[align=c,width=1.5cm]{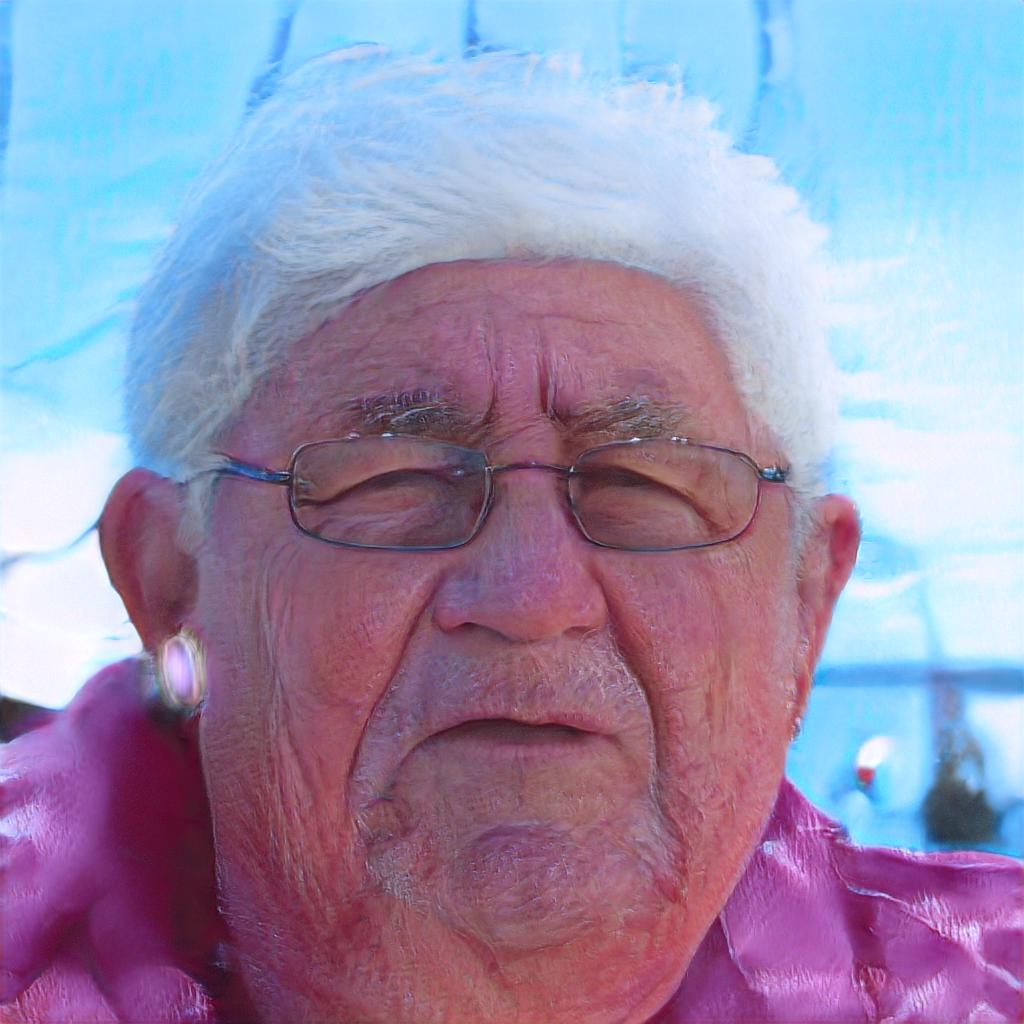} & \includegraphics[align=c,width=1.5cm]{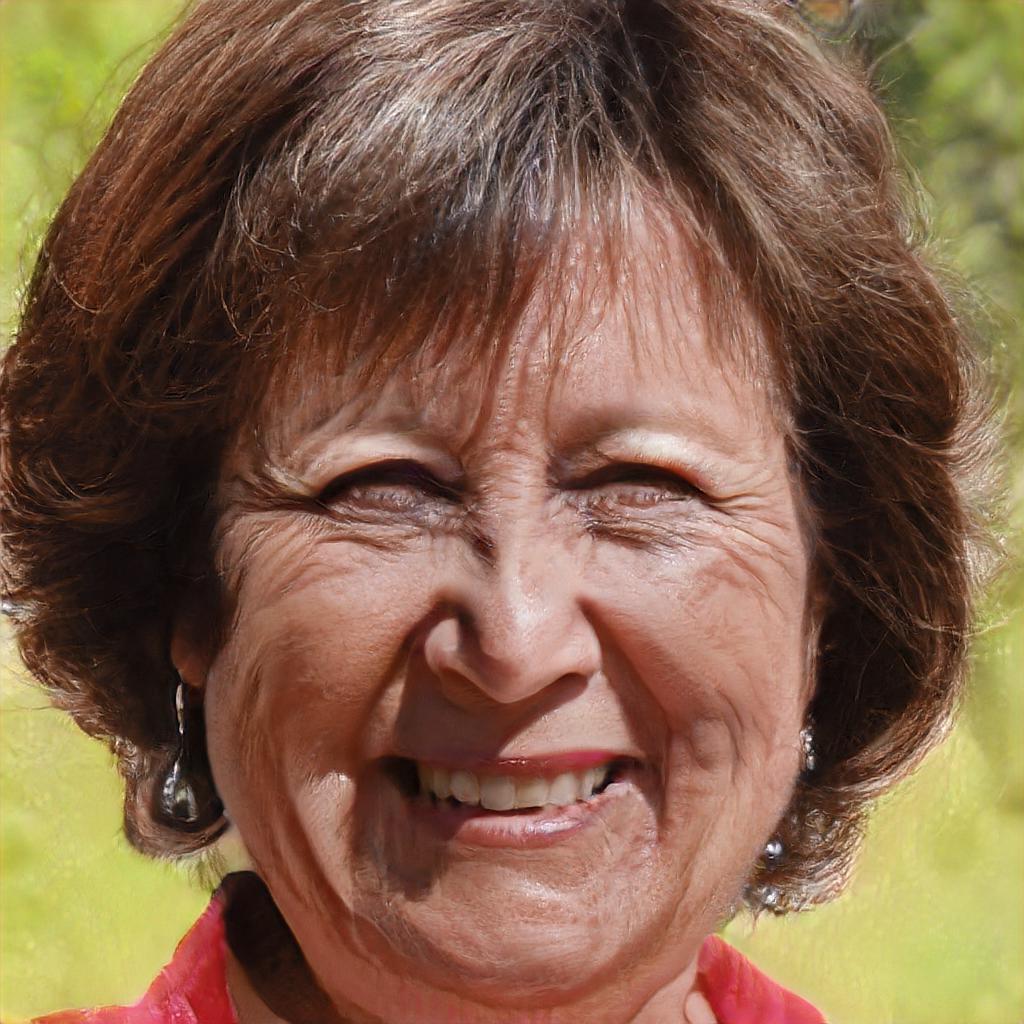} & \includegraphics[align=c,width=1.5cm]{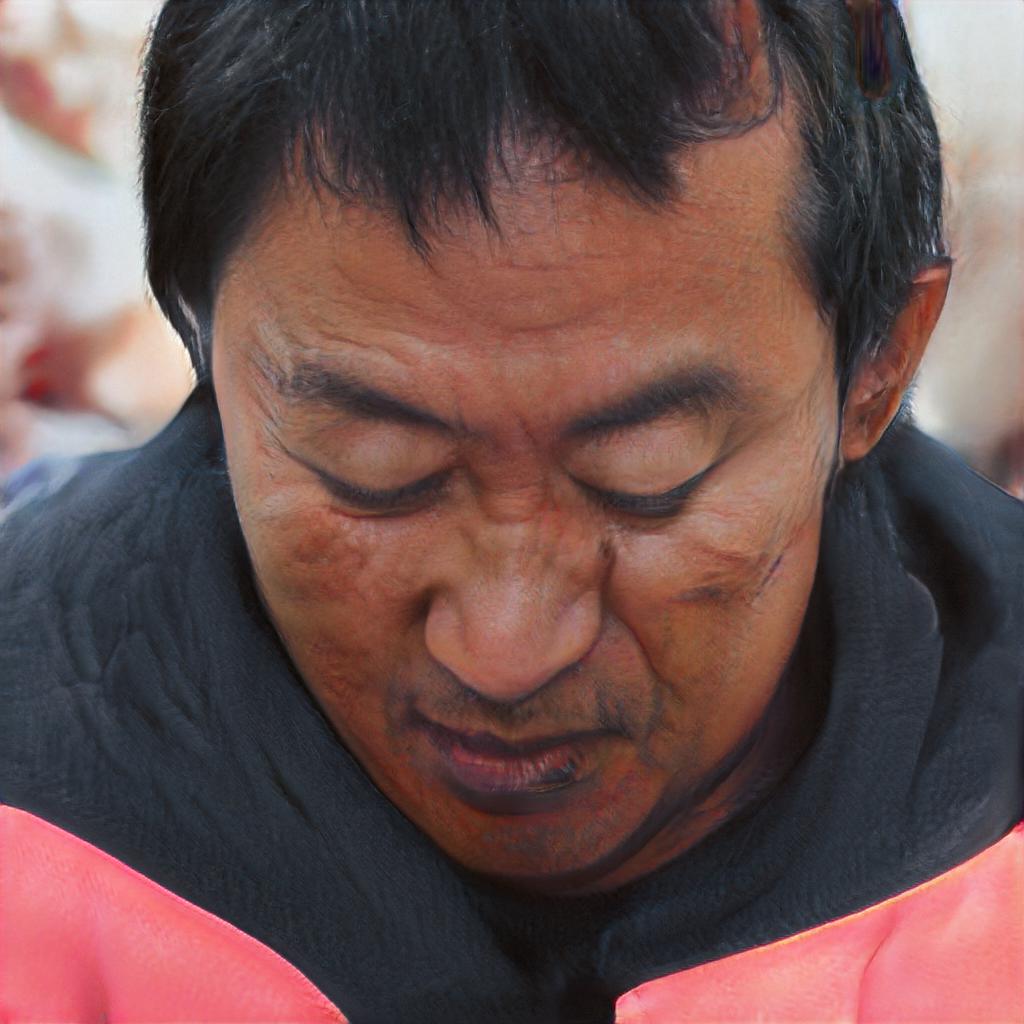} &  \includegraphics[align=c,width=1.5cm]{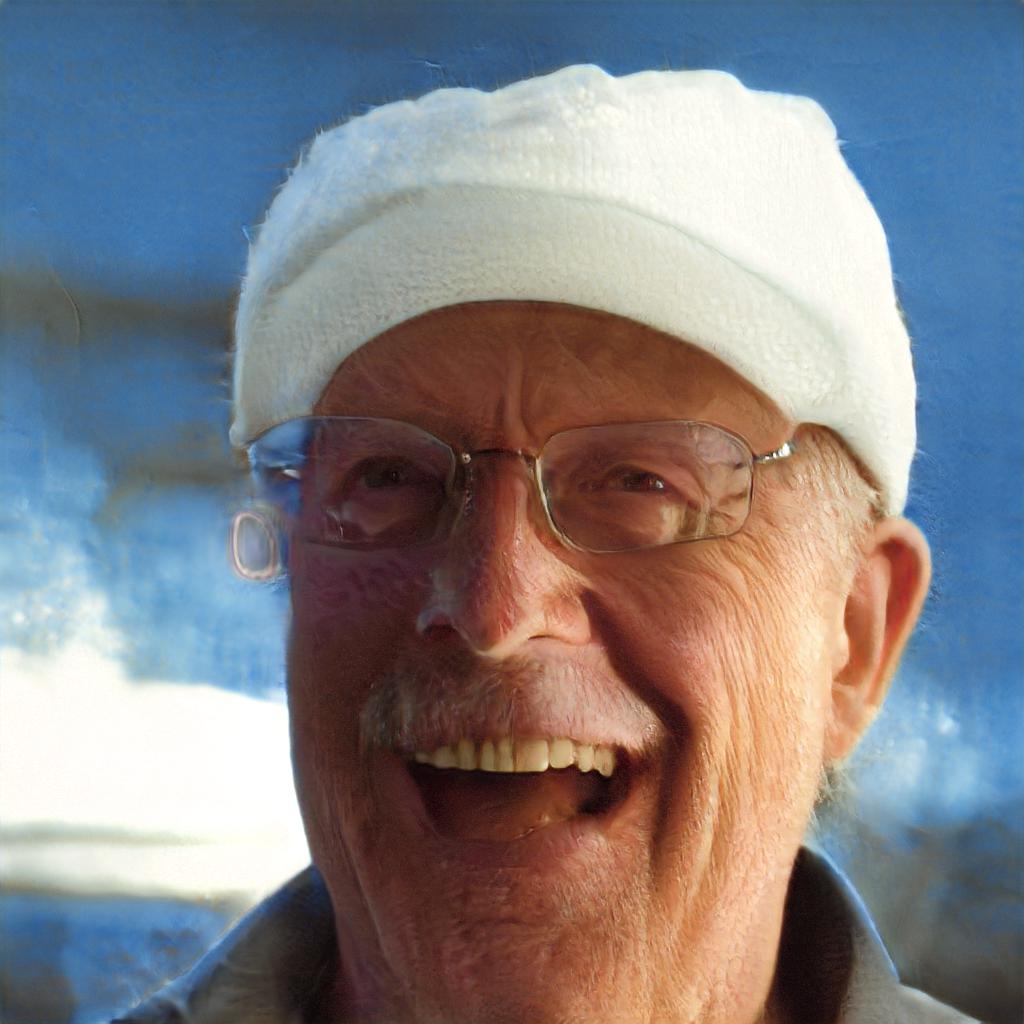} &
    \includegraphics[align=c,width=1.5cm]{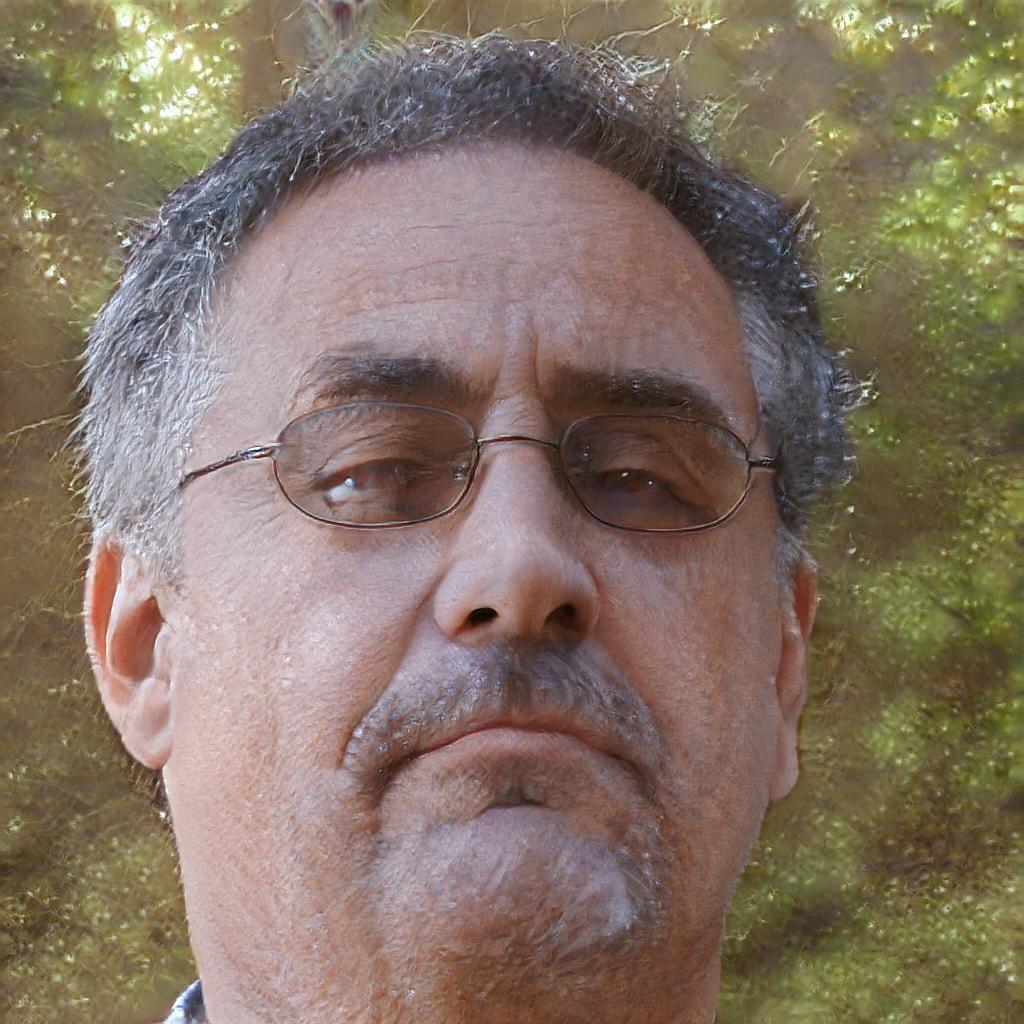} & \includegraphics[align=c,width=1.5cm]{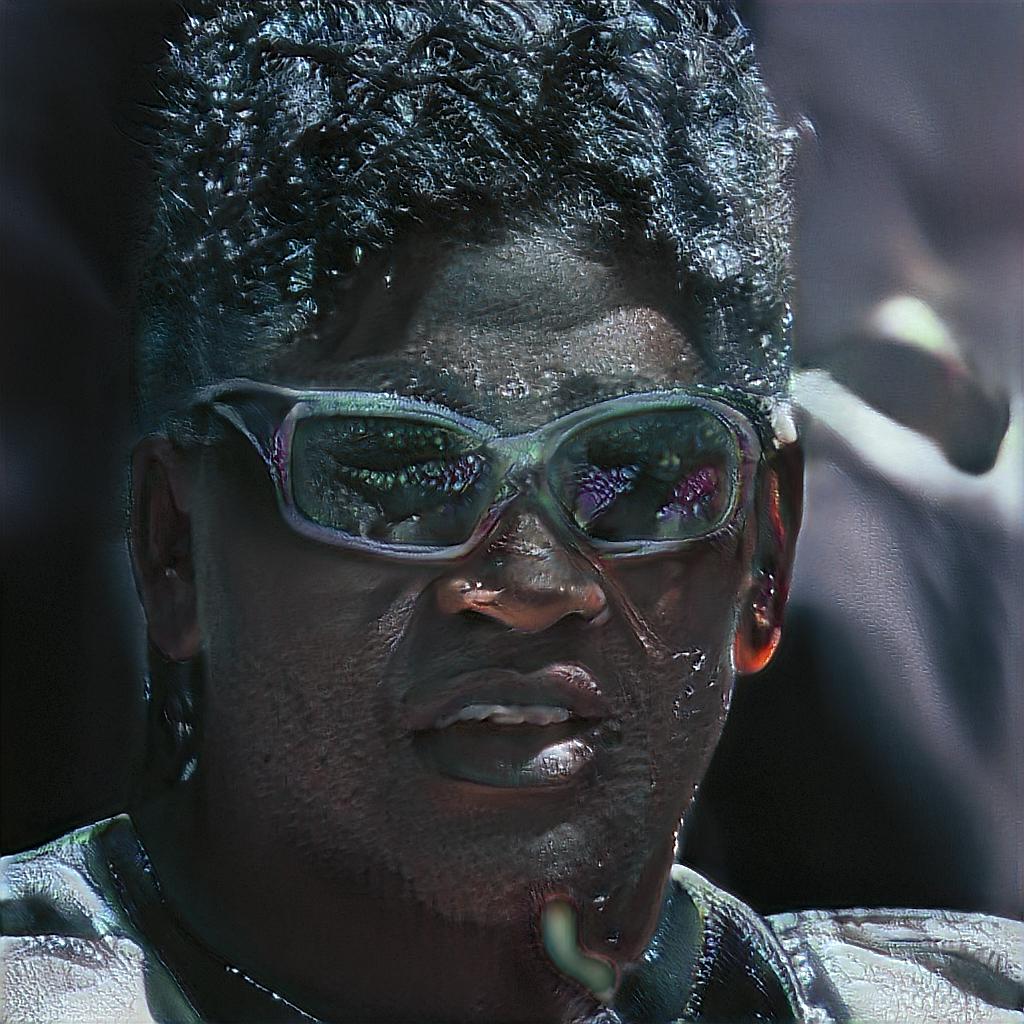} & \includegraphics[align=c,width=1.5cm]{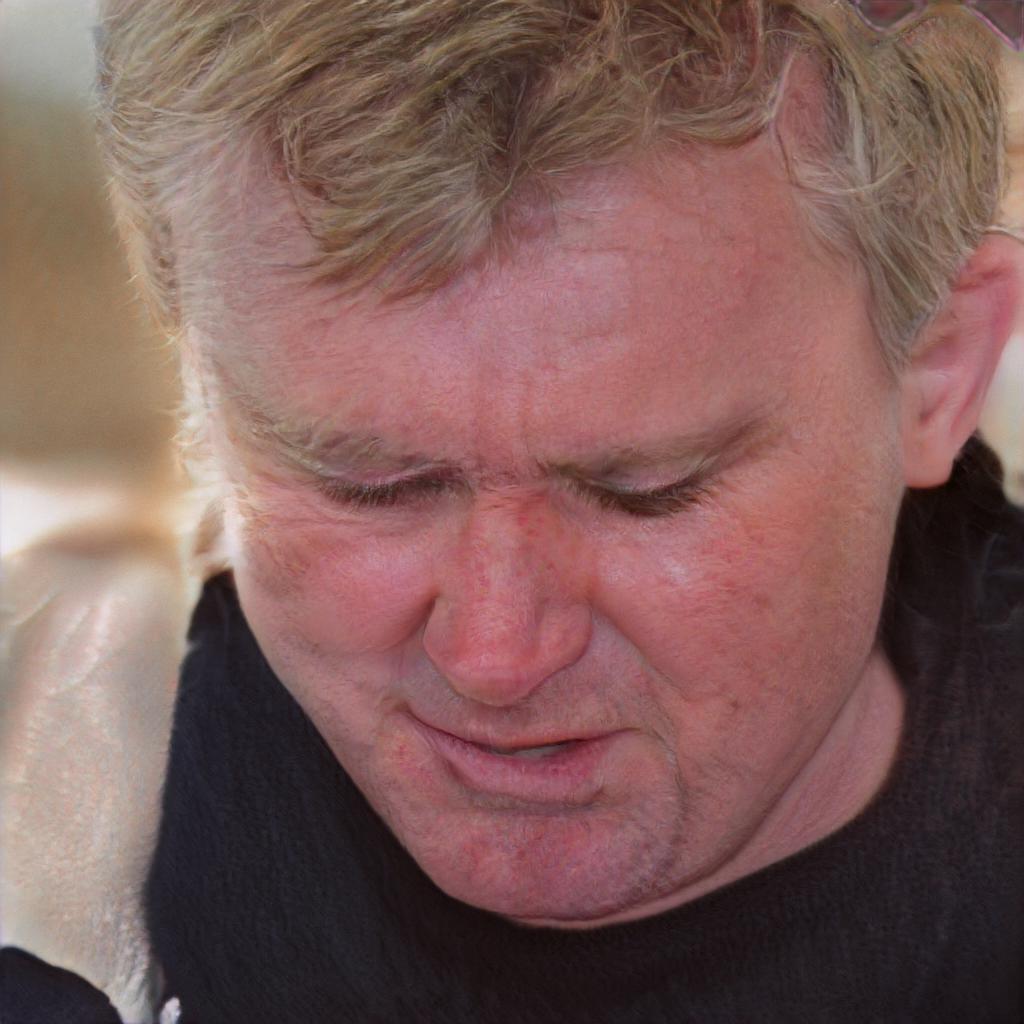} &  \includegraphics[align=c,width=1.5cm]{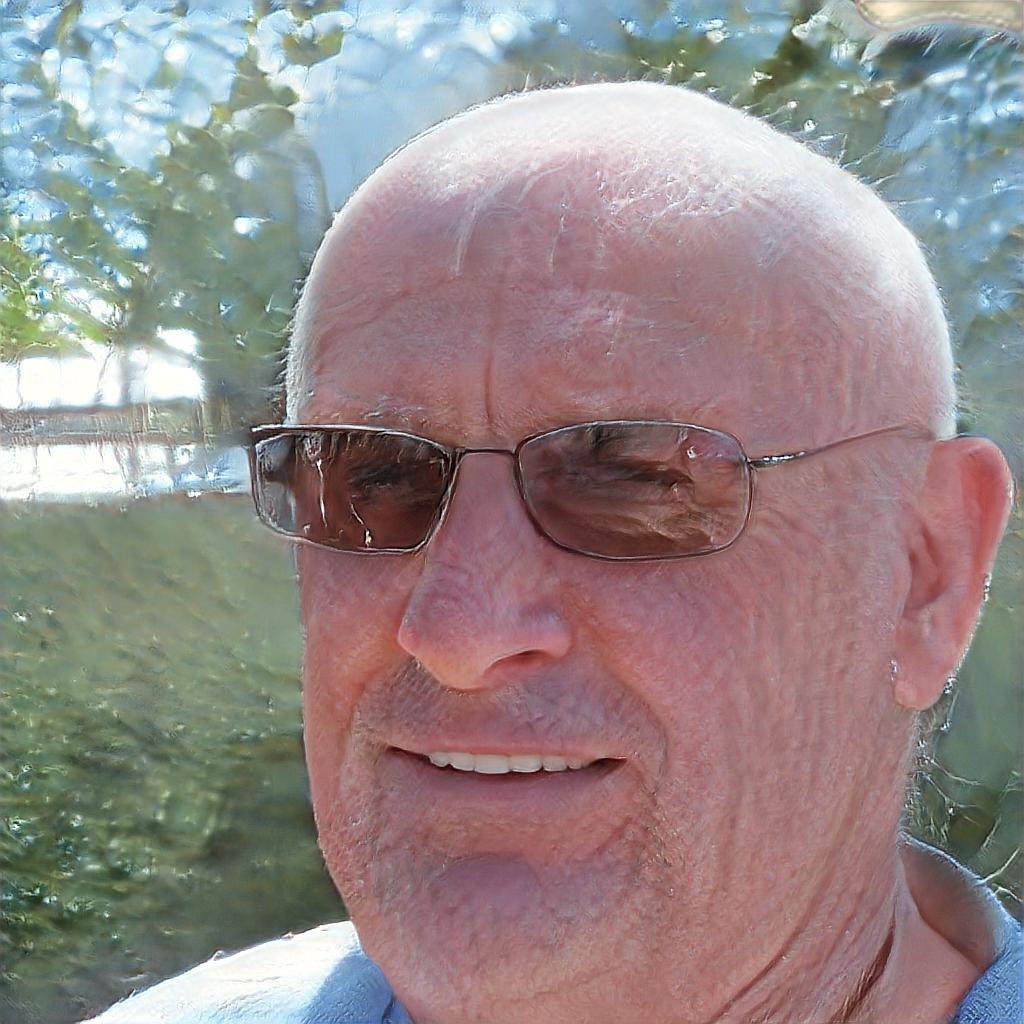} &
    \includegraphics[align=c,width=1.5cm]{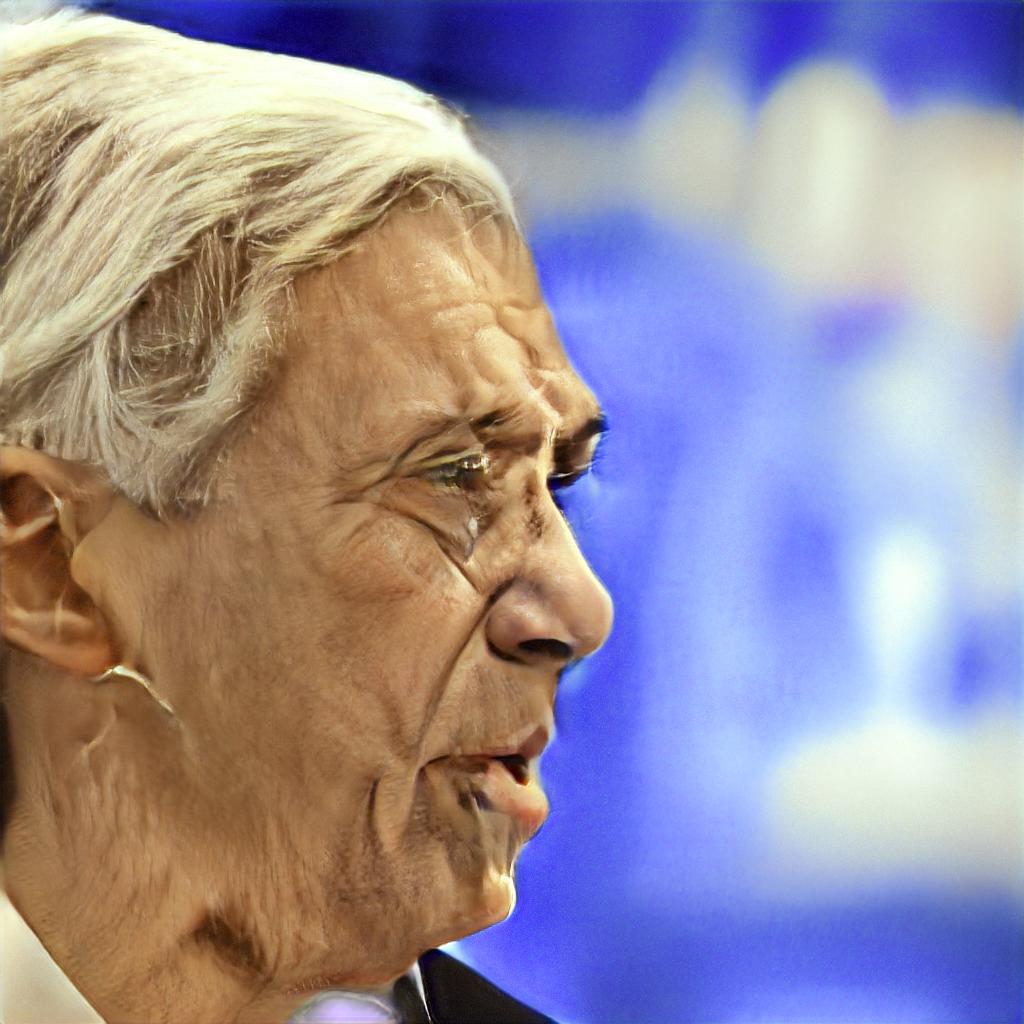} \\
    \small & $17.17\%$ & $4.75\%$ & $4.63\%$ & $4.45\%$ & $3.65\%$ & $3.23\%$ & $2.20\%$ & $2.10\%$ & $1.97\%$ \\
      (h) &
    \includegraphics[align=c,width=1.5cm]{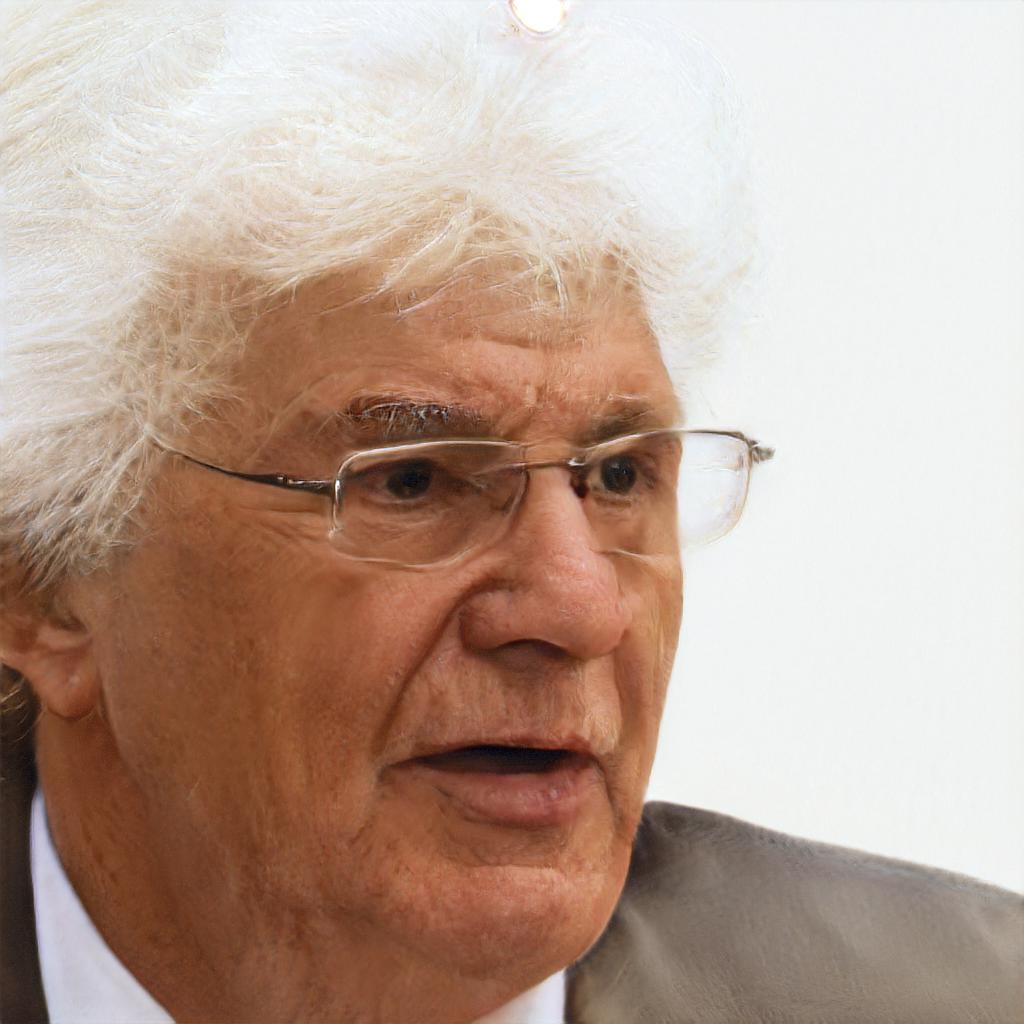} & \includegraphics[align=c,width=1.5cm]{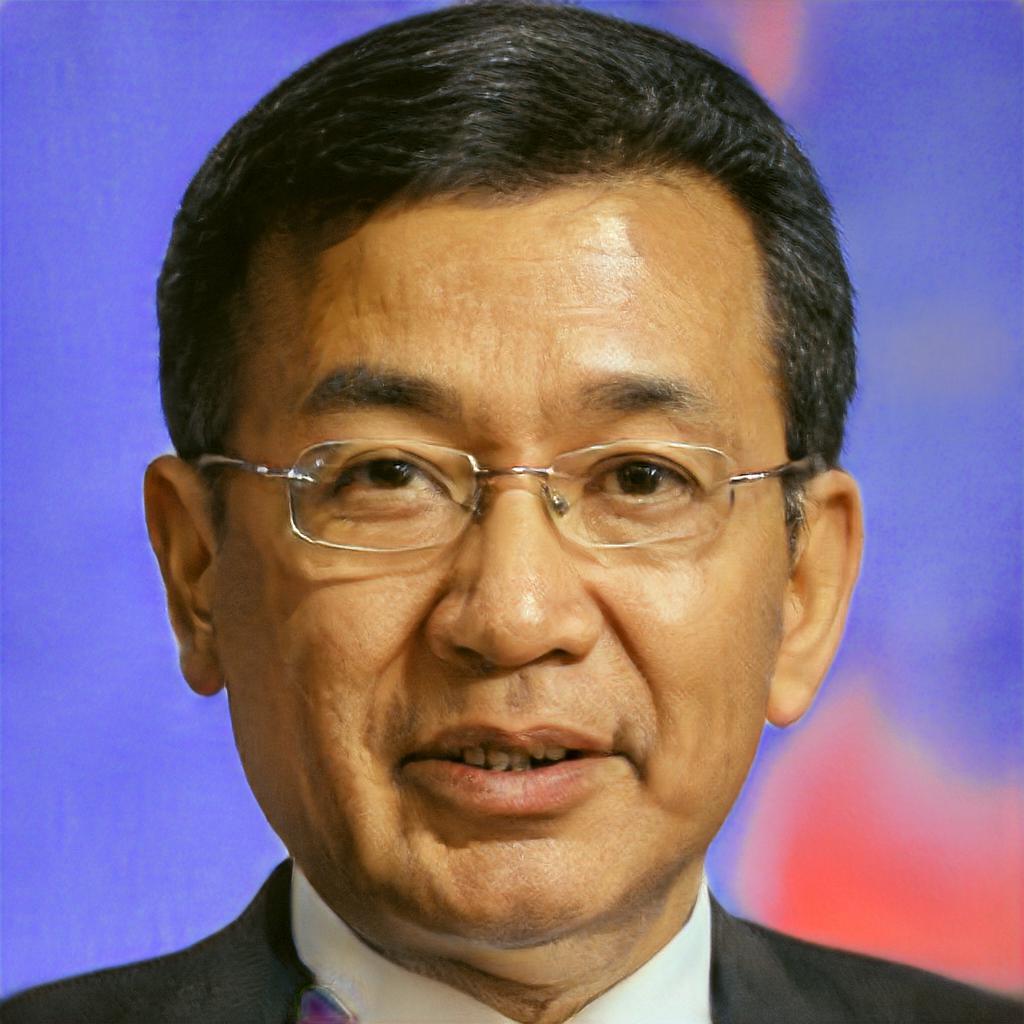} & \includegraphics[align=c,width=1.5cm]{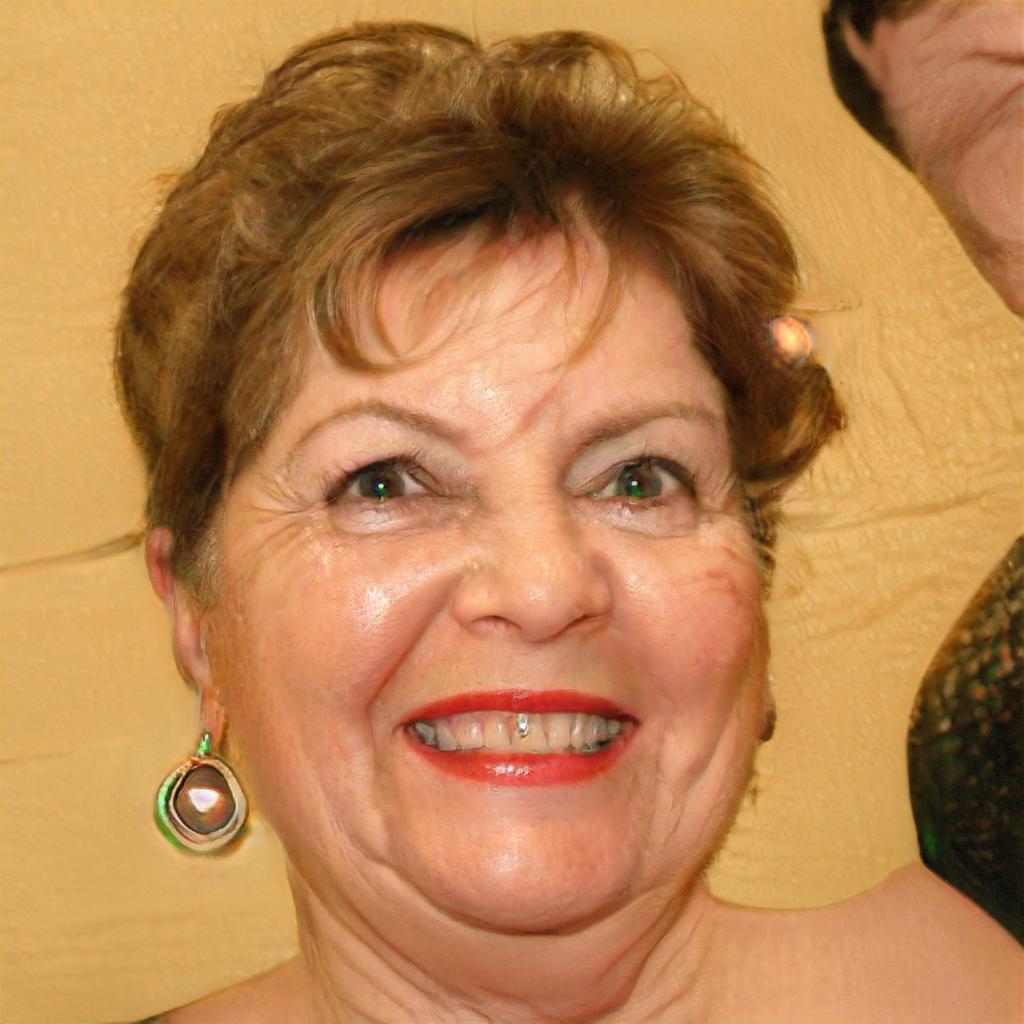} &      \includegraphics[align=c,width=1.5cm]{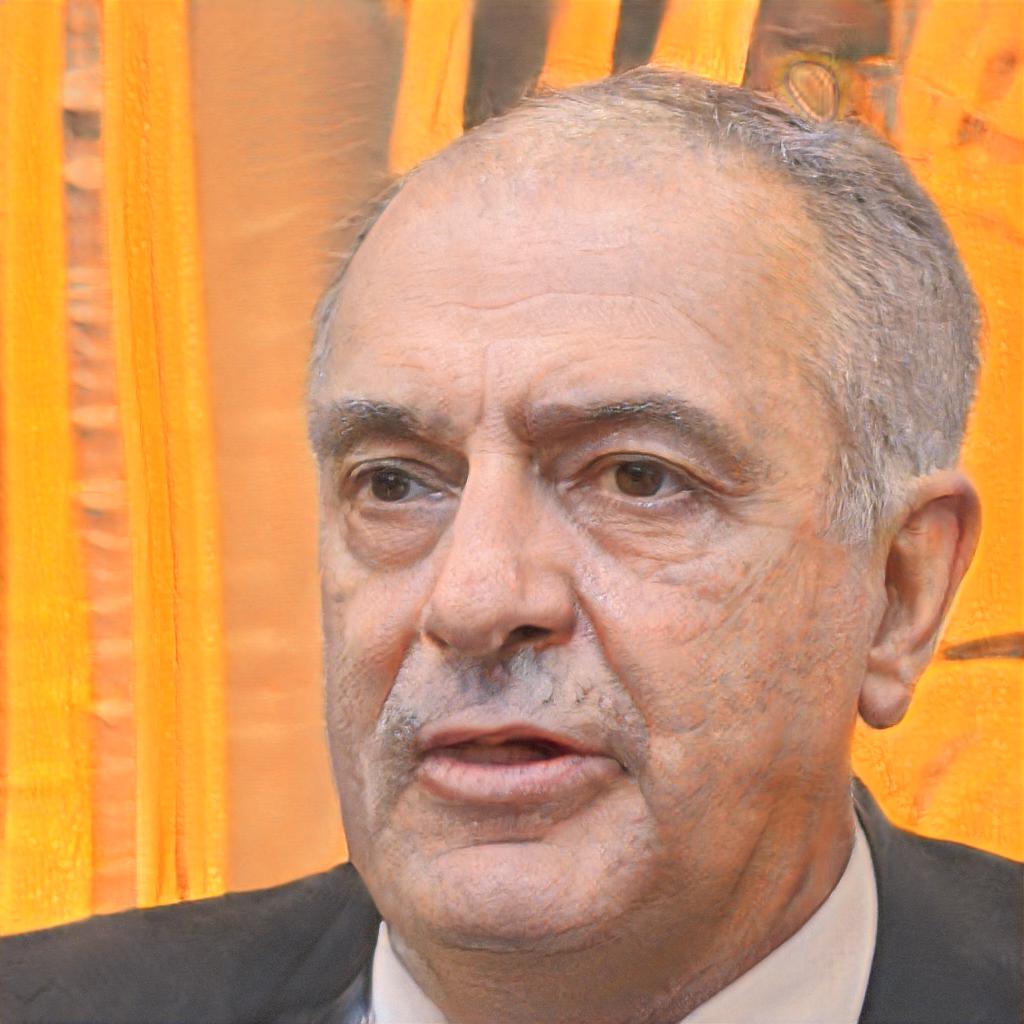} &   
    \includegraphics[align=c,width=1.5cm]{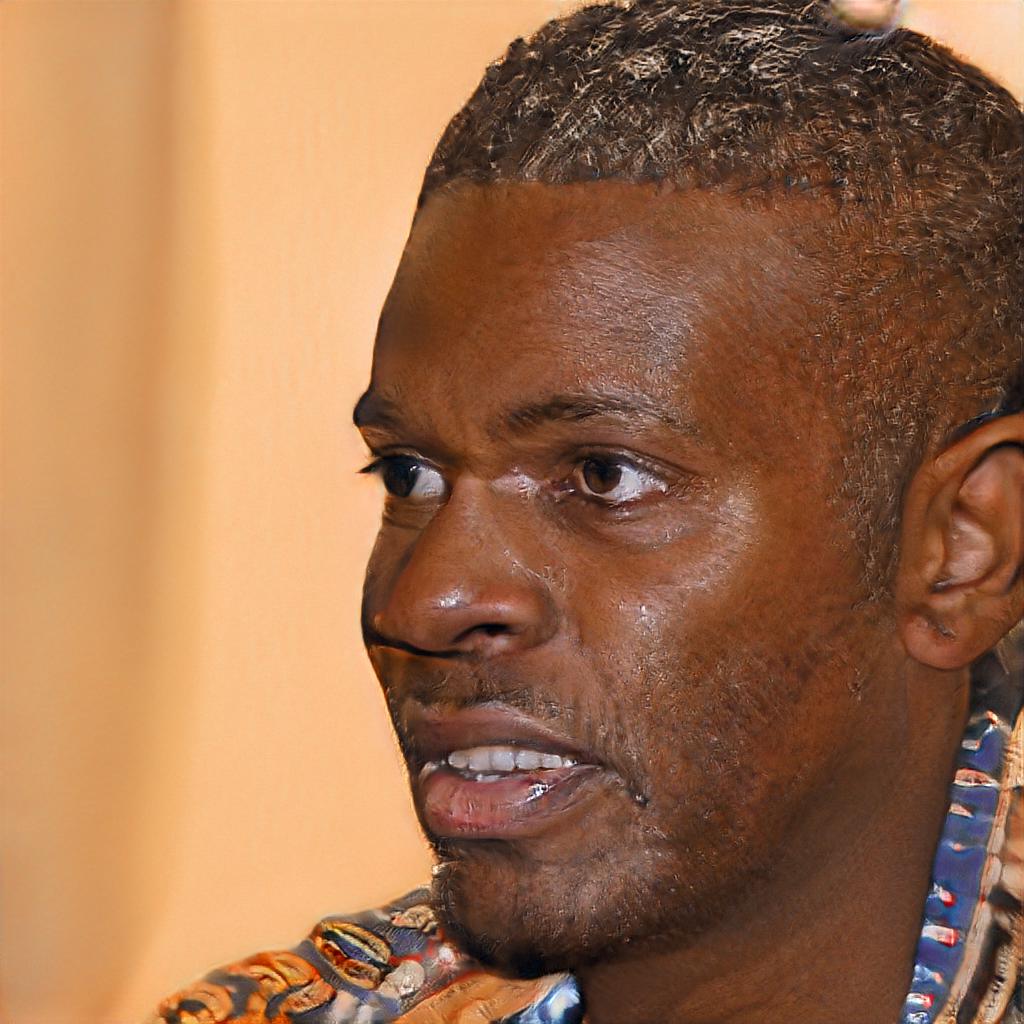} & \includegraphics[align=c,width=1.5cm]{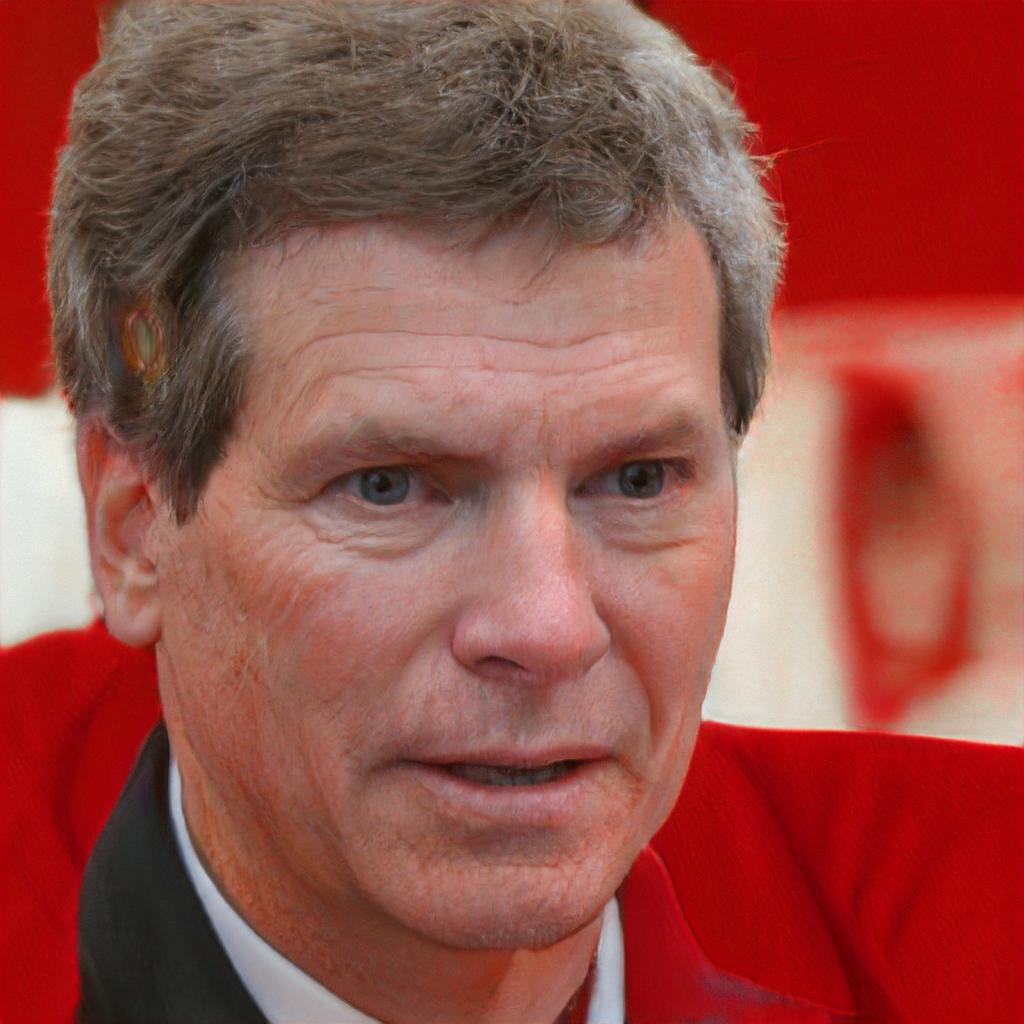} & \includegraphics[align=c,width=1.5cm]{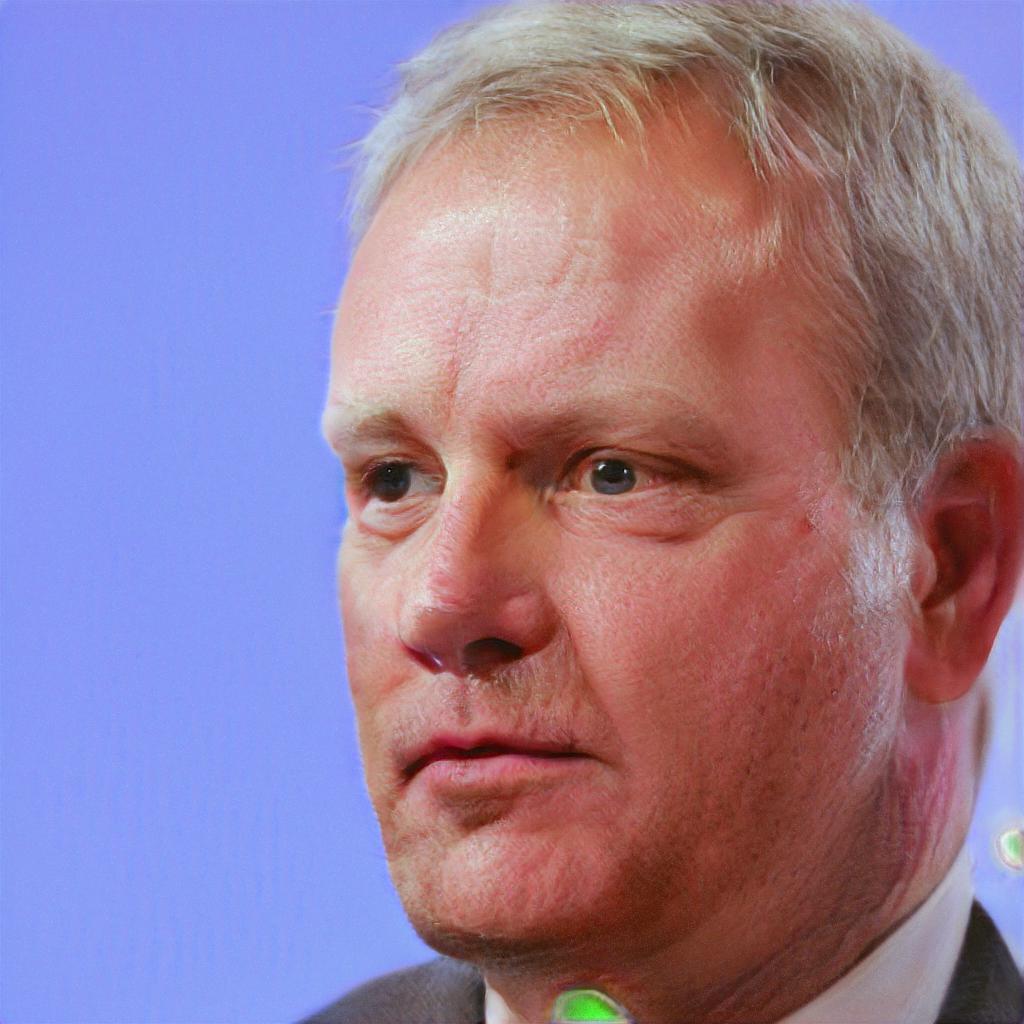} &     \includegraphics[align=c,width=1.5cm]{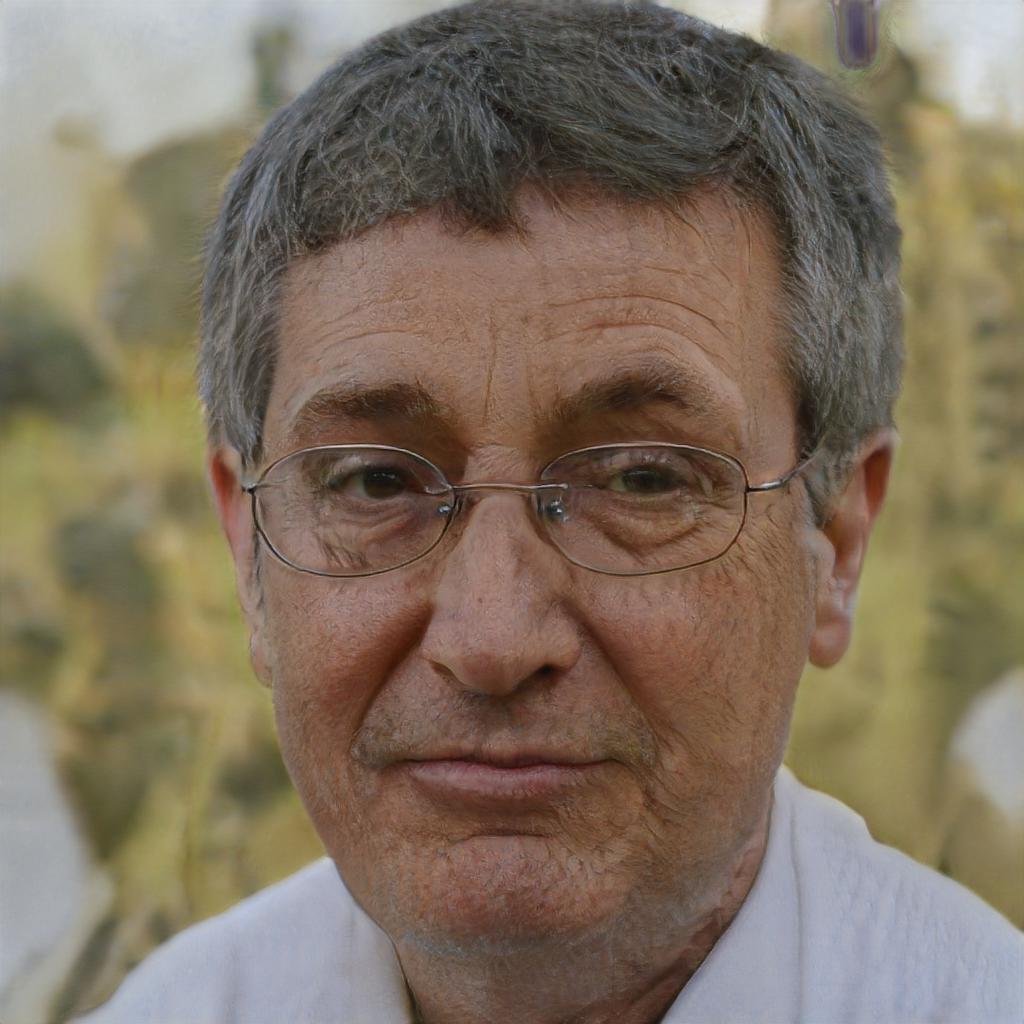} &   
    \includegraphics[align=c,width=1.5cm]{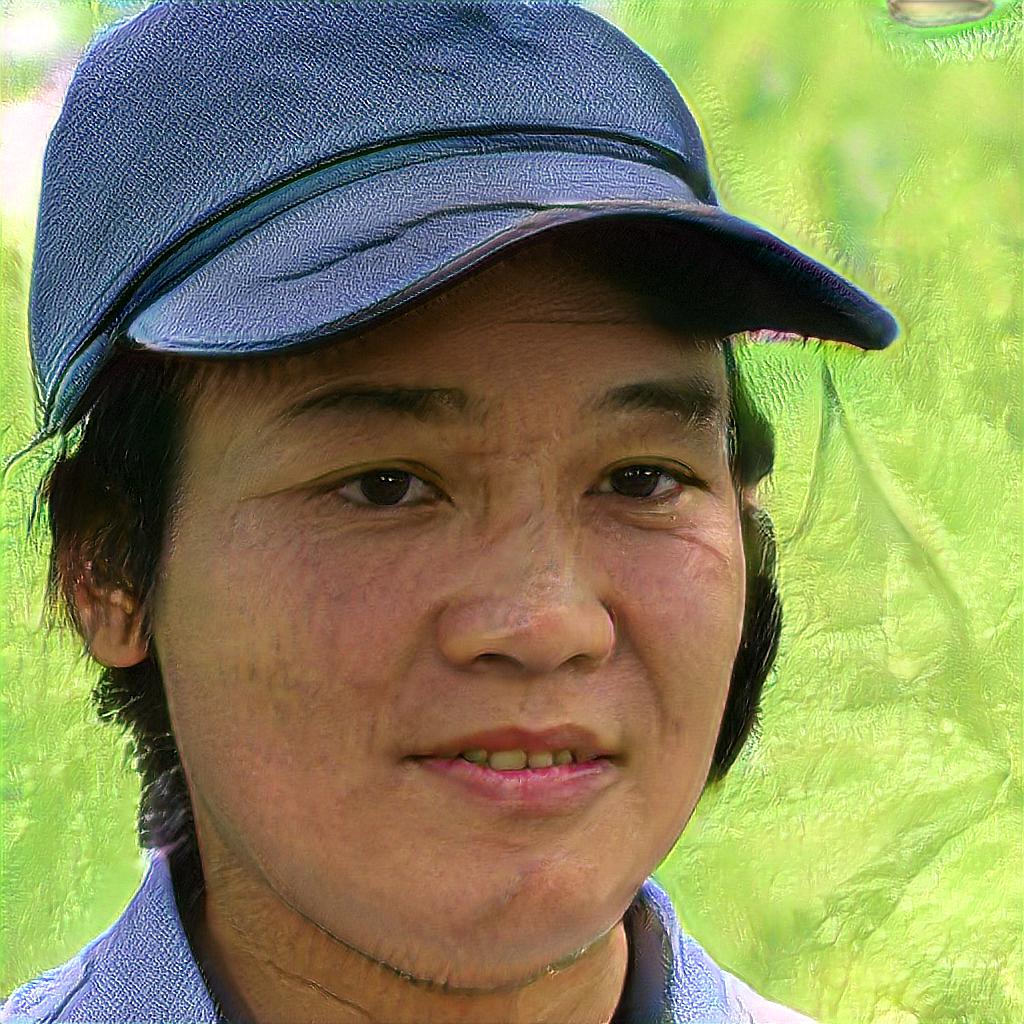} \\
      \small & $15.85\%$ & $5.05\%$ & $4.85\%$ & $4.00\%$ & $3.84\%$ & $3.67\%$ & $3.35\%$ & $1.61\%$ & $1.58\%$ \\
      (i) &
    \includegraphics[align=c,width=1.5cm]{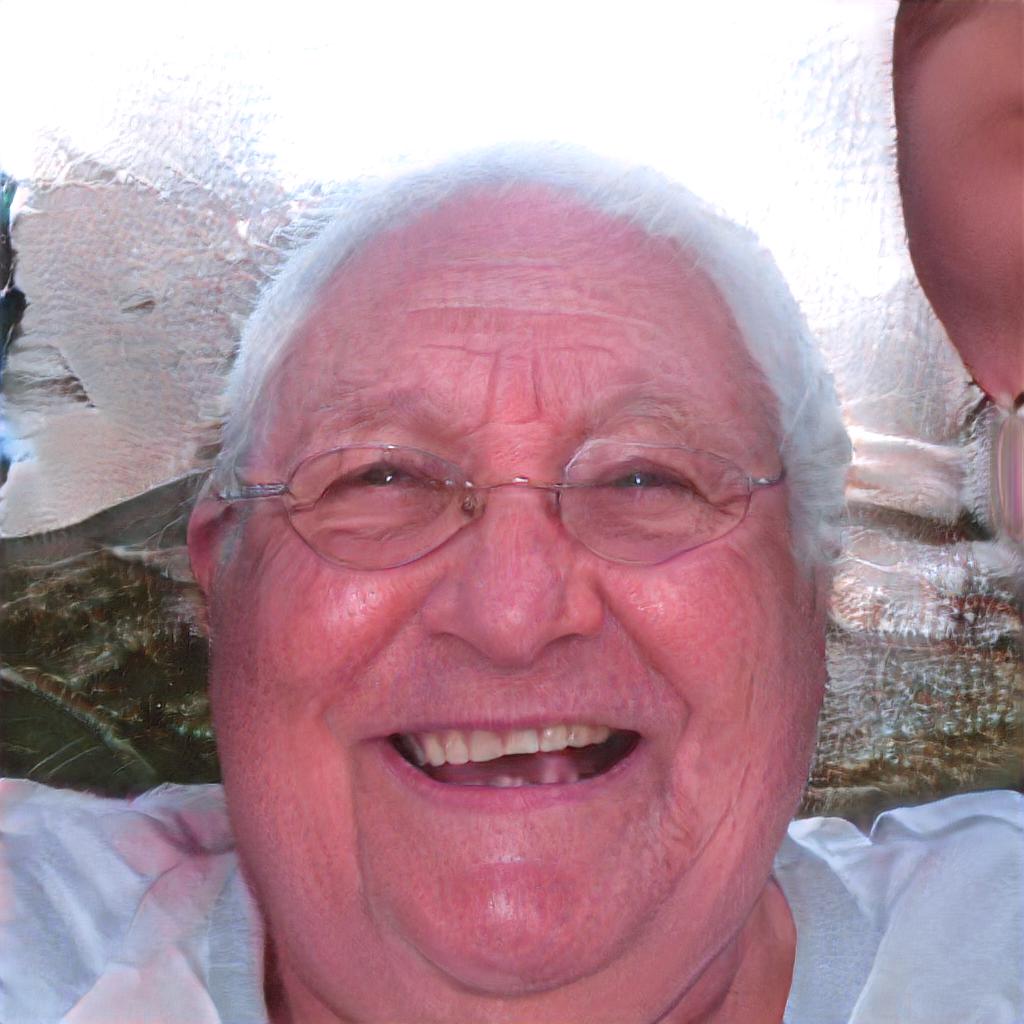} & \includegraphics[align=c,width=1.5cm]{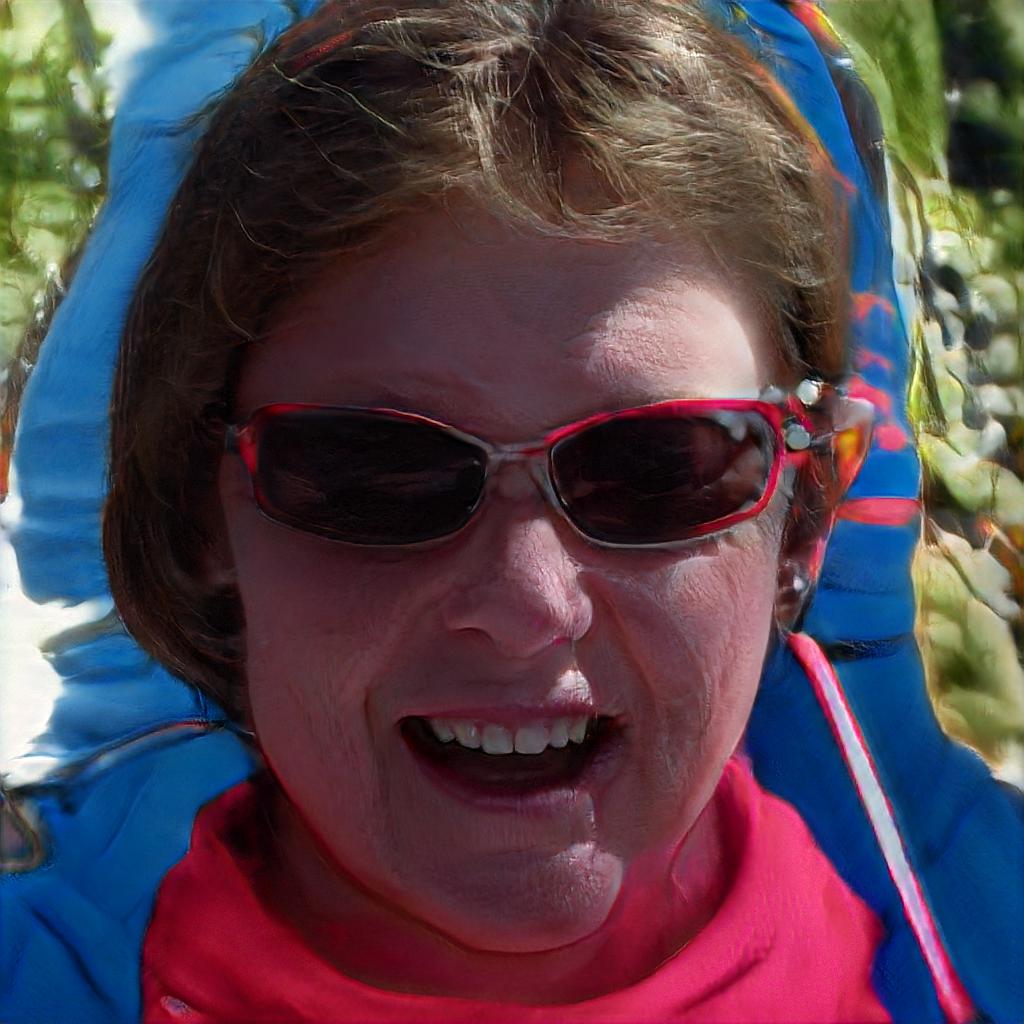} & \includegraphics[align=c,width=1.5cm]{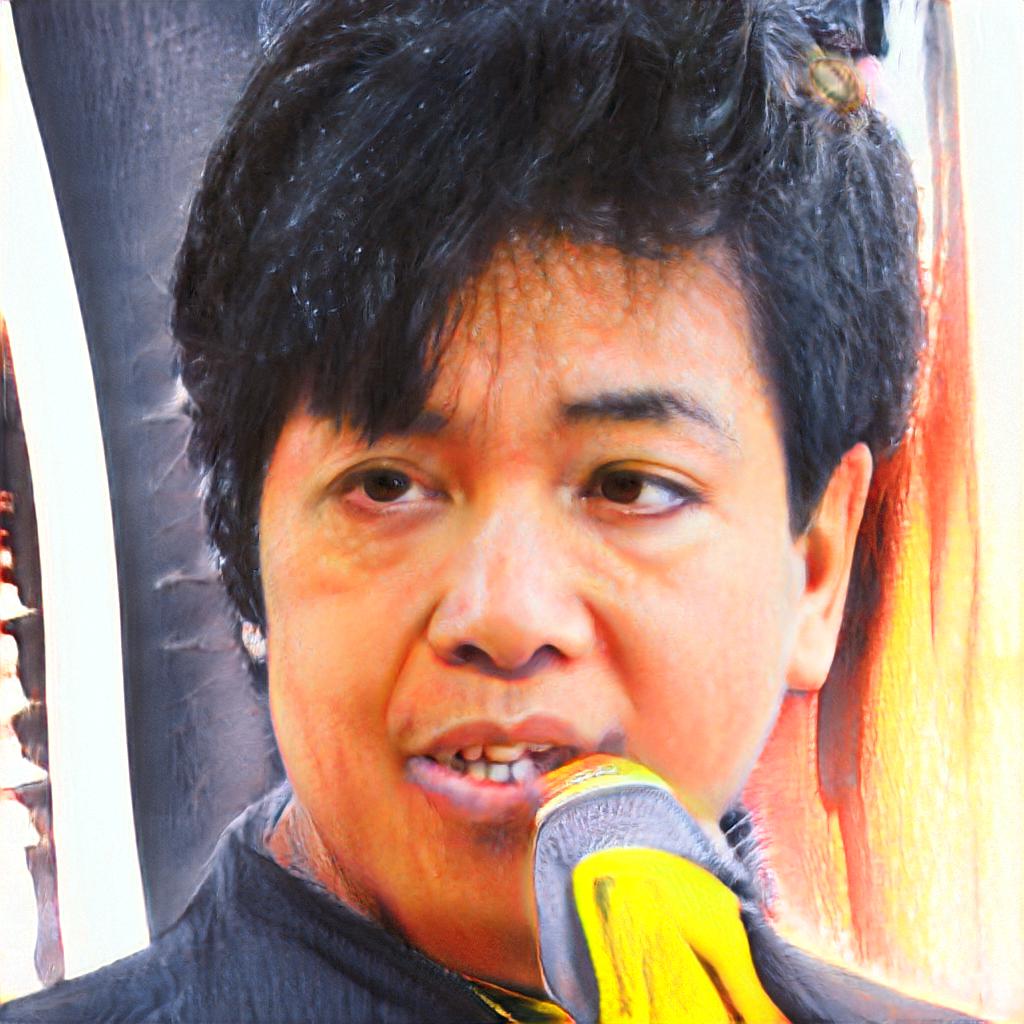} &      \includegraphics[align=c,width=1.5cm]{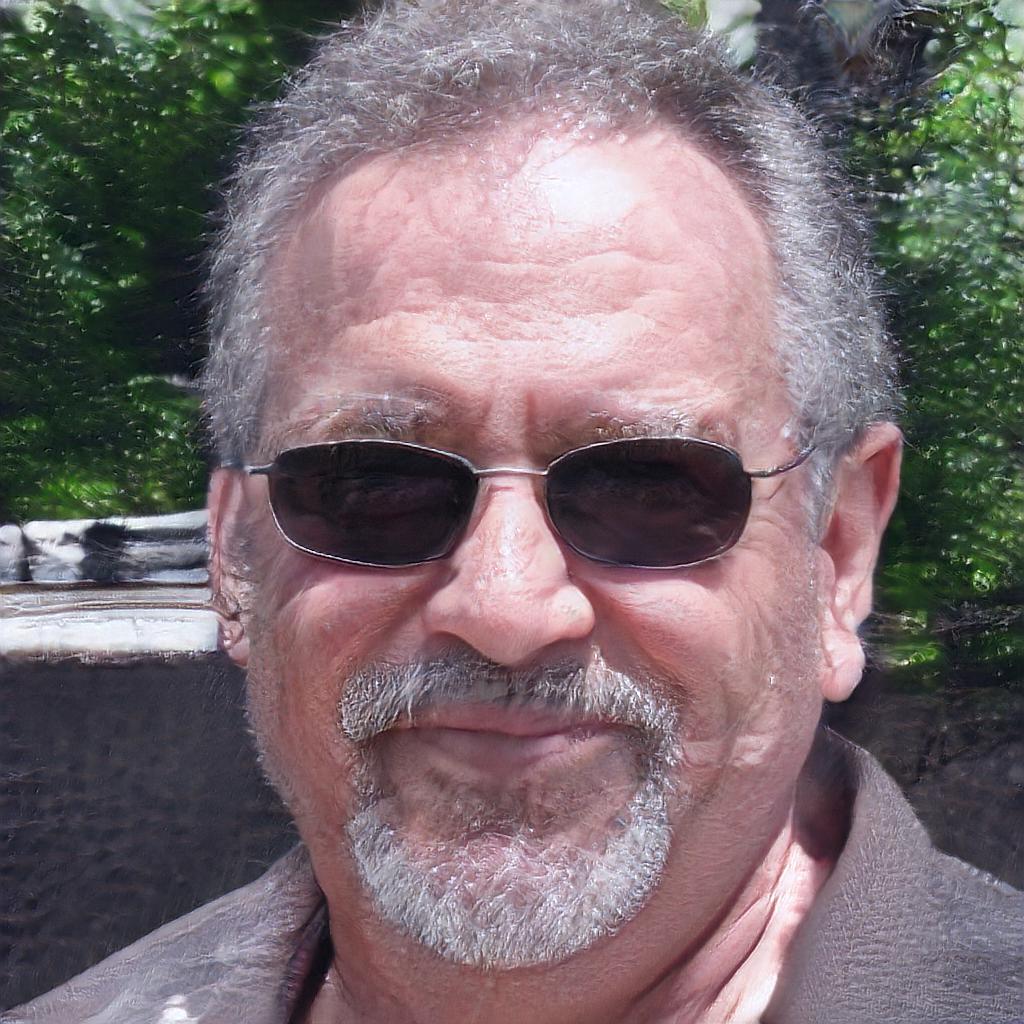} &   
    \includegraphics[align=c,width=1.5cm]{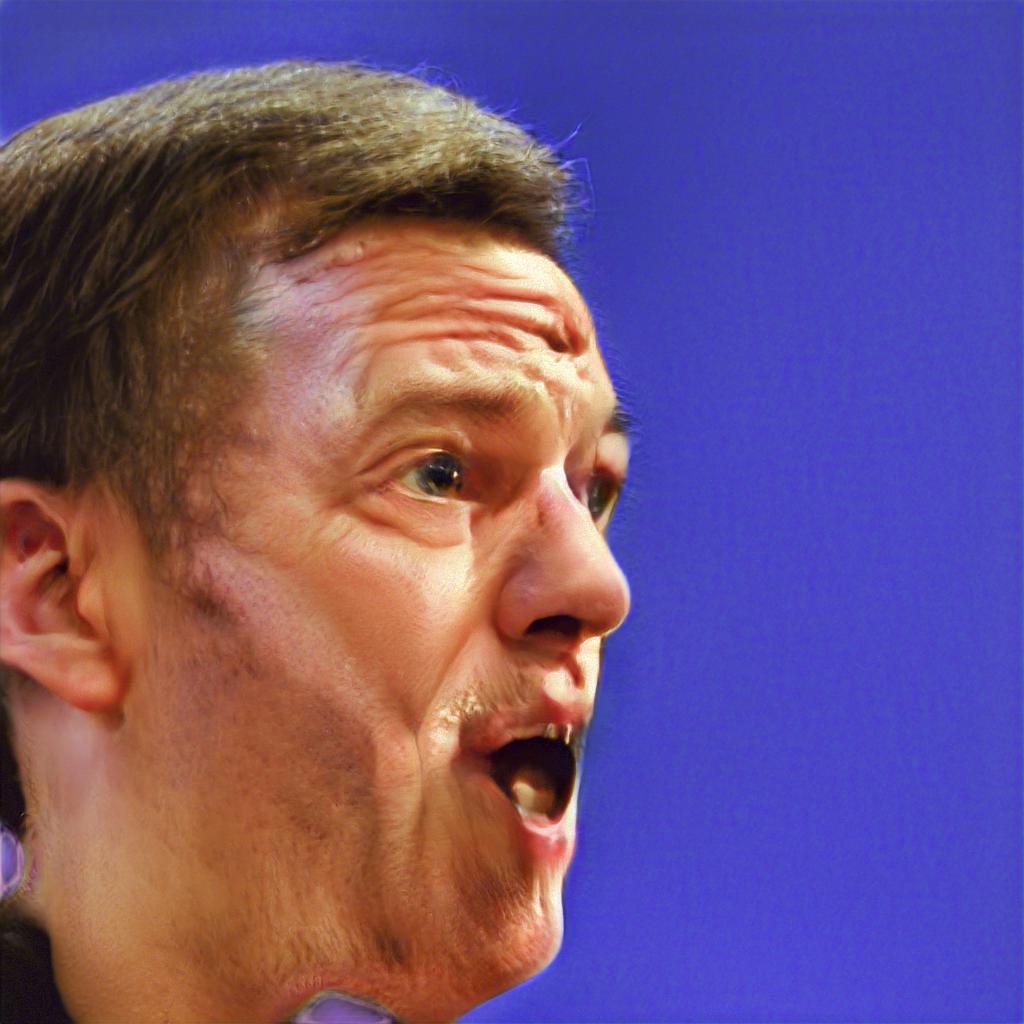} & \includegraphics[align=c,width=1.5cm]{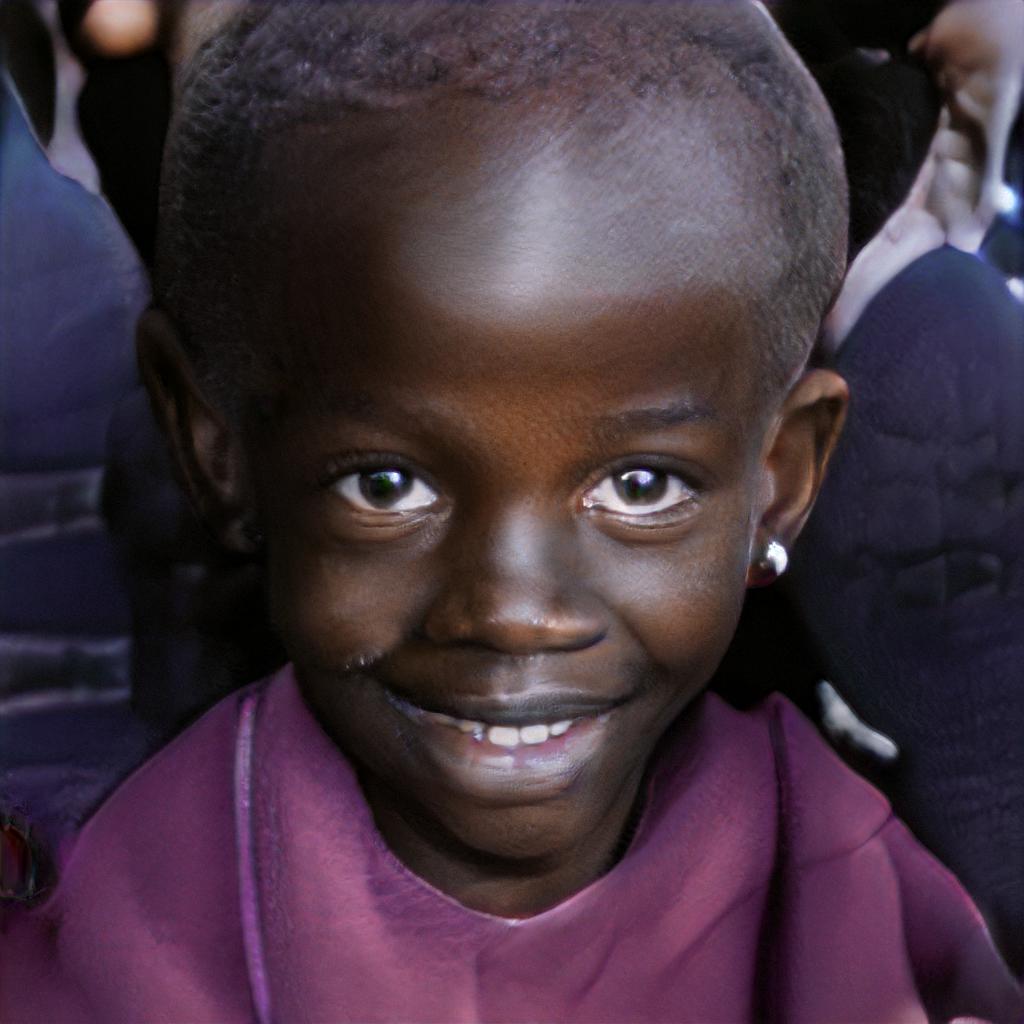} & \includegraphics[align=c,width=1.5cm]{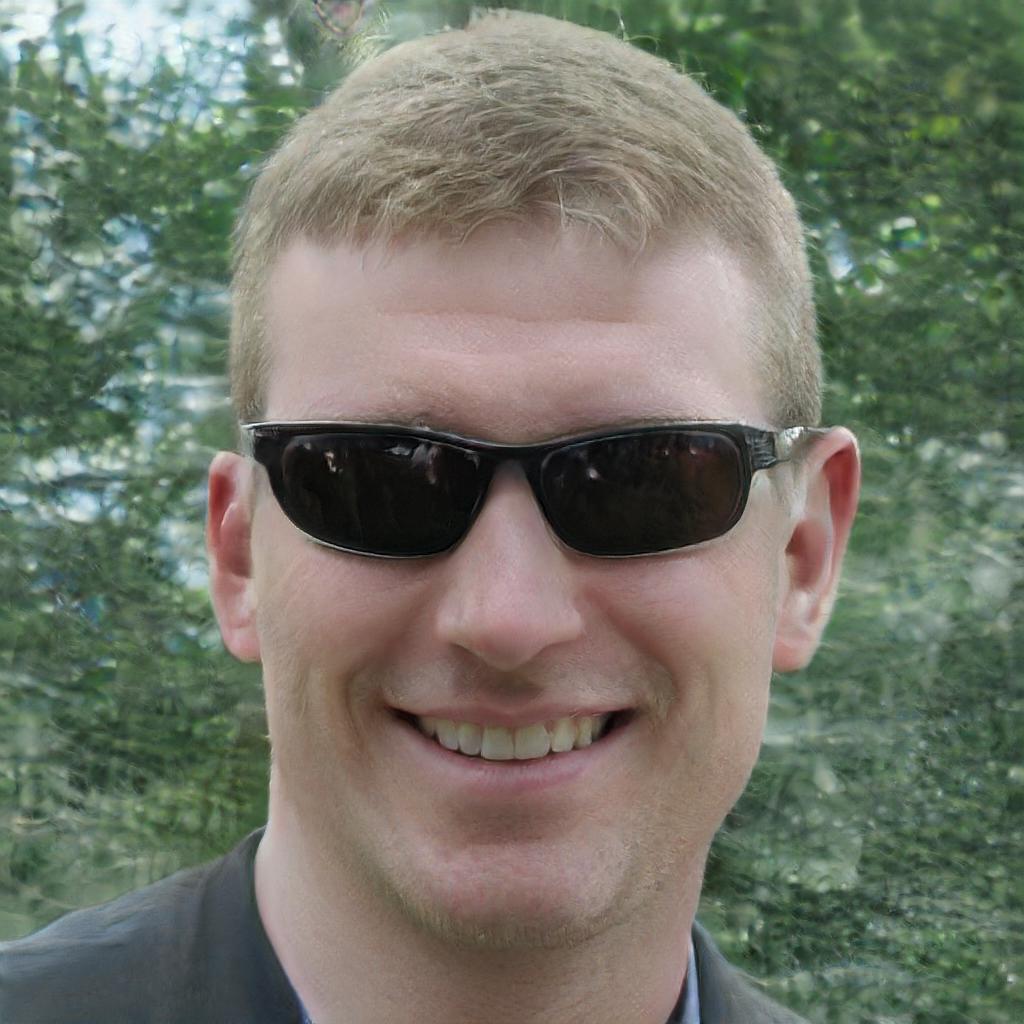} &      \includegraphics[align=c,width=1.5cm]{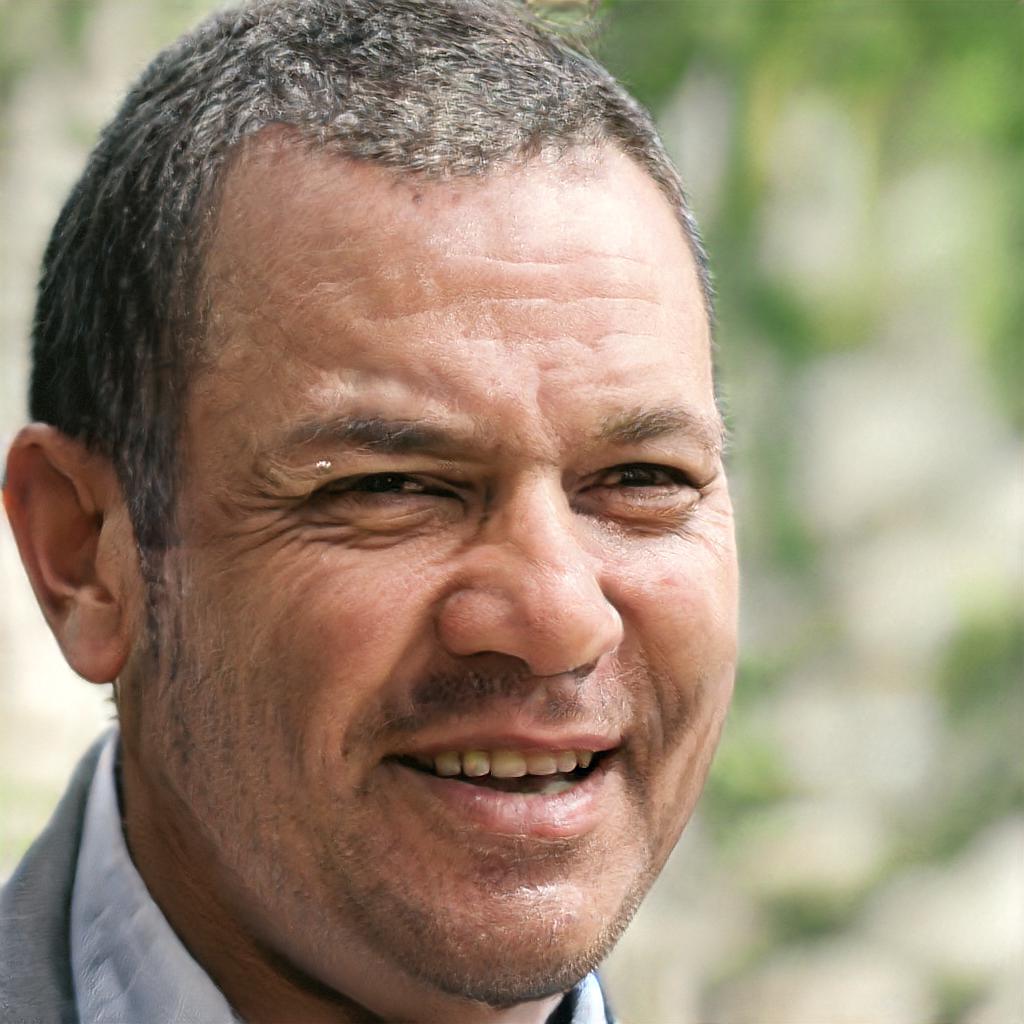} &   
    \includegraphics[align=c,width=1.5cm]{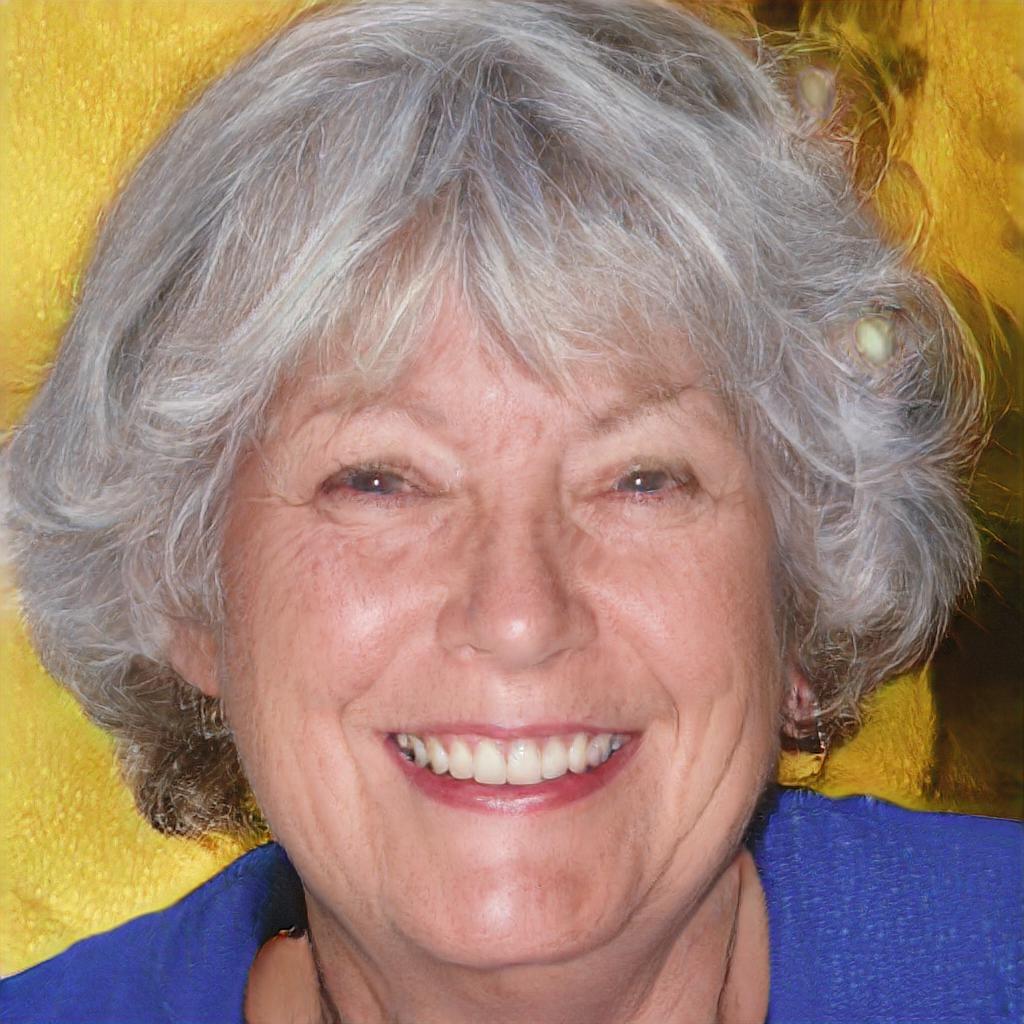} \\
      \small & $23.92\%$ & $7.60\%$ & $6.60\%$ & $6.00\%$ & $5.53\%$ & $5.34\%$ & $5.06\%$ & $2.80\%$ & $1.59\%$ \\
\end{tabular}
\caption{Set of nine master face images generated with each of the Coverage Search methods: LM-MA-ES Coverage Search on clustered data (a-c), greedy-Coverage Search methods (d-f) LM-MA-ES, (g-i) LM-MA-ES+Success Predictor. Embedded using either (a,d,g) SphereFace, (b,e,h) FaceNet ,or the (c,f,i) Dlib face descriptors. Below each image, the MSC score for that image is listed.}\label{fig:cover_image}
\end{figure*}

\subsection{Dataset Coverage}
\label{subsec:dataset_covarage}
We run Alg.~\ref{alg:greedy} in order to find a minimal number of images that match the largest number of faces in the entire dataset $D$, where $|D|=5,749$. 

As a baseline, we clustered the embedding faces to nine clusters. Based on these clusters, we split the dataset into nine disjoint datasets. We run the single image generation method (Fig~\ref{fig:alg_flow}) on each of the datasets. Our $greedy$ method (Alg.~\ref{alg:greedy}) is also run for nine iterations in order to generate a number of faces that is equal to the number of clusters. We run it twice, once with the LM-MA-ES optimization method and once with our LM-MA-ES+Success Predictor.

Table~\ref{table:ds_cover} lists the results. Evidently, the greedy method outperforms the per-cluster solution. The greedy method is able to create nine master face images that cover 42\%-64\% of the dataset and by applying the Success Predictor the results slightly improve for all three face descriptors. The coverage results for FaceNet and SphereFace models are lower than for the Dlib. The higher embedding dimension of these two face descriptors allows better separability between the faces, which in turn makes the coverage task more difficult. 
In Fig.~\ref{fig:cover_image} nine master face images are presented for each of the face descriptors, dataset $greedy$-Coverage Search algorithms as well as the MSC score of each of the master faces.

\section{CONCLUSIONS AND FUTURE WORKS}

Our results imply that face-based authentication is extremely vulnerable, even if there is no information on the target identity. This is true for all three face recognition methods, with some differences in the obtained success probability.

Interestingly, the obtained faces are not blurry and their pose is mostly frontal. There is a tendency to observe older faces in the generated images. Since the face recognition methods are trained on faces from different ages, the representation has some age invariance. It is possible that the methods make use of this fact. However, we do not observe such a tendency toward facial hair or glasses. 

{We further note that according to \cite{han2014age}, the group of 60+ years-old white males is the third most common group in the LFW dataset, while younger groups of white males (40-60, 20-40) are even more common. Nevertheless, the group of 60+ years-old white males is usually less varied in comparison to the younger groups, so a single older master face can cover a larger portion of its group. As the iterations of the coverage algorithm proceed, the model covers less represented groups in the dataset since each iteration is performed on the reduced dataset. Eventually master faces of several ethnicities and several ages are generated. The lower number of female faces out of the nine master faces generated by our method, matches the much lower frequency of female faces (22\%) in the LFW dataset according to \cite{han2014age}.}

{In order to provide a more secure solution for face recognition systems, anti-spoofing~\cite{antispoofing_survey} methods are usually applied. Our method might be combined with additional existing methods to bypass such defenses. For example, DeepFake methods \cite{DeepFake_attack} can be used to animate the generated master faces and overcome liveness detection methods. We leave this research extension as future work.}

It is also interesting to explore the possibility of using master faces faces in order to help protect the face recognition systems against dictionary attacks, as well as a way to reduce the overall false positive rate. Anecdotally, although unreported in publications as far as we can ascertain, the same set of faces appear multiple times in the false matches.

\section{ACKNOWLEDGMENTS}
This project has received funding from the European Research Council (ERC) under the European Unions Horizon 2020 research and innovation programme (grant ERC CoG 725974). This research was partially
supported by The Yandex Initiative for Machine Learning.
{\small
\bibliographystyle{ieee}
\bibliography{faceevolution}
}

\end{document}